\documentclass[lettersize,journal]{IEEEtran}
\usepackage{amsmath,amsfonts}
\usepackage{algorithmic}
\usepackage{algorithm}
\usepackage{array}
\usepackage[caption=false,font=normalsize,labelfont=sf,textfont=sf]{subfig}
\usepackage{textcomp}
\usepackage{stfloats}
\usepackage{url}
\usepackage{verbatim}
\usepackage{graphicx}
\usepackage{cite}
\usepackage{color}
\usepackage{enumitem}
\usepackage{array}
\usepackage{booktabs}

\usepackage{multirow}
\usepackage{float}

\usepackage {rotating}

\usepackage{bbding}

\usepackage{amssymb}

\usepackage{amsmath} 
\allowdisplaybreaks[4]

\newcommand{\subsubsubsection}[1]{\paragraph{#1}}
\begin{document}

% \title{Towards Socialized Learning in Edge Intelligence: A Contemporary Survey}
% \title{Integrating Socialized Learning into Edge Networks: A Contemporary Survey}
% \title{Socialized Learning in Edge Intelligence \\ -Empowered Communication Systems and Networks: A Contemporary Survey}
\title{Socialized Learning: A Survey of the Paradigm Shift for Edge Intelligence in Networked Systems}

%\thanks{This paper was produced by the IEEE Publication Technology Group. They are in Piscataway, NJ.}% <-this % stops a space
%\thanks{Manuscript received April 19, 2021; revised August 16, 2021.}
\author{
{Xiaofei Wang,~\IEEEmembership{Senior Member,~IEEE,}
Yunfeng Zhao,~\IEEEmembership{Student Member,~IEEE,}
        Chao Qiu,~\IEEEmembership{Member,~IEEE,}
        \\
        Qinghua Hu,~\IEEEmembership{Senior Member,~IEEE,}
        Victor C. M. Leung,~\IEEEmembership{Life Fellow,~IEEE}
        }% <-this % stops a space
% XXXX\\
% YYYY
\IEEEcompsocitemizethanks{

This work was supported in part by the National Science Foundation of China under Grant 62072332; in part by the Beijing-Tianjin-Hebei Basic Research Cooperation Special Project, Research on Key Technologies for Efficient Crowd Intelligence Understanding and Situation Deduction in Intelligent Connected Vehicle Environments under Grant F2024201070; in part by the National Natural Science Foundation of China under Grant No. U23B2049; in part by the Tianjin Natural Science Foundation General Project under Grant 23JCYBJC00780; and in part by the Tianjin Xinchuang Haihe Lab under Grant 22HHXCJC00002.
\emph{(Corresponding authors: Chao Qiu; Qinghua Hu.)}

\IEEEcompsocthanksitem Xiaofei Wang, Yunfeng Zhao, Chao Qiu and Qinghua Hu are with the College of Intelligence and Computing, Tianjin University, Tianjin 300072, China. (e-mail: xiaofeiwang@tju.edu.cn; yfzhao97@tju.edu.cn; chao.qiu@tju.edu.cn; huqinghua@tju.edu.cn).

% Victor C. M. Leung is with the College of Computer Science and Software Engineering, Shenzhen University, Shenzhen 518060, China, and also with the Department of Electrical and Computer Engineering, The University of British
% Columbia, Vancouver V6T 1Z4, Canada. E-mail: vleung@ieee.org.

Victor C. M. Leung is with the Artificial Intelligence Research Institute, Shenzhen MSU-BIT University, Shenzhen 518172, China, the College of Computer Science and Software Engineering, Shenzhen University, Shenzhen 518060, China, and also the Department of Electrical and Computer Engineering, The University of British Columbia, Vancouver V6T 1Z4, Canada. E-mail: vleung@ieee.org.

% \IEEEcompsocthanksitem XXXXXXXX, XXXXXXXX, XXXX, XXXX, XXXX, XXXXXXXXX and XXXXXXXX are with the College of XXXXXXXXXXXXXXX, XXXXXXXXXX University, XXXXXXXXXX, China. 
% XXXXXXXX and XXXXXXXX are also with XXXXXXXXXXXXXX, XXXXXXXXXXXX, China.
% E-mail: XXXXXXX@tju.edu.cn, 
% XXXXXXX@tju.edu.cn, XXXXX@tju.edu.cn,
% XXXXXXXXXX@tju.edu.cn,
% XXXXXXXXX@tju.edu.cn, XXXXXXXX@tju.edu.cn,
% XXXXXX@tju.edu.cn.
}% <-this % stops a space
%\thanks{(Corresponding author: Chao Qiu)}
}

\markboth{IEEE Communications Surveys \& Tutorials, October~2024}%
{Shell \MakeLowercase{\textit{et al.}}: A Sample Article Using IEEEtran.cls for IEEE Journals}

\maketitle

\begin{abstract}
Amidst the robust impetus from artificial intelligence (AI) and big data, edge intelligence (EI) has emerged as a nascent computing paradigm, synthesizing AI with edge computing (EC) to become an exemplary solution for unleashing the full potential of AI services. Nonetheless, challenges in communication costs, resource allocation, privacy, and security continue to constrain its proficiency in supporting services with diverse requirements. In response to these issues, this paper introduces socialized learning (SL) as a promising solution, further propelling the advancement of EI. SL is a learning paradigm predicated on social principles and behaviors, aimed at amplifying the collaborative capacity and collective intelligence of agents within the EI system. SL not only enhances the system's adaptability but also optimizes communication, and networking processes, essential for distributed intelligence across diverse devices and platforms.
Therefore, a combination of SL and EI may greatly facilitate the development of collaborative intelligence in the future network.
This paper presents the findings of a literature review on the integration of EI and SL, summarizing the latest achievements in existing research on EI and SL. Subsequently, we delve comprehensively into the limitations of EI and how it could benefit from SL. Special emphasis is placed on the communication challenges and networking strategies and other aspects within these systems, underlining the role of optimized network solutions in improving system efficiency. Based on these discussions, we elaborate in detail on three integrated components: socialized architecture, socialized training, and socialized inference, analyzing their strengths and weaknesses. Finally, we identify some possible future applications of combining SL and EI, discuss open problems and suggest some future research, with the hope of arousing the research community’s interest in this emerging and exciting interdisciplinary field.% We anticipate that this survey will ignite further discussion on the synergistic operation of EI and SL, offering guidance and insights for research in EI, SL, and other relevant domains.
\end{abstract}


\section{Introduction}
\IEEEPARstart{W}{ith} the rapid advancement of information technology and the proliferation of wireless communications and Internet of Things (IoT) devices, human society is confronted with both the challenges and opportunities presented by vast quantities of data \cite{DBLP:journals/comsur/BalkusWCMNF22}. 
{\color{black} Data serves not only as a critical resource but also as a potent catalyst, propelling progress and innovation in the field of artificial intelligence (AI) \cite{xiao2019edge}.} AI has emerged as a transformative technology in today's world, finding extensive application across diverse domains, including healthcare, education, transportation, manufacturing, and entertainment, among others \cite{xie2018survey}. However, AI's evolution imposes heightened demands on data processing, necessitating substantial bandwidth and robust computational resources to achieve efficient, precise, and real-time data analysis and decision-making.

\vspace{-0.8em}
\subsection{From Cloud Computing to Edge Intelligence (EI)}
% The conventional cloud computing paradigm involves transmitting data from network peripheries (such as sensors, smartphones, and similar devices) to the cloud for processing and storage, subsequently returning the results to users or applications.
Traditional cloud computing sends data from devices like sensors and smartphones to the cloud for processing, then returns results to users.
Although this approach can provide abundant computational and storage resources, it also includes difficulties such as latency in the network, bandwidth usage, and privacy and security issues. In response to these issues, a novel computing paradigm, edge computing (EC), has acquired extensive attention in recent years \cite{wang2019edge}. EC shifts computational services to the network edge, reducing data transmission and enhancing processing efficiency and user experience. The transition from centralized cloud computing to EC underscores the necessity of robust communication frameworks, network design, and resource allocation, catering to the unique demands of edge environments for efficient system operation and data transmission.

\begin{figure}[pt]
  \centering
  \includegraphics[width=\linewidth]{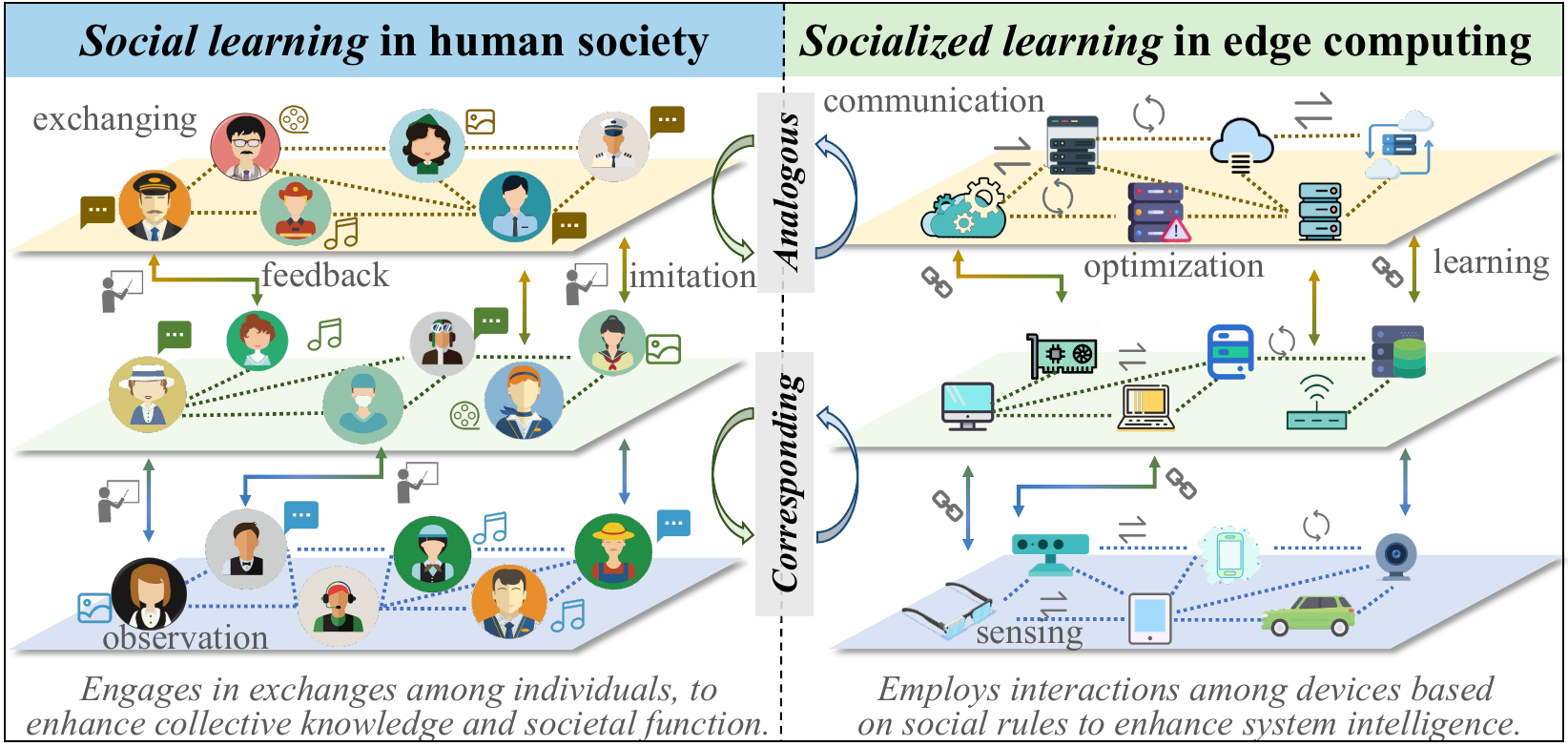}
  \caption{Social learning in human society versus SL in EC.}\label{Social learning vs SL}
  % \vspace{-0.5em}
\end{figure}

Building upon the foundation of EC, EI has been introduced as an emergent paradigm, {\color{black}integrating AI with EC to facilitate intelligent data processing and decision-making at the network edge \cite{yang2021edge,zhang2019edge,an2020edge}.} EI, employing AI algorithms and models on edge devices, can rapidly perceive and train on vast troves of local data \cite{deng2020edge}. {\color{black}This enables rapid adaptation to evolving environments, reduced latency, and enhanced computing efficiency.} Consequently, EI is regarded as a formidable driving force in the future development of networks. EI's reliance on AI algorithms at the network edge also amplifies the need for efficient communication, computing, resource allocation, and other aspects to manage complex data flows and interactions among various devices.

Primarily, EI can fulfill critical requirements such as agile connectivity, real-time operations, data optimization, intelligent applications, and security and privacy \cite{xu2021edge}. For instance, in autonomous driving scenarios, EI enables rapid communication and collaborative control between vehicles or between vehicles and roadside infrastructure, offering timely responses in emergencies \cite{wang2022integrating}. The collaboration of communication, networking, computing, resource, and system optimization plays a pivotal role in ensuring that EI operates at peak efficiency.
Furthermore, EI is anticipated to reach a phase of stability in the coming years and find widespread application in numerous burgeoning domains. For example, in the metaverse, EI can utilize artificial intelligence-generated content (AIGC) technologies on edge devices to construct and render virtual worlds, providing immersive experiences in virtual reality (VR) and augmented reality (AR).

\vspace{-0.8em}
\subsection{Combination of EI and Socialized Learning (SL)}
{\color{black}
Despite the immense potential and prospects of EI, its evolution still confronts numerous challenges and dilemmas \cite{liao2020cognitive}. For instance, users inevitably generate analogous machine learning (ML) tasks, potentially necessitating identical categories of data or even anticipating congruent training outcomes. This massive repetition of model training can lead to a substantial redundancy in computations across the network and a squandering of finite edge resources. Moreover, smaller data sizes can lead to model overfitting. Consequently, prevalent meaningless model training results in diminished training efficiency and a degradation in quality of service (QoS).

%分段
Another challenge encompasses achieving efficient communication, resource allocation, and privacy protection within EI. Given the involvement of various types and devices, pivotal issues that demand resolution include coordinating collaboration and competition among them, balancing mutual interests and costs, and ensuring security in interactions \cite{li2020optimizing}.}

To address these challenges, the concept of SL is introduced as an innovative solution. As shown in Fig. \ref{Social learning vs SL}, SL is the adaptation or mapping of social learning into machine interactions \cite{Bandura1977}. It is aimed at creating a cohesive machine society where social rules are implemented to facilitate collaborative interactions among devices, thereby enhancing overall system intelligence. { \color{black}SL is inspired by the concept of social learning in human societies. In these societies, individuals acquire knowledge, skills, attitudes, and values through observation, imitation, communication, and feedback, integrating these learnings into their behavior and thought processes.}

\begin{figure*}[pt]
  \centering
  \includegraphics[width=6.5in]{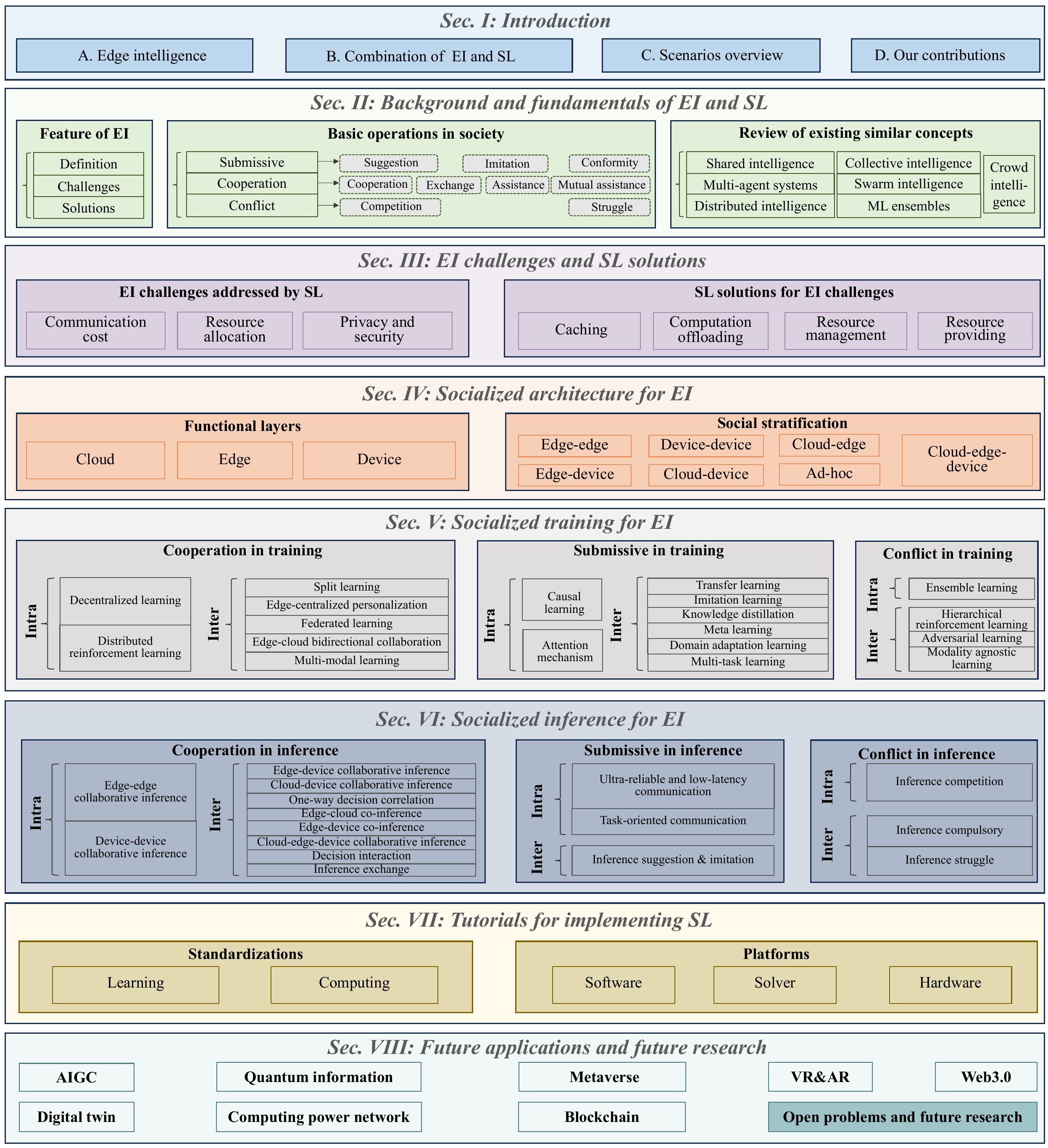}
  \caption{\color{black}The taxonomy graph on the integration of EI and SL.}\label{framework}
  % \vspace{-0.5em}
\end{figure*}

Hence, SL is a learning paradigm based on social principles and behaviors, aimed at enhancing the collaborative capabilities and collective intelligence of various agents within the EI system. SL draws inspiration from a myriad of social interaction modalities present in human societies, such as cooperation, submissive and conflict interactions, and applies them within the EI system. SL perceives each device as an independent, autonomous entity with social awareness. These devices can learn not only from their own data and experiences but also from the data and experiences of others. Through this method of social interaction, SL can effectively utilize data resources within the EI, reduce redundant computations, and enhance training efficiency and QoS. Moreover, SL can effectively allocate computational resources within the EI system, optimize communication costs, and safeguard privacy and security. SL promises to endow the EI system with greater intelligence, adaptability, and robustness.

{\color{black}
\textbf{A real-life application of SL in EI.}
In the context of intelligent transportation systems, the integration of EI and SL enables autonomous vehicles to learn collaboratively and adapt to dynamic road conditions. A network of self-driving vehicles, equipped with EI capabilities and SL algorithms, continuously gathers and processes real-time data about traffic, road conditions, and weather using onboard sensors and EC resources. This decentralized approach allows for swift responses to immediate situations \cite{mao2017survey}.
The true potential of this system lies in the application of SL principles, enabling vehicles to learn from their own experiences and the collective wisdom of the entire system. When a vehicle encounters a novel situation, such as an unexpected obstacle or congestion, it shares this information with other vehicles in the network \cite{lim2020federated}. This shared knowledge is incorporated into the collective intelligence, benefiting all vehicles.

For instance, if a vehicle detects a hazard, it can alert other vehicles to take alternate routes or proceed with caution. This real-time information sharing contributes to the long-term learning and optimization of the entire transportation network. As more vehicles encounter and share similar situations, the system develops a comprehensive understanding of the city's traffic dynamics and potential hazards \cite{zhang2018openvdap}.
This collaborative learning approach allows vehicles to proactively adapt to changing conditions based on insights from the fleet. Over time, this learning process enhances the safety, efficiency, and resilience of the autonomous transportation system, leading to smoother traffic flow, reduced congestion, and minimized accidents.
The combination of EI and SL exemplifies the transformative potential of these technologies in revolutionizing transportation networks. By leveraging EC and SL, vehicles make intelligent decisions in real time while contributing to the collective intelligence of the system, ultimately improving urban transportation.
}

\vspace{-0.8em}
\subsection{Scenarios Overview of SL in EI}\label{I-C}
The complex synergy of EI and SL has been integrated and reflected in various fields. This fusion leverages the strengths of both EI and SL, enabling systems to process data at the edge efficiently and enhance system intelligence. Additionally, it allows systems to learn and adapt through shared experiences and social principles, strengthening their adaptability to complex environments and user experiences.

{\color{black}
Across diverse application scenarios such as autonomous driving, IoT, smart healthcare, recommendation systems, smart transportation, and smart cities, the integration of SL in EI exhibits certain common characteristics. These include collaborative learning among edge devices, improved adaptability to dynamic conditions, and enhanced privacy and security due to the decentralized nature of SL. However, the specific objectives and implementation details of SL in EI may vary depending on the unique requirements and constraints of each application domain.
In the following contents, we will briefly explore how the integration of SL and EI manifests in various current application scenarios, showcasing the potential of this powerful combination to drive innovation and transform diverse sectors. A more detailed discussion on the future applications of SL in EI and a comparative analysis of these applications will be provided in Section \ref{section9.A}.}

\textbf{Autonomous Driving.}
The integration of EI and SL primarily manifests in the use of edge devices for real-time data processing and decision-making. The deployment on vehicles enables collaborative learning among vehicles, sharing road condition information, thereby enhancing overall driving safety and efficiency. Vehicles can react based on their own sensory inputs and learn from other vehicles' experiences through communication and data sharing.

\textbf{IoT.}
The integration of EI and SL enables devices to process data and share knowledge and experiences through SL algorithms. In smart home systems, this facilitates accurate environmental control and user behavior prediction, elevating IoT system intelligence through device interconnectivity and data sharing.

\textbf{Smart Healthcare.}
The combination of EI and SL improves efficiency and accuracy in medical data processing. By deploying edge resources for preliminary data processing and analysis, the burden on central servers is alleviated, while patient data privacy is protected. Edge-assisted medical systems allow for cost savings and improved healthcare service delivery, facilitated by shared knowledge and experiences among medical devices.

\textbf{Recommendation Systems.}
EI and SL integration is exemplified in the use of EC for personalized recommendations and user data processing, safeguarding privacy. Personalization is achieved through edge-based user behavior analysis and model training, ensuring privacy while offering tailored recommendations. SL algorithms further optimize the system based on user interactions and feedback.

\textbf{Smart Transport and Unmanned Aerial Vehicles (UAVs).}
EI and SL integration is evident in the use of UAVs for data collection and processing. Deploying EC and SL algorithms on UAVs facilitates real-time monitoring and data processing, improving system response time and efficiency. UAVs sharing flight experiences and data processing strategies through SL enhance the performance of the entire system.

\textbf{Smart City.}
The convergence of EI and SL aids in more effective city data management and processing. Deploying edge devices across different city areas and utilizing SL for data and experience sharing facilitate intelligent and automated city management. For instance, data sharing and collaborative learning among traffic monitoring devices can lead to more effective traffic flow management and accident prevention.

{\color{black}
The above examples provide a concise overview of how SL and EI are currently integrated across various application domains, highlighting the common characteristics and potential benefits of this synergistic combination. However, it is important to note that this section serves as a brief introduction to the current state of SL in EI applications. A more comprehensive analysis of the future applications of SL in EI, along with a detailed comparative study of these application scenarios, will be presented in Section \ref{section9.A}.}

% It is evident that the integration of EI and SL in various domains not only enhances system efficiency and intelligence but also strengthens adaptability to complex environments and user experiences. 
% In subsequent sections, we will explore in greater depth the principles of EI and SL integration and their potential and challenges in future development.

\vspace{-0.5em}
\subsection{Our Contributions}
% Despite the immense potential and promising prospects of EI, its evolution still confronts numerous challenges and dilemmas. Consequently, the primary objective of this article is to probe the integration of SL with EI, alongside offering several pragmatic tutorials.
% Despite the immense potential and promising prospects of EI, its evolution still confronts numerous challenges and dilemmas. 
This article aims to explore the integration of SL with EI and offers practical tutorials. More specifically, we enumerate the following principal contributions to this paper:
\begin{itemize}
\item We first outline the basic definition and architecture of EI and SL and discuss the necessity of SL in the presence of EI. We also present the problems addressed by EI.
\item We discuss the motivations for combining SL and EI from two perspectives. SL can be instrumental in addressing several challenges faced by EI, thereby compensating for the shortcomings of conventional methods. Further, the integration of SL within EI plays a pivotal role across various facets, optimizing system performance.
% We discuss the motivations for combining SL and EI from two perspectives: 
%     \begin{itemize}
%         \item SL  can be instrumental in addressing several challenges faced by EI, compensating for the deficiencies of conventional methods; 
%         \item The integration of SL within EI plays a pivotal role across various facets, optimizing system performance.
%     \end{itemize}
\item We summarize three ideas for deploying SL training and inference tasks in the EI architecture based on existing studies and analyze their advantages and disadvantages. 
\item We also review the future applications for applying SL to other fields under the EI architecture. Additionally, we identify some open problems and discuss future research on combining SL and EI.
\end{itemize}

% A taxonomy graph of this paper is presented as Fig. \ref{framework}.
% Specifically, we first give the background on EI and SL in Section \ref{section2}. Further, we review some existing and similar concepts in Section \ref{section3}. Then, the motivation for integrating EI and SL is explained in Section \ref{section4}. 
% Next, we describe in detail the three integration parts, i.e., socialized architecture (Section \ref{section5}), socialized training (Section \ref{section6}) and socialized inference (Section \ref{section7}). In Section \ref{section8}, we investigate applications and tutorials of SL in EI. Lastly, the conclusion is drawn in Section \ref{section9}. 
{\color{black}This paper's structure, depicted in Fig. \ref{framework}, commences with an introduction to the backgrounds of EI and SL, followed by a review of relevant concepts in Sec. \ref{section2}.
The rationale for integrating EI and SL is in Sec. \ref{section4}, followed by an exposition on the integration points: socialized architecture, training, and inference in Secs. \ref{section5}–\ref{section7}.} Tutorials of SL in EI are presented in Sec. \ref{section8}. Applications, open problems, and future research are discussed in Sec. \ref{section9}, concluding with Sec. \ref{section10}.

% A taxonomy graph of this paper is presented as Fig. \ref{framework}. Specifically, this paper begins with a background on EI and SL and reviews related concepts in Section \ref{section2}. The rationale for merging EI and SL is in Section \ref{section4}, followed by an exposition on the integration points: socialized architecture, socialized training, and socialized inference in Sections \ref{section5}–\ref{section7}. Tutorials of SL in EI are presented in Section \ref{section8}. Applications, open problems, and future research of SL in EI are discussed in Section \ref{section9}, concluding with Section \ref{section10}.

\section{Background and fundamentals of EI and SL}\label{section2}

\subsection{EI: Definition, Challenges, and Solutions}
% With the exponential growth in the number of connected devices and the rise of AI capabilities, the demands placed on cloud infrastructure have become untenable for real-time applications. 
The surge in connected devices and AI capabilities has strained cloud infrastructure for real-time applications.
EI, characterized by the integration of computation at the network's edge, offers a viable solution to these evolving challenges. %In this section, we delve deeper into the definition of EI, elucidate the key challenges hindering its widespread adoption, and explore the cutting-edge solutions developed to address them. 
We summarize the important definitions and related abbreviations listed in Table \ref{table:fuhao}.

\subsubsection{\textbf{Definition of EI}}
EI is the convergence of AI and EC, as detailed below:
\begin{itemize}[leftmargin=*]
\item \textbf{AI:} 
%Traditionally understood as the replication or simulation of human intelligence, AI encompasses a vast range of techniques, methods, and technologies, with deep learning being a prominent contributor. 
AI simulates human intelligence, encompassing techniques like deep learning (DL).
The main objective of AI is to model intelligent human behavior by discerning patterns and insights from extensive data repositories.

\item \textbf{EC:} {\color{black}This refers to a computational paradigm that extends beyond traditional cloud architectures. It offers computational services closer to the data sources or the `edge' of the network \cite{shi2016edge}.} The `edge' encompasses a spectrum of resources ranging from data origins to cloud-based centers, predominantly excluding the cloud itself. Unlike fog computing, which has an inherent hierarchical structure, EC emphasizes the significance of individual nodes that may not necessarily be interconnected in a network \cite{naranjo2019focan}.

\item \textbf{Interplay of AI and EC:} EI emerges naturally from the synergy of AI and EC. AI provides advanced techniques, while EC supplies varied, real-time application platforms \cite{deng2020edge}.
% EC's ability to process data at its source aids AI's data-driven decision-making by minimizing latency. This is especially valuable when large datasets, such as audio or video, need prompt analysis.
EC's processing at the data source reduces latency, which is critical for analyzing large datasets such as audio and video. 
These rapid insights power applications like public transport planning \cite{chen2015deepdriving}, urban monitoring \cite{DBLP:journals/csur/NingHWGGWG24} and early forest fire alerts.
\end{itemize}

\begin{table*}[t]
\centering
\caption{IMPORTANT ABBREVIATIONS LIST IN ALPHABETICAL ORDER}
\begin{tabular}{|c l c l c l|}
\hline
\textbf{Abbr.} & \textbf{Definition}     & \textbf{Abbr.} & \textbf{Definition}                          & \textbf{Abbr.} & \textbf{Definition}     \\ \hline
AI             & artificial intelligence& 
EI             & edge intelligence &  
IoT             & Internet of Things \\ \hline
EC             & edge computing & 
FL             & federated learning &  
ES             & edge server \\ \hline
CPU               &      central processing unit&    
LAN         &    local area network&    
QoS             &         quality of service\\ \hline
ShI              &     shared intelligence&    
DI         &    distributed intelligence&    
MAS             &         multi-agent system\\ \hline
CI             &     collective intelligence&
ML             &    machine learning&    
D2D            &         device-to-device\\ \hline
KD             &     knowledge distillation&
QoE             &    quality of experience&    
RL            &         reinforcement learning\\ \hline
GAN             &     generative adversarial network&
DNN             &    deep neural network&    
MEC            &         mobile edge computing\\ \hline
DL             &     deep learning&
MARL             &    multi-agent reinforcement learning&    
CrI            &        crowd
intelligence\\ \hline
DA            &     domain adaptation&
MTL            &     multi-task learning&    
NLP            &        natural language processing\\ \hline
HRL            &     hierarchical reinforcement learning&
DAG            &     directed acyclic graph&    
CPN            &        computing power network\\ \hline
AIGC            &     artificial intelligence-generated content&
AR            &     augmented reality&    
VR            &        virtual reality\\ 
\hline
SL            &     socialized learning&
SI           &     swarm intelligence&    
         DRL& distributed reinforcement learning     \\ 
\hline
MPC            &     multi-party computation&
SMMA           &     secure multi-party model aggregation&    
IRS         &    intelligent reflecting
surface  \\ 
\hline
\end{tabular}
\label{table:fuhao}
\end{table*}

\subsubsection{\textbf{Challenges in Implementing EI}}
%While EI presents a transformative approach to data processing and computation, it is not devoid of challenges:
EI offers a new approach to data processing but faces the following challenges:
\begin{itemize}[leftmargin=*]
\item \textbf{Trustworthiness:} Privacy, security, and regulatory compliance, such as general data protection regulation (GDPR), are crucial in EI's decentralized framework, which heightens exposure to security risks \cite{DBLP:journals/network/DongCHL19}.
% Ensuring privacy, security, and adhering to regulations such as the general data protection regulation (GDPR) is paramount. The decentralized nature of EI can potentially expose the network to various security threats, with data transmissions becoming more vulnerable to interceptions and breaches \cite{DBLP:journals/network/DongCHL19}.

\item \textbf{Scalability:} The limited computational and storage capacities of edges can hinder the efficient deployment of AI models. Striking a balance between latency, energy consumption, and model accuracy becomes a critical concern in EI.

\item \textbf{Resource Constraints:} Edges are often constrained in terms of computation, storage, and communication bandwidth. It can be difficult to effectively manage these resources while maintaining the best possible AI performance \cite{DBLP:journals/jsac/LiuS21}.
\end{itemize}

\subsubsection{\textbf{Solutions to Overcome Challenges}}
To address the aforementioned challenges, the research community and industry have proposed several innovative solutions:

\begin{itemize}[leftmargin=*]
\item \textbf{Advanced Learning Models and Architectures:} Techniques such as federated learning (FL), swarm intelligence (SI), and split learning are used to enhance data privacy and training efficiency at the edge \cite{Swarm2021, DBLP:journals/pieee/ParkSEKBKD21}. %Reinforcement learning (RL) strategies and trustworthy learning methods further offer dynamic adaptability in adversarial environments
Methods such as reinforcement learning (RL) offer dynamic adaptability in adversarial environments \cite{DBLP:journals/tcns/ChenZGB22, DBLP:conf/icml/ZhangYL0B18}.

\item \textbf{Security Enhancements:} Incorporating cryptographic techniques like differential privacy and Lagrange-coded computing can strengthen EI architectures against potential breaches. The use of quantum computing and blockchain further ensures data integrity and confidentiality \cite{ DBLP:journals/jnca/AzzaouiSP22}.

\item \textbf{Task-oriented Communication:} A shift from data-centric communication approaches to task-oriented paradigms can address the scalability concerns. By focusing on optimizing tasks over maximizing data rates, networks can efficiently handle the demands of EI applications.
\end{itemize}

The role of advanced communication protocols, robust networking strategies, efficient computing, and resource allocation becomes increasingly pivotal, especially for ensuring efficient collaboration. By understanding its challenges and actively developing innovative solutions, EI can reshape the landscape of communications and AI deployments.

\vspace{-0.5em}
\subsection{The Basic Operation Rules in Society}
There are three basic forms of social interaction: cooperative, confrontational, and submissive interactions \cite{hoppler2022six}. These interactions are crucial in shaping collective evolution.
Each form contributes uniquely to the complex structure of interpersonal relationships, reflecting the intricacies and diverse outcomes of our social environment. 
%By understanding these interactions, we can gain deeper insights into the underlying dynamics that shape our society and drive its evolution.

\subsubsection{\textbf{Submissive Interaction}}
Submissive interactions, characterized by one party yielding to another, play a crucial role in achieving social equilibrium, evident across diverse settings like workplaces, families, and educational institutions. This dynamic, while promoting social stability \cite{hoppler2022six}, may also inadvertently lead to power imbalances. \textbf{Suggestion} is a subtle but pervasive form of influence where individuals or entities use authority or trust to shape the actions or beliefs of others. \textbf{Imitation} serves as a critical mechanism for social relations, reflecting observed behaviors either deliberately or innately, thereby bridging the gap between external actions and personal behaviors \cite{meeker2020navigation}. This process can enhance social cohesion but also risks reinforcing undesirable actions. Meanwhile, \textbf{conformity} emerges from the pursuit of social acceptance \cite{audet2010neuroendocrine}, manifesting in behaviors ranging from strict adherence to social rules to more subtle adjustments in conduct.

\subsubsection{\textbf{Cooperation Interaction}} 
Cooperative interaction is fundamental in shaping an interconnected world, representing the amalgamation of individual or group efforts towards common goals, thereby fostering social progress. At its core, \textbf{cooperation} involves voluntary coordination of resources, knowledge, and skills, emphasizing synergistic collaboration for shared objectives or mutual interests without direct resource exchange \cite{castaner2020collaboration}. In contrast, \textbf{exchange} is defined by the principle of reciprocity, focusing on a balanced give-and-take of resources, information, or services to preserve equitable relationships and trust \cite{dale2020matters}. This differs from \textbf{assistance}, which is characterized by one-way support provided without the expectation of immediate return, often driven by altruism, empathy, or a sense of social responsibility \cite{shahidi2019impact}. Lastly, \textbf{mutual assistance} combines elements of cooperation, exchange, and assistance, highlighting bidirectional support and shared responsibility based on interdependence among participants \cite{becker2003creating}.

\subsubsection{\textbf{Conflict Interaction}}
Conflict interaction, in contrast to cooperative dynamics, encompasses disagreements and rivalries arising from divergent beliefs or interests, ranging from subtle disagreements to open conflicts. Such interactions can be a double-edged sword. \textbf{Competition}, as a form of conflict interaction, involves entities competing for limited resources or recognition. While it can drive innovation and excellence, unbalanced competition may lead to discord and inequity \cite{dimenichi2015power}. \textbf{Compulsory} interaction, on the other hand, is characterized by one party imposing its will on another, creating a power dynamic that can maintain social order or, if unchecked, may lead to resentment and power imbalances \cite{hem2014ethical}. The third subtype, \textbf{struggle}, represents a more intense form of conflict where entities actively contend over differing beliefs, values, or interests. Although initially disruptive, struggles can signify underlying needs for change, potentially leading to long-term social improvements \cite{overton2013conflict}.

\subsubsection{\textbf{Deciphering SL-emergence and Essence}}
% Understanding the nature of SL requires delving deep into its roots and the influences that have shaped its definition and application. 
% To understand SL, we examine its origins and influences.
At its core, SL borrows heavily from the rules that govern human social interaction, rendering it akin to an intelligent society. %To elucidate the nature and significance of SL, it becomes beneficial to draw an analogy with SI. 

\textbf{SI: A Preliminary Analogy.}
SI arises from the collective behavior of decentralized, self-organized systems often inspired by natural phenomena, like ant colonies or bird flocks \cite{beni1993swarm}. It is characterized by local interactions among individuals without centralized control. For example, in an ant colony, tasks such as nest building, larvae care, defense, and foraging are distributed without centralized command, despite a queen's presence. Ants rely on local communication, like pheromone trails, for complex problem-solving and colony survival \cite{dorigo1999ant}.

\textbf{From Swarm to Society: The Leap to SL.}
While SI is rooted in mimicking natural systems, SL moves a step further by integrating principles inherent in human society. SL is concerned with both artificial and natural systems wherein multiple entities coordinate using self-organization and decentralized control, much like SI. However, SL distinguishes itself by more heavily drawing from the norms, behaviors, and interactions prevalent in human society.

% In SL, individual agents act autonomously, devoid of a central command or global plan, akin to SI. However, these agents, while independent, are influenced by social norms, imitating complex interplays of submissive, cooperative, and conflicting interactions found in human relationships. 
% This gives SL its multifaceted character, combining the decentralized nature of SI with rich, layered intricacies of social dynamics.

\textbf{SL: Social Norms Meeting Intelligence.}
At its core, SL is about emergent collective intelligence. It encapsulates systems that encompass multiple independent devices whose behaviors and interactions are reminiscent of human social operations. By integrating the lessons from human society, SL aims to address real-world challenges in more sophisticated, nuanced ways, transcending the capabilities of traditional SI systems.

While SI offers an illustrative backdrop to understand the decentralized and self-organizing aspects of SL, \textbf{it is the addition of human social principles that truly defines and differentiates SL}. This integration, drawing from the richness of social norms and behaviors, allows SL to offer more holistic and adaptive solutions.
Furthermore, the implementation of SL in EI systems necessitates sophisticated computing techniques and system designs to facilitate effective communication and coordination among devices, optimizing system performance.

{\color{black}\textbf{The relationship between SL and SI.} Although SL and SI share fundamental principles like decentralized control and self-organization, they differ in scope and complexity. SI focuses on simple, local interactions among homogeneous agents, while SL incorporates diverse social norms, behaviors, and heterogeneous interactions found in human societies. SL emphasizes individual autonomy and social awareness in shaping collective behavior, enabling it to tackle complex, real-world challenges in EI systems. SI, in contrast, may be more applicable to specific EI scenarios involving homogeneous devices or simple coordination tasks. By leveraging social principles and learning mechanisms, SL fosters the development of socially-aware, adaptive, and collaborative EC systems, addressing a broader range of challenges compared to traditional SI approaches.}

{\color{black}
\subsubsection{\textbf{Formal Definition of SL}}
SL is a paradigm that integrates the principles and dynamics of human social interactions into the learning processes of intelligent systems. It aims to create a cohesive and adaptive learning environment where multiple agents, such as devices or algorithms, can learn collaboratively and share knowledge, experiences, and resources, guided by social norms and behaviors. SL draws inspiration from various social interaction modalities, including cooperation, competition, and hierarchical relationships, to optimize learning outcomes, resource allocation, and collective decision-making. By incorporating social principles, SL enables systems to tackle complex, real-world challenges more effectively, adapting to dynamic environments and leveraging the collective intelligence of the participating agents.}

\vspace{-0.5em}
\subsection{SL: A Function incorporating Multiple Intelligence}
% As we transition from understanding the foundational aspects of SL to exploring its functional architecture, it's crucial to conceptualize SL not just as a theoretical construct but as a multifaceted function capable of encapsulating a range of intelligences. 
% Moving from SL's foundation to its functional architecture, we view SL as a function encompassing various intelligences. This transition mirrors the evolution from simple, nature-inspired swarm intelligence to sophisticated systems influenced by the complexities of human social interaction.

We will introduce the function form of SL. The goal is to adjust different variables of function to achieve different SL schemes that can adapt to various application scenarios and requirements in EI.
The SL function can be represented as:
% \vspace{-0.2em}
\begin{equation}
    F(\mathbf{A}, \mathbf{R}_w, \mathbf{R}_a, \mathbf{I}, \mathbf{L}),
    % \vspace{-0.2em}
    \label{SL-function-define}
\end{equation}
where $\mathbf{A}$ represents the architecture of the network and other related information. $\mathbf{R}_w$ and $\mathbf{R}_a$ represent the methods of showing relationships within and across layers, respectively. $\mathbf{I}$ is an identifier that distinguishes inference and training operations. $\mathbf{L}$ indicate the specific operation object.

We can choose appropriate variable values according to different application scenarios and requirements to achieve different SL schemes. For example:
\begin{itemize}[leftmargin=*]
\item If we want to implement FL in EI, we can set $\mathbf{A}$ as a two-layer structure, where the first layer is a server and the other is a set of devices. We can set $\mathbf{R}_a$ as cooperation, where each device cooperates with others in the same layer to share models. $\mathbf{I}$ sets as a training operation, and $\mathbf{L}$ sets as DNNs.
\item If we want to implement edge-device collaborative inference in EI, we define a two-layer structure with edge servers (ESs) and devices. $\mathbf{R}_a$ is set to assistance, signifying a dynamic where one layer offers or seeks help from another. Inference operations are denoted as $\mathbf{I}$, and tasks as $\mathbf{L}$.
\end{itemize}
With this function form, we can flexibly express and design various SL schemes and match them with EI.
{\color{black} A visual summary of these schemes and their corresponding parameter values will be presented in the taxonomy graph in the end of Section \ref{section7}, as shown in Fig. \ref{SL-function}.}
%In the following sections, we will introduce different types of SL schemes in detail and analyze their advantages and challenges in EI environments.
This flexibility also extends to the optimization of network architectures and computational resource allocation, ensuring that the integration of SL and EI is not only intelligent but also systemically efficient and communication-coherent.

\vspace{-0.5em}
\subsection{Opportunities Brought by the Integration of EI and SL}
% The convergence of EI and SL, with their respective strengths and applications, presents a groundbreaking amalgamation that holds the potential to redefine the realms of AI and EC. 
% The integration of EI and SL can redefine the fields of AI and EC. The ensuing discussion explores these innovations and applications.

Several recent surveys have investigated the integration of EC with advanced technologies. Wang \emph{et al.} \cite{DBLP:journals/comsur/WangHLNYC20} discussed the convergence of EC and DL, including application scenarios and enabling technologies. Taleb \emph{et al.} \cite{DBLP:journals/comsur/TalebSMFDS17} examined 5G network edge cloud architecture. Yang \emph{et al.} \cite{8624417} explored the integration of blockchain with EC, addressing key motivations and challenges. Lim \emph{et al.} \cite{lim2020federated} reviewed FL in EC, focusing on communication and privacy challenges. Lastly, Ning \emph{et al.} \cite{DBLP:journals/csur/NingHWGGWG24} analyzed EC and ML applications for the Internet of UAVs.
While these surveys offer valuable insights, they do not organize and present state-of-the-art research in EI from an SL perspective, providing an innovative and comprehensive understanding of the challenges, techniques, and applications in EI systems.

\begin{figure*}[pt]
  \centering
  \includegraphics[width=7in]{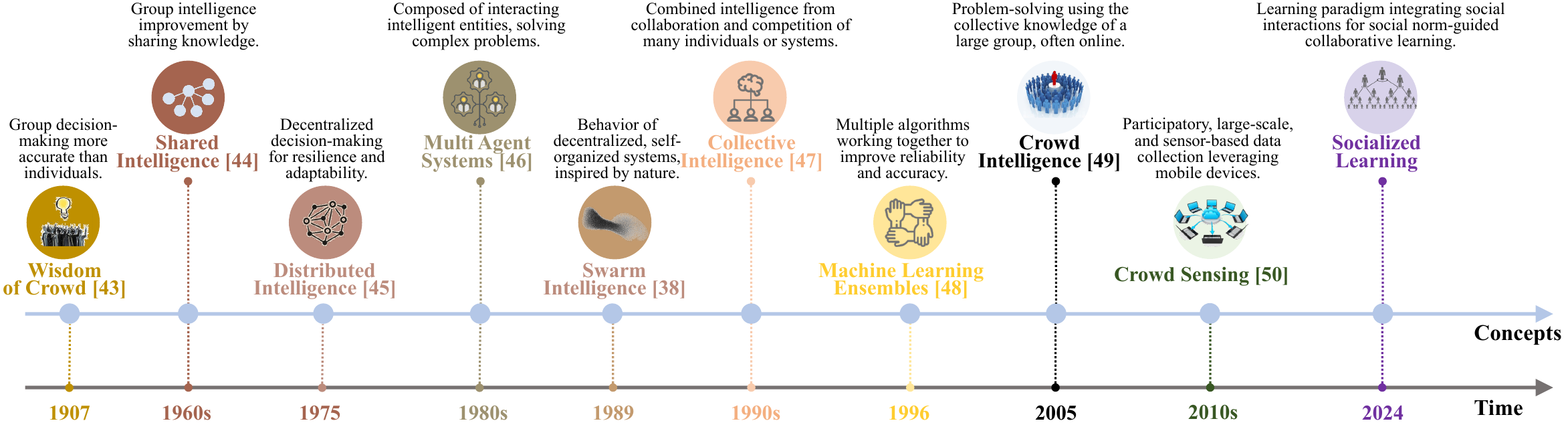}
  \caption{\color{black}The timelines of related concepts.}\label{timeline}
  % \vspace{-1em}
\end{figure*}

\begin{table*}[t]
\centering
\color{black}
\caption{{\color{black}Comparison of Various Types of Concepts}}
\begin{tabular}{|c|c|c|c|c|c|c|}
\hline
\textbf{Concept}                                                            & \textbf{\begin{tabular}[c]{@{}c@{}}Entity \\ Type\end{tabular}}      & \textbf{\begin{tabular}[c]{@{}c@{}}Representative \\ Algorithms\end{tabular}}                   & \textbf{\begin{tabular}[c]{@{}c@{}}Decision\\ Structure\end{tabular}} & \textbf{\begin{tabular}[c]{@{}c@{}}Main\\ Advantages\end{tabular}}                                 & \textbf{\begin{tabular}[c]{@{}c@{}}Main \\ Challenges\end{tabular}}                & \textbf{\begin{tabular}[c]{@{}c@{}}Result\\ Sharing\end{tabular}} \\ \hline
\textit{\begin{tabular}[c]{@{}c@{}}Wisdom of \\ the crowd  \cite{galton1907wisdom}\end{tabular}}     & \begin{tabular}[c]{@{}c@{}}People, esp.,\\ non-experts\end{tabular} & \begin{tabular}[c]{@{}c@{}}Aggregation methods, \\ Diversity-based methods\end{tabular}         & Distributed                                                           & \begin{tabular}[c]{@{}c@{}}Collective judgment, \\ error cancellation\end{tabular}                 & \begin{tabular}[c]{@{}c@{}}Independent opinions,\\ unbiased judgment\end{tabular}        & $\checkmark$                                                      \\ \hline
\textit{\begin{tabular}[c]{@{}c@{}}Shared \\ Intelligence \cite{engelbart1962augmenting}\end{tabular}}     & \begin{tabular}[c]{@{}c@{}}Humans,\\ Computers\end{tabular}          & \begin{tabular}[c]{@{}c@{}}Collaborative tools, \\ Content management Systems\end{tabular}      & Distributed                                                           & \begin{tabular}[c]{@{}c@{}}Effective knowledge \\ utilization, enhanced \\ collaboration\end{tabular} & \begin{tabular}[c]{@{}c@{}}Trust,\\ privacy issues\end{tabular}                    & $\checkmark$                                                      \\ \hline
\textit{\begin{tabular}[c]{@{}c@{}}Distributed \\ Intelligence \cite{chaib1992trends}\end{tabular}}     & \begin{tabular}[c]{@{}c@{}}Humans,\\ Computers\end{tabular}          & \begin{tabular}[c]{@{}c@{}}Distributed computing, \\ distributed gradient descent\end{tabular}           & Distributed                                                           & \begin{tabular}[c]{@{}c@{}}Scalability, \\ fault tolerance\end{tabular}                            & \begin{tabular}[c]{@{}c@{}}Coordination,\\ communication \\ overhead\end{tabular}     & $\checkmark$                                                      \\ \hline
\textit{\begin{tabular}[c]{@{}c@{}}Swarm \\ Intelligence \cite{beni1993swarm}\end{tabular}}      & \begin{tabular}[c]{@{}c@{}}Robots,\\ Algorithms\end{tabular}         & \begin{tabular}[c]{@{}c@{}}Particle Swarm Optimization, \\ Ant Colony Optimization\end{tabular} & Distributed                                                           & \begin{tabular}[c]{@{}c@{}}Solves complex problems, \\ good adaptability\end{tabular}              & \begin{tabular}[c]{@{}c@{}}Coordination,\\ scalability issues\end{tabular}         & $\times$                                                          \\ \hline
\textit{\begin{tabular}[c]{@{}c@{}}Multi-agent \\ Systems \cite{wooldridge2009introduction}\end{tabular}}     & AI Agents                                                            & \begin{tabular}[c]{@{}c@{}}Coordination Graphs, \\ Q-Learning\end{tabular}                      & Distributed                                                           & \begin{tabular}[c]{@{}c@{}}Solves complex problems, \\ good resilience\end{tabular}                & \begin{tabular}[c]{@{}c@{}}Coordination,\\ collaboration issues\end{tabular}       & $\checkmark$                                                      \\ \hline
\textit{\begin{tabular}[c]{@{}c@{}}Collective \\ Intelligence \cite{levy1997collective}\end{tabular}} & \begin{tabular}[c]{@{}c@{}}Humans,\\ Computers\end{tabular}          & \begin{tabular}[c]{@{}c@{}}Social Network Analysis, \\ Collaborative Filtering\end{tabular}     & Distributed                                                           & \begin{tabular}[c]{@{}c@{}}Strong innovation, \\ rich knowledge\end{tabular}                       & \begin{tabular}[c]{@{}c@{}}Quality control,\\ coordination issues\end{tabular}     & $\checkmark$                                                      \\ \hline
\textit{\begin{tabular}[c]{@{}c@{}}ML \\Ensembles \cite{breiman1996bagging} \end{tabular}}                                                      & ML Models                                                            & \begin{tabular}[c]{@{}c@{}}Random Forests, \\ Boosting\end{tabular}                             & Ensemble                                                              & \begin{tabular}[c]{@{}c@{}}High accuracy, \\ good robustness\end{tabular}                          & \begin{tabular}[c]{@{}c@{}}High computational,\\ storage requirements\end{tabular} & $\checkmark$                                                      \\ \hline
\textit{\begin{tabular}[c]{@{}c@{}}Crowd \\ Intelligence \cite{howe2006rise}\end{tabular}}      & \begin{tabular}[c]{@{}c@{}}People, esp.,\\ online users\end{tabular} & \begin{tabular}[c]{@{}c@{}}Crowdsourcing algorithms, \\ Reputation systems\end{tabular}         & Distributed                                                           & \begin{tabular}[c]{@{}c@{}}Solves complex problems, \\ high diversity\end{tabular}                 & \begin{tabular}[c]{@{}c@{}}Quality control,\\ incentive issues\end{tabular}        & $\checkmark$                                                      \\ \hline
\textit{\begin{tabular}[c]{@{}c@{}}Crowd \\ Sensing \cite{ganti2011mobile}\end{tabular}}      & \begin{tabular}[c]{@{}c@{}}People with \\ mobile devices\end{tabular} & \begin{tabular}[c]{@{}c@{}}Data aggregation, \\ Sensor fusion\end{tabular}         & Distributed                                                           & \begin{tabular}[c]{@{}c@{}}Large-scale data collection, \\ real-time insights\end{tabular}                 & \begin{tabular}[c]{@{}c@{}}Privacy concerns,\\ data quality\end{tabular}        & $\checkmark$                                                      \\ \hline
\end{tabular}
\label{tab:comparison}
\end{table*}

SL deepens EI's understanding of cooperation and improves EI systems' intelligence and collaboration through inspiration from human social interaction. This understanding fosters improved decision-making, addressing gaps that traditional AI might overlook. In decentralized EI, SL harmonizes entities to optimize resources for collective goals. Furthermore, SL's emphasis on consensus and collective intelligence yields adaptable, resilient systems. This boosts trust and ensures efficient configurations in EI systems.
%Additionally, by leveraging the nuances of human social behavior, SL can provide insights to simplify complex network dynamics, resulting in systems that can predict, adapt, and collaborate in unprecedented ways.

Conversely, EI's resources and intelligence offer fertile ground for SL. EI's distributed architecture enables seamless SL integration. SL entities leverage edge resources to optimize computation and resource allocation. SL models in edge contexts access EI capabilities, creating unique, society-inspired services. Lastly, integrating EI with SL enhances compatibility, scalability, and security, boosting performance.

{\color{black}
The integration further facilitates a more dynamic resource allocation and network management, enabling the EI system to adapt and respond to varying computational demands and communication needs, thereby exemplifying a comprehensive system-level approach to modern AI and EC challenges. This survey emphasizes the synergistic potential of EI and SL, aiming to construct a potent platform for maximizing the value of information, stimulating resource and intelligence sharing at the edge, and enhancing system efficiency.
}

\vspace{-0.5em}
\subsection{Review of Existing Similar Concepts}
{\color{black}In the dynamic AI area, SL emerges as a distinctive advancement, integrating aspects of various established paradigms. Illustrated in Fig. \ref{timeline} and Table \ref{tab:comparison}, SL's evolution builds upon these concepts, differentiating itself by focusing on the social aspects of learning and intelligence.} Key concepts that have influenced the development of SL include:

\textbf{Wisdom of the Crowds} and \textbf{Shared Intelligence (ShI)}: Documented by Galton in 1907 \cite{galton1907wisdom}, wisdom of the crowds underscores the superiority of collective judgment over individual expertise. Its relevance in computing is demonstrated through platforms like Wikipedia, aligning with SL’s emphasis on collaborative and socially driven knowledge construction.
ShI \cite{engelbart1962augmenting}, which gained prominence with the growth of computer networks in the 1960s, emphasizes the distribution of intelligence through collaboration. This forms a foundational aspect of SL, where social interactions play a crucial role in the learning process.

\begin{table*}[t]
\color{black}
\centering
\caption{{\color{black}Comparative Differentiation of SL from Existing Paradigms}}
\begin{tabular}{|m{0.08\linewidth}|m{0.4\linewidth}|m{0.43\linewidth}|}
\hline
\textbf{\textbf{Paradigm}} & \textbf{\textbf{Core Concept}}                                                                                                       & \textbf{\textbf{SL's Differentiation}}                                                                                                                                \\ \hline
\emph{CrI v.s. SL}         & Aggregates many independent inputs, assuming collective wisdom leads to optimal outcomes \cite{galton1907wisdom}. & Promotes genuine collaboration and shared learning. Interactions\  in SL are processed and synthesized, not just aggregated.\\ \hline
\emph{ShI v.s. SL}         & Emphasizes only distributing tasks and information \cite{engelbart1962augmenting}.                                                                                     & Harmonizes distribution with dynamic social contexts, \ making learning socially contextualized and adaptive.              \\ \hline
\emph{SI v.s. SL}          & Uses simple decentralized entities for collective goals \cite{beni1993swarm}.                                                                           & Infuses decentralized model with intricate learning \ algorithms, adding a social awareness component.                    \\ \hline
\emph{MAS v.s. SL}         & Systems wherein agents engage to accomplish tasks \cite{wooldridge2009introduction}.                                                                                 & Agents in SL not only interact but learn from these interactions, creating a richer knowledge repository.                \\ \hline
\emph{CI v.s. SL}          & Focuses on agent collaboration for mutual goals \cite{levy1997collective}.                                                                                   & Integrates social dynamics, emphasizing learning, growing, and evolving through collective experiences. \\ \hline
\end{tabular}                   
\label{tab:SLComparison}
\end{table*}

\textbf{Distributed Intelligence (DI)} and \textbf{SI}: These concepts, originating from Hutchins' theory \cite{chaib1992trends} and Beni and Wang's work \cite{beni1993swarm}, emphasize decentralized and collective behavior in systems. DI explores how cognition extends beyond individuals to networks, resonating with SL's focus on collective processes. SI, inspired by natural phenomena like ant colonies, illustrates how simple rules at the individual level can lead to complex, intelligent behavior in groups.

\textbf{Multi-agent Systems (MAS)}, \textbf{Collective Intelligence (CI)} and \textbf{ML Ensembles}: Originating from distributed AI \cite{wooldridge2009introduction}, MAS encompasses autonomous agents collaborating, an idea further expanded in SL to incorporate social dynamics. CI, popularized by Lévy \cite{levy1997collective}, focuses on collaborative efforts for shared objectives, mirroring SL's emphasis on social collaboration.
Introduced by Breiman \cite{breiman1996bagging}, these methods in ML, such as random forests and boosting algorithms, parallel SL’s approach by integrating diverse models for enhanced learning, akin to SL's focus on integrating diverse social interactions.

\begin{figure}[pt]
  \centering
  \includegraphics[width=3.in]{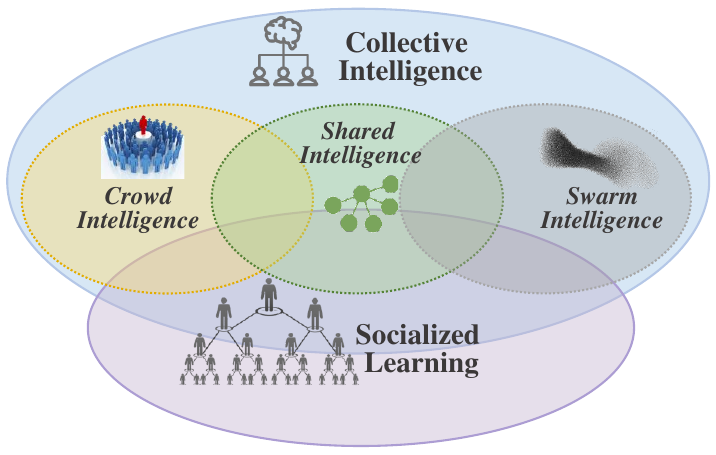}     
  \caption{Relationships among related AI paradigms.}\label{butonggainianduibi}
  \vspace{-1em}
\end{figure}

\textbf{Crowd Intelligence and Crowd Sensing}: Crowd Intelligence, emerging around 2005, harnesses the collective wisdom of large-scale online communities to solve complex problems \cite{howe2006rise}. It is based on the idea that aggregated intelligence of non-experts can rival or surpass experts. Crowdsourcing involves distributing tasks to a large number of people via the Internet, requiring effective management and incentive strategies. Crowd sensing, introduced in the 2010s, focuses on large-scale data collection using mobile devices and sensors carried by individuals, enabling the gathering of rich, context-aware information about environments and activities \cite{ganti2011mobile}. While Crowd Sensing faces challenges in data processing and privacy protection, both approaches leverage mass participation for generating insights and knowledge.

SL’s integration of social context into AI is further detailed in Table \ref{tab:comparison}, contrasting it with these existing paradigms. Fig. \ref{butonggainianduibi} visually represents the relationships among these paradigms, highlighting SL's innovative approach to embedding social dynamics into collective intelligence. In summary, SL represents a shift towards socialized aspects of AI, emphasizing collaborative and socially-aware problem-solving. This shift necessitates the development of sophisticated communication systems and network architectures, as discussed in our comparative analysis in Table \ref{tab:SLComparison}. Such advancements are vital for realizing the full potential of socialized AI paradigms like SL.

% \vspace{-0.5em}
% \section{Challenges of SL and EI}
\section{EI Challenges and SL Solutions}
\label{section4}
% As we contemplate the integration of SL into EI, it becomes imperative to first scrutinize the challenges currently inherent in EI. Issues surrounding efficient communication, optimal resource allocation, and data privacy and security, in particular, demand rigorous attention. 
To integrate SL into EI, we must first address inherent EI challenges such as communication efficiency, resource allocation, and data security, as shown in Fig. \ref{fig:motivation}.

\vspace{-0.5em}
\subsection{EI Challenges Addressed by SL}
% \subsection{SL at EI} 
% EI has experienced considerable growth. While offering unparalleled advantages over traditional cloud computing, its deployment also poses challenges, as follows:
Despite the advantages of EI over traditional cloud computing, it presents several challenges:

\subsubsection{\textbf{Communication Cost}} %Why does EI need SL? 为什么EI需要SL
% EI intensifies communication efficiency challenges, paramount for evaluating distributed processing performance and scalability. 
In EI, model synchronization during training, particularly with larger models and uncertain convergence, intensifies communication challenges for distributed processing and scalability. Differences in edge devices' capabilities and data amplify these costs. Resource-constrained devices, such as smartphones, face hardware and bandwidth limitations that hinder model transmissions. 

%分段
Furthermore, disparate data sizes and non-independent and identically distributed (non-IID) distributions complicate communication. Harnessing SL within EI requires addressing these communication complexities \cite{ranaweera2021survey}. To address these complexities effectively, exploring innovative collaborative methods for efficient network designs and data transmission is essential. Such methods should align with the unique demands of EI environments, aiming to enhance model synchronization and reduce bandwidth constraints.

\subsubsection{\textbf{Resource Allocation}}
% EI signifies a transformative departure from traditional cloud computing, pivoting to distributed processing and localized resource management. 
EI’s transition to efficient distributed processing and local resource management presents challenges in managing a varied landscape of edge systems \cite{xu2021edge}, from powerful servers to simple IoT devices, requiring strategic resource allocation. Especially when dynamic edge environments are constantly changing as devices connect and disconnect \cite{lim2021decentralized}. 
SL integration within EI can address these challenges, enhancing both efficiency and sustainability. This strategic allocation requires sophisticated algorithms capable of dynamically managing computational and communication resources, which is vital for maintaining system performance in the diverse and changing landscape of edge environments.

\begin{figure*}[pt]
  \centering
  \includegraphics[width=7.2in]{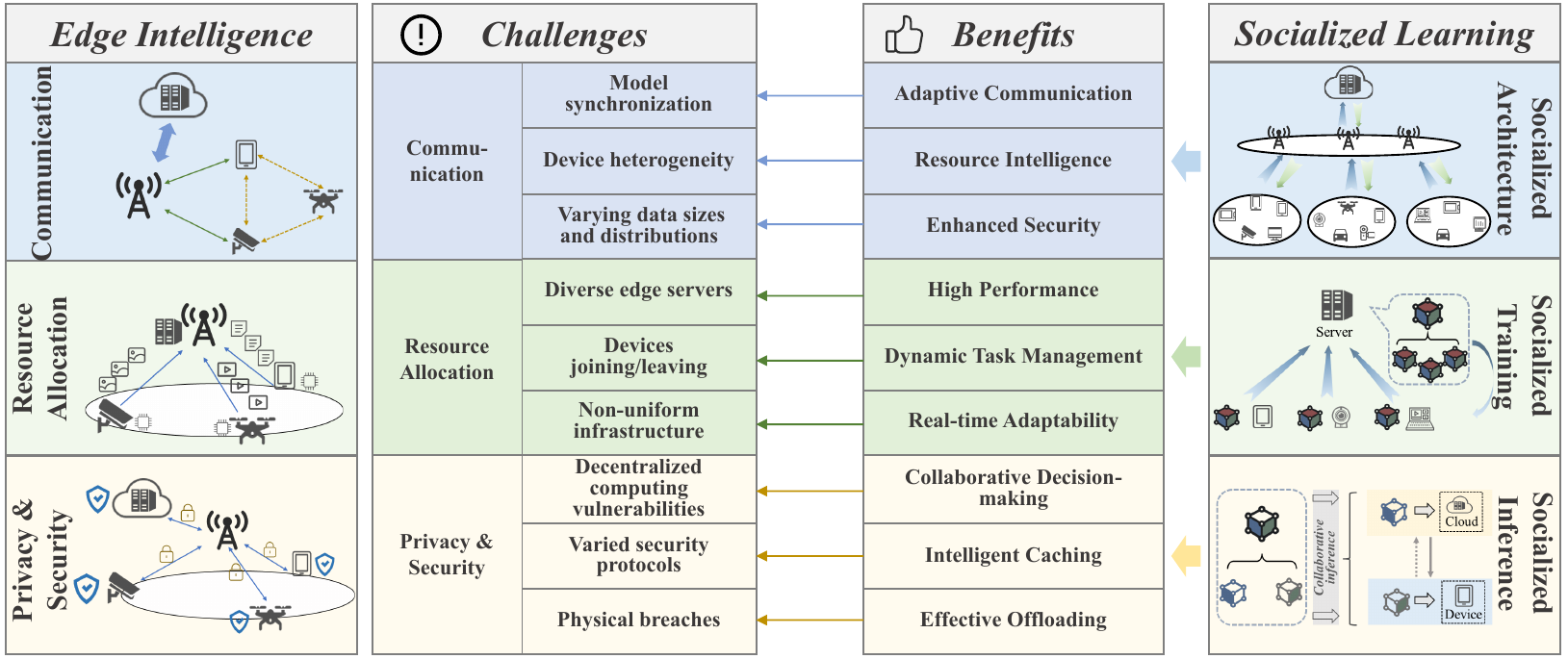}
  \caption{The motivation for integrating EI and SL.}\label{fig:motivation}
  % \vspace{-1em}
\end{figure*}

\subsubsection{\textbf{Privacy and Security}}
The increased exposure to diverse security protocols and edge heterogeneity necessitates enhanced privacy and security measures against multiple vulnerabilities. Data processed and stored at the edge faces increased exposure from diverse security protocols \cite{ranaweera2021survey}. The distributed nature of EI introduces multiple vulnerabilities. Diverse threats and edge heterogeneity complicate uniform security measures. 

%分段
SL integration in EI could enable adaptive security models that detect and evolve with network threats \cite{DBLP:journals/tsc/AlDoghmanMKSTZ23}. Maintaining a balance between EI's vulnerabilities and security solutions is crucial. To achieve this balance, developing and integrating adaptive security protocols is imperative. These protocols should cater to the varied and evolving security needs of decentralized networks, ensuring data integrity and privacy across diverse edge environments.

% While existing traditional methods, e.g., resource allocation and computing offloading \cite{gong2022edge}, have demonstrated promise in EI optimization, they often fall short of fully addressing the complex challenges, leading to sub-optimal solutions.

\vspace{-0.5em}
\subsection{SL Solutions for EI Challenges}
% \subsection{SL in EI}
% Given the identified challenges and the recognized potential of SL, it's natural to ask: How can SL be harnessed to enhance EI capabilities? 
%Considering these challenges, how can SL enhance EI capabilities?
We explore the ways through which SL can be integrated into the EI system to address its inherent challenges. By integrating the principles of SL, we can adapt and optimize EI systems, making them more resilient, efficient, and intelligent. %SL brings about adaptive and decentralized solutions well-suited for EI's complexities.

\subsubsection{\textbf{Communication Efficiency through SL}}
SL introduces a cooperative, adaptive approach to enhance communication efficiency in EI, addressing challenges posed by diverse edge devices and varying data distributions. Leveraging shared knowledge across devices, SL optimizes transmission, ensuring only pertinent information is exchanged. Such an intelligent, collective approach paves way for reduced communication overhead, especially in complex, non-IID scenarios \cite{ranaweera2021survey}.

\subsubsection{\textbf{SL-driven Resource Allocation}}
In the intricate landscape of EI, resource management not only involves communication but also encompasses caching, computing, and control. In EI's intricate landscape, SL aids in managing the diverse demands of computation, radio resources, and task allocation by leveraging collective knowledge to anticipate and optimize resource use. {\color{black}This socialized approach, backed by real-time adaptability, empowers EI systems to meet the dynamic requirements of modern tasks and applications efficiently \cite{li2023tapfinger}.}

\subsubsection{\textbf{Enhancing Privacy and Security with SL}}
The decentralized structure of EI presents privacy and security challenges. SL addresses these by emphasizing knowledge sharing over direct data exchange. SL enhances privacy and security by encouraging devices to share insights and learning experiences, rather than raw data, making the network more robust against security threats. This not only safeguards individual data but also strengthens the network's collective intelligence, protecting against data breaches and misuse \cite{guo2021lightfed,DBLP:journals/tsc/AlDoghmanMKSTZ23}.

\subsubsection{\textbf{Optimized Caching using SL}}
% EI seeks to bring forth efficient caching solutions to accommodate the ever-growing demands for swift data access and real-time processing.
EI aims for efficient caching solutions to meet the growing demands for fast data access. Traditional centralized caching methods struggle to address the dynamic and diverse needs of edges. SL improves caching in EI by enabling devices to cooperatively predict and adaptively cache data based on collective behaviors, enhancing performance in real-time. Rather than working in isolation, devices share insights on frequently accessed data, intelligently populating caches and reducing unnecessary transfers to ensure pertinent data is available \cite{liu2016caching}.

\subsubsection{\textbf{SL for Effective Computation Offloading}}
% Computation offloading is a critical aspect of EI, ensuring that tasks are allocated to devices that can handle them efficiently.
Computation offloading in EI ensures efficient task allocation across devices. SL facilitates informed offloading decisions by allowing devices to share knowledge, enabling them to determine the most suitable device for a task's requirements based on operational factors. The insights from SL ensure offloading strategies evolve and adapt to task and environmental changes. Hence, SL optimizes computing efficiency, contributing to energy conservation and enhanced user experience \cite{pham2019coalitional}.

% As we embarked on this exploration of integrating SL within the realm of EI, it was evident that this synergy harbored the potential to revolutionize the EI landscape. SL, with its emphasis on collaborative learning and shared experiences, offers the much-needed intelligence and adaptability to address the inherent complexities of EI. From enhancing communication efficiency to safeguarding data, SL emerges as an indispensable asset for the evolving demands of EI.

SL, with its emphasis on collaborative learning and shared experiences, offers the much-needed intelligence and adaptability to address the inherent complexities of EI. 
To better support EI deployment, we tailor SL to socialized architecture, training, and inference, as shown on the right side of Fig. \ref{fig:motivation}. SL is detailed in Sec. \ref{section5}, \ref{section6}, and \ref{section7}.

% \vspace{-0.5em}
\subsection{Lessons Learned}

The convergence of EI and SL illuminates synergistic gains while highlighting critical challenges such as agent coordination and privacy concerns in collaborative settings. The \textbf{symbiotic potential} of EI and SL is evident in its capacity to alleviate communication overhead, streamline resource management and maintain privacy. SL's approach to collaborative learning navigates through EI's complexities, offering adaptive problem-solving and efficient communication where traditional methods may falter. This cooperative paradigm not only refines information exchange but also underscores the need for robust and adaptable communication networks and computing architectures, essential for dynamic resource allocation and processing in EI systems.

\textbf{Resource optimization} extends to computing and control processes, and this necessitates harnessing SL's approach to optimize resources, leveraging shared intelligence to improve usage efficiency. SL also protects privacy by sharing insights over data, enhancing network intelligence and stopping data misuse. Beyond providing immediate solutions, the integration of SL into EI aims to create inherently \textbf{adaptive} systems, poised to evolve with emerging technological demands and ensure long-term relevance.

\section{Socialized Architecture for EI}\label{section5}
We establish the foundational concepts underpinning SL. Then, we delve into the social stratification that characterizes different architectural choices.
In addressing these architectural choices, it's crucial to emphasize the role of advanced communication technologies and robust network designs. These aspects are integral in ensuring efficient inter-layer communication and resource allocation, especially when dealing with complex system interconnections inherent in EI.

\vspace{-0.5em}
\subsection{Functional Layers in Socialized Architecture}
The EI architecture's socialized structure includes device, edge, and cloud layers, tackling system interconnection challenges through optimized operations and resource use.

\subsubsection{\textbf{The Main Role in the SL Architecture}}
In the SL, three key layers operate collaboratively. The \emph{Device Layer} includes diverse end devices like UAVs, AR tools, health sensors, IoT devices, and smart home gadgets \cite{sun2018double,ren2019edge,li2020energy,deng2019parallel}. These devices at the Device Layer, crucial for data production and collection, process data locally or offload it based on Quality of Experience (QoE) and QoS requirements.
\emph{Edge Layer} provides computing and storage, mainly through ESs and gateways, orchestrating services and tasks for efficient data handling and resource optimization \cite{dziyauddin2021computation, dolati2022layer}.
\emph{Cloud Layer} enhances traditional cloud computing with advanced capabilities for handling complex computations and large-scale storage, integrating closely with edge layer \cite{costa2022orchestration}.

\begin{table}[t]
\color{black}
\centering
\caption{{\color{black}Comparative Analysis in Different Aspects}}
\begin{tabular}{|c|c|c|c|c|}
\hline
\textbf{Aspect}                                                                          & \textbf{Criteria}                                                & \textbf{Device} & \textbf{Edge} & \textbf{Cloud} \\ \hline
\multirow{5}{*}{\textbf{\begin{tabular}[c]{@{}c@{}}Resource \\ Allocation \\ \cite{luo2021resource}\end{tabular}}} & Heterogeneity                                                    & High            & Moderate      & Low            \\ \cline{2-5} 
     & \begin{tabular}[c]{@{}c@{}}Algorithm \\ Requirement\end{tabular} & High            & Moderate      & Low            \\ \cline{2-5} 
     & Computing                                                        & Limited         & Moderate      & High           \\ \cline{2-5} 
     & Storage                                                          & Limited         & Moderate      & Abundant       \\ \cline{2-5} 
     & \begin{tabular}[c]{@{}c@{}}Energy \\ Efficiency\end{tabular}     & High            & Moderate      & Low            \\ \hline
\multirow{3}{*}{\textbf{\begin{tabular}[c]{@{}c@{}}Data \\ Management\\ \cite{DBLP:journals/jnca/SadriRSH21}\end{tabular}}}     & \begin{tabular}[c]{@{}c@{}}Local \\ Collection\end{tabular}      & High            & Moderate      & Low            \\ \cline{2-5} 
     & Encryption                                                       & Low             & Moderate      & High           \\ \cline{2-5} 
     & Analytics                                                        & Basic           & Moderate      & Advanced       \\ \hline
\multirow{3}{*}{\textbf{\begin{tabular}[c]{@{}c@{}}Task \\ Execution \\ \cite{almutairi2021novel}\end{tabular}}}      & \begin{tabular}[c]{@{}c@{}}Real-time \\ Need\end{tabular}        & High            & High          & Moderate       \\ \cline{2-5} 
     & \begin{tabular}[c]{@{}c@{}}Energy \\ Consideration\end{tabular}  & High            & Moderate      & Low            \\ \cline{2-5} 
     & Complexity                                                       & Simple          & Moderate      & Complex        \\ \hline
\multirow{3}{*}{\textbf{\begin{tabular}[c]{@{}c@{}}Computing \\ Power \\  \cite{DBLP:journals/tcc/CozzolinoTMDO23}\end{tabular}}}     & Speed                                                            & Slow            & Moderate      & Fast           \\ \cline{2-5} 
     & \begin{tabular}[c]{@{}c@{}}Real-time \\ Capability\end{tabular}  & High            & Moderate      & Low            \\ \hline
\end{tabular}
\label{Table:Comparative Analysis in Different Aspects}
\end{table}

\subsubsection{\textbf{Comparative Analysis in Different Aspects}} 
We analyze differences between the device, edge, and cloud layers, focusing on resource allocation, data management, task execution, and computing power to guide efficient solution design.

\textbf{Resource Allocation}:
resource allocation differs notably among the three layers. The device layer, limited by its heterogeneity, has scarce resources, necessitating optimal algorithms for managing computation, storage, and energy \cite{luo2021resource}. While the edge layer has more resources, it's still less capable than the cloud, emphasizing low-latency service and local data processing. The cloud layer, resource-rich, is mainly used for tasks demanding substantial computing power.

\textbf{Data Management}:
The scope and sophistication of data management vary among the layers. The device layer centers on local data gathering with basic processing \cite{DBLP:journals/jnca/SadriRSH21}. The edge layer handles advanced data tasks like preprocessing, encryption, and limited analytics. In contrast, the cloud layer is the epicenter for sophisticated data analytics and ML, providing the most extensive data solutions.

\textbf{Task Execution}:
Task handling differs among the layers. The device layer manages simple, immediate tasks like data collection and preliminary analysis. The edge layer takes on tasks offloaded from devices, focusing on real-time actions with critical latency and energy concerns \cite{almutairi2021novel}. The cloud layer addresses complex tasks, such as in-depth data analysis and ML, leveraging sufficient resources without tight time limits.

\textbf{Computing Power}:
The device, edge, and cloud layers have varied computing capabilities. The device layer is geared towards simpler tasks due to its limited power \cite{DBLP:journals/tcc/CozzolinoTMDO23}. The edge layer, more powerful than devices, ensures swift processing of offloaded tasks but is still limited compared to the cloud. The cloud layer has abundant resources for intricate tasks but may lag in real-time processing because of its location. Each layer caters to distinct needs across applications. 

Table \ref{Table:Comparative Analysis in Different Aspects} provides a summarized comparison. To effectively cater to these distinct needs, the integration of scalable computing resources and dynamic network topologies is essential. This integration facilitates a harmonious flow of data and tasks across layers, optimizing system performance and enabling more effective management of computational and communication resources.

%和人类社会的关系
\subsubsection{\textbf{Analogies to Human Society}}
In analogy with human society, devices are like frontline workers handling tasks with limited resources. The edge layer, similar to local officials, focuses on local needs and policies. The cloud layer, akin to top-level leaders, offers vast resources but may face challenges due to its distance. Each layer, with its distinct role, plays a vital part in optimizing resources and cooperation, thereby enhancing the overall performance of the system.

\vspace{-0.5em}
\subsection{Social Stratification in Socialized Architecture} 
Social stratification in SL reveals the resource and task hierarchy in computing layers. This section discusses each architecture's strengths, challenges, and overall performance.

\subsubsection{\textbf{Non-Layer Architecture}}  
Social stratification in SL begins with the basic non-layer architecture, forming a foundation for understanding more complex, centralized structures.

% \usepackage{multirow}
\begin{table*}[t]
\color{black}
\caption{\color{black}Comparison of Various Commonly used Architectures}
\centering
\begin{tabular}{|c|c|c|c|c|c|c|c|c|}
\hline
\textbf{\begin{tabular}[c]{@{}c@{}}Layer\\ Type\end{tabular}} & \textbf{\begin{tabular}[c]{@{}c@{}}Specific\\  Architecture\end{tabular}} & \textbf{Latency}                                           & \textbf{\begin{tabular}[c]{@{}c@{}}Energy \\ Efficiency\end{tabular}} & \textbf{Scalability} & \textbf{\begin{tabular}[c]{@{}c@{}}Fault \\ Tolerance\end{tabular}} & \textbf{\begin{tabular}[c]{@{}c@{}}Application \\ Suitability\end{tabular}}       & \textbf{Advantages}                                                                        & \textbf{Limitations}                                                         \\ \hline
\textbf{Non}                                                  & \textbf{\begin{tabular}[c]{@{}c@{}}Centralized\\ \cite{guo2018computation,meskar2018fair,kaffes2019centralized} \end{tabular}}                                                    & Low                                                        & High                                                                  & Low                  & Low                                                                 & High determinism                                                                  & Predictable                                                                                & \begin{tabular}[c]{@{}c@{}}Single point \\ of failure\end{tabular}           \\ \hline
\multirow{2}{*}{\textbf{Intra}}                               & \textbf{\begin{tabular}[c]{@{}c@{}}Edge-edge\\ \cite{huang2018parked,na2018frequency,miao2020intelligent,thai2019workload} \end{tabular}}                                                        & Moderate                                                   & Moderate                                                              & High                 & Moderate                                                            & \begin{tabular}[c]{@{}c@{}}Geographically\\  dispersed\end{tabular}               & \begin{tabular}[c]{@{}c@{}}Scalability and \\ resource sharing\end{tabular}                & \begin{tabular}[c]{@{}c@{}}Complex \\ coordination\end{tabular}              \\ \cline{2-9} 
                                                              & \textbf{\begin{tabular}[c]{@{}c@{}}Device-device\\ \cite{naranjo2019focan,9296560,ranji2020eedos,liu2021novel} \end{tabular}}                                                     & High                                                       & Moderate                                                              & High                 & High                                                                & \begin{tabular}[c]{@{}c@{}}High-mobility \\ scenarios\end{tabular}                & \begin{tabular}[c]{@{}c@{}}High scalability, \\ low latency\end{tabular}                   & \begin{tabular}[c]{@{}c@{}}Limited \\ device \\ resources\end{tabular}       \\ \hline
\multirow{5}{*}{\textbf{Inter}}                               & \textbf{\begin{tabular}[c]{@{}c@{}}Cloud-edge\\ \cite{forecast2019cisco}\end{tabular}}                                                               & \begin{tabular}[c]{@{}c@{}}Low to\\  Moderate\end{tabular} & Moderate                                                              & High                 & Moderate                                                            & \begin{tabular}[c]{@{}c@{}}Media streaming,\\  IoT\end{tabular}                   & \begin{tabular}[c]{@{}c@{}}Load \\ distribution\end{tabular}                               & \begin{tabular}[c]{@{}c@{}}Limited \\ capabilities\end{tabular}              \\ \cline{2-9} 
                                                              & \textbf{\begin{tabular}[c]{@{}c@{}}Edge-device \\ \cite{gao2020edge4sys,li2019edge}\end{tabular}}                                                      & Low                                                        & High                                                                  & Moderate             & Moderate                                                            & \begin{tabular}[c]{@{}c@{}}Real-time \\ applications\end{tabular}                 & \begin{tabular}[c]{@{}c@{}}Low latency,\\ high energy\\  efficiency\end{tabular}           & \begin{tabular}[c]{@{}c@{}}Complex resource\\  management\end{tabular}       \\ \cline{2-9} 
                                                              & \textbf{\textbf{\begin{tabular}[c]{@{}c@{}}Cloud-device\\
                                                              \cite{10133831,duan2021joint}\end{tabular}} }                                                     & Moderate                                                   & \begin{tabular}[c]{@{}c@{}}Moderate \\ to High\end{tabular}           & High                 & Moderate                                                            & Hybrid scenarios                                                                  & \begin{tabular}[c]{@{}c@{}}Scalability, \\ flexible \\ resource \\ allocation\end{tabular} & High latency                                                                 \\ \cline{2-9} 
                                                              & \textbf{\begin{tabular}[c]{@{}c@{}}Cloud-edge\\ -device \cite{DBLP:journals/cloudcomp/MorshedJSGVR18}\end{tabular}}     & \begin{tabular}[c]{@{}c@{}}Low to \\ Moderate\end{tabular} & High                                                                  & High                 & High                                                                & \begin{tabular}[c]{@{}c@{}}Complex, multi-tier \\ applications\end{tabular}       & \begin{tabular}[c]{@{}c@{}}Highly \\ flexible\end{tabular}                                 & \begin{tabular}[c]{@{}c@{}}High management \\ complexity\end{tabular}        \\ \cline{2-9} 
                                                              & \textbf{Ad-hoc \cite{wang2023starling,wang2021unsupervised}
}                                                           & High                                                       & \begin{tabular}[c]{@{}c@{}}Low to\\ Moderate\end{tabular}             & High                 & High                                                                & \begin{tabular}[c]{@{}c@{}}Dynamic and \\ unpredictable \\ scenarios\end{tabular} & \begin{tabular}[c]{@{}c@{}}self-\\ organizing\end{tabular}                                 & \begin{tabular}[c]{@{}c@{}}Suboptimal\\  resource \\ allocation\end{tabular} \\ \hline
\end{tabular}
\label{table:Comparison of Various Commonly used Architectures}
\end{table*}

\subsubsubsection{Centralized Architecture}
In contrast to other collaborative approaches, centralized strategies operate under a single governing entity, either a cloud or edge. Such a centralized system takes responsibility for resource management across all nodes within that layer. This finds particular favour when:
% \vspace{-1.5em}
\begin{itemize}[leftmargin=*]
\item Resources are homogeneous across entities.
\item The processing capabilities of the entity are similar.
\item There is a high degree of uncertainty in predicting resource demands across individual entities.
% \vspace{-0.2em}
\end{itemize}
Centralized architecture yields more efficient and reliable outcomes compared to decentralized ones. Guo \emph{et al.} addressed computational offloading with a centralized heuristic algorithm in dense mobile networks \cite{guo2018computation}. Erfan \emph{et al.} introduced DRF-ER for fair multi-resource allocation in EC, adapting the dominant resource fairness (DRF) model to include external resources \cite{meskar2018fair}. Additionally, Kostis \emph{et al.} proposed a centralized scheduler to balance queues and minimize core interference, enhancing server efficiency \cite{kaffes2019centralized}.
    
\subsubsection{\textbf{Intra-layer Architecture}}   
% As the name suggests, Intra-layer Architectures focus on optimizing resource allocation and workload distribution within a single layer of the system. 
Two primary architectures, namely edge-edge and device-device, have evolved as significant shifts from traditional centralized models.

\subsubsubsection{Edge-edge Architecture} 
In EI, edge-edge architecture marks a shift from conventional centralized models. It uses edge interactions for better resource and workload distribution, which is beneficial for geographically dispersed networks.
A key study is Huang \emph{et al.}'s introduction of parked vehicle EC (PVEC), which lets vehicle EC (VEC) servers utilize parked vehicles' (PV) idle resources for workload allocation \cite{huang2018parked}. Similarly, Na \emph{et al.} used edge gateways to reduce edge workloads and maximize IoT efficiency \cite{na2018frequency}. Miao \emph{et al.} also explored dynamic task offloading between edges for user QoE \cite{miao2020intelligent}. Thai \emph{et al.} proposed a tiered collaboration model for cost-minimizing horizontal and vertical offloading \cite{thai2019workload}.

\subsubsubsection{Device-device Architecture}     
While edge-edge architectures have been well-investigated, the potential of device-device collaborations, especially for executing deep neural networks (DNNs), remains largely underexplored. Device-device architecture offers a paradigm shift with advantages:
\begin{itemize}[leftmargin=*]
\item \textbf{Enhanced Local Resources:} Collaboration allows for more efficient utilization of local computing power.
\item \textbf{Improved Privacy:} Local processing minimizes the need for external data transmission.
\item \textbf{Reduced Bandwidth Dependency:} Localized operations lessen the burden on network resources.
\end{itemize}
These advantages become particularly salient in high-mobility or remote environments, where traditional infrastructure might be lacking \cite{naranjo2019focan}.

In terms of empirical studies, Zeng \emph{et al.} developed CoEdge, a distributed DNN computing system that partitions workloads to adapt to device capabilities and network conditions \cite{9296560}. Ranji \emph{et al.} focused on improving energy efficiency and latency in device-to-device (D2D) architectures \cite{ranji2020eedos}. Liu \emph{et al.} offered a novel cellular D2D-MEC system, highlighting effective task offloading and resource allocation strategies \cite{liu2021novel}.

\subsubsection{\textbf{Inter-layer Architecture}} 
Inter-layer architectures, encompassing cloud-edge, edge-device, and cloud-edge-device collaborations, leverage multi-layer resources to improve latency, energy use, and computational load.

\subsubsubsection{Cloud-edge Architecture} The cloud-edge architecture minimizes latency and enhances user QoE, moving beyond the slower cloud-only systems. For instance, edge caching can immediately update product views in mobile shopping applications, providing faster responses to customer requests.
In the domain of multimedia streaming, video transcoding also is an indispensable, yet resource-intensive operation typically relegated to clouds \cite{forecast2019cisco}. This approach, while effective, has the drawback of added latency. Hence, researchers like Yoon \emph{et al.} suggest offloading these tasks to edge to achieve efficient and scalable solutions \cite{DBLP:conf/edge/YoonLB16}. Similarly, studies such as those by DMRA \emph{et al.} propose to expand edge layer's role, focusing on contractual arrangements between service providers and edges, aiming to optimize resource use and minimize latency \cite{zhang2019dmra}.

\subsubsubsection{Edge-device Architecture} 
The edge-device architecture in EI focuses on minimizing latency and improving energy efficiency by leveraging closer computational resources. This approach enables data processing at the edge and on the device itself, avoiding the need for data to travel to remote clouds. Such an architecture is especially beneficial for applications where real-time performance is crucial.
This architecture particularly excels in environments where rapid response times are essential for functionality or user experience. Many researchers have explored various facets of this collaborative architecture. For instance, the Edge4Sys framework suggests that the edge-device synergy can yield significant reductions in both network traffic and computational latency \cite{gao2020edge4sys}. On a related note, advanced techniques such as `early exit' have been developed to make this collaboration more efficient \cite{li2019edge}. 

\subsubsubsection{Cloud-device Architecture} 
The architecture emerges to alleviate latency and energy concerns associated with inference in cloud-only systems. By enabling a hybrid approach that optimizes task allocation, these architectures facilitate collaborative processing and reduce unnecessary data transmissions, thereby balancing computational needs with efficiency.
Research into cloud-device architectures has led to innovations such as Hu \emph{et al.}'s object detection system, which delegates complex tasks to cloud and simple ones to local devices \cite{10133831}, and Duan \emph{et al.}'s binary search algorithm for optimal DNN partitioning \cite{duan2021joint}. Systems tend to be more effective in low-mobility settings like surveillance systems, with a focus on minimizing latency. For scenarios with high mobility, different approaches such as SPINN are necessary \cite{laskaridis2020spinn}.
    
\subsubsubsection{Cloud-edge-device Architecture}
The cloud-edge-device framework combines the computing power of cloud computing with the edge's low latency. This cloud-edge-device framework offers refined task and data distribution, thereby improving scalability and efficiency. While it can alleviate network bandwidth stress and reduce transmission latency, the edge's limited computing capabilities for large-scale models can hinder its full potential.
Research, such as the studies conducted by Morshed \emph{et al.} analyze the complexities of developing cloud-edge-device collaborative DL algorithms \cite{DBLP:journals/cloudcomp/MorshedJSGVR18}. These algorithms are designed to be resource and data-aware, considering a multitude of factors like heterogeneity data models and resource availability. Yet, with an increase in the number of involved entities, focus optimization shifts to overall system cost and stability.
    
\subsubsubsection{Ad-hoc Architecture} Ad-hoc architecture, which allows for spontaneous resource allocation, is crucial in volatile or dynamic settings. It allows spontaneous resource allocation based on current system needs and limitations. In emergencies, resources might be allocated based on the event's severity and available assets. Wang \emph{et al.} suggest this strategy can be vital when traditional structures fall short due to fluctuating needs \cite{wang2023starling}. Ad-hoc architecture in SL offers flexibility for dynamic settings but faces challenges in efficient resource allocation, balancing responsiveness with system efficiency. Despite existing the above challenges, their adaptive nature renders them particularly relevant in certain contexts where responsiveness is paramount, like the unsupervised learning scenarios emphasized by Wang \emph{et al.} \cite{wang2021unsupervised}.

% In Systems of Layers' social stratification context, diverse architectures allow for resource and task optimization. 
Table \ref{table:Comparison of Various Commonly used Architectures} outlines the attributes of architectures. Centralized architectures offer control and consistency, while intra- and inter-layer ones offer flexibility for dynamic needs. The selection of an appropriate architecture should be based on the specific requirements of the EI system.

\vspace{-0.5em}
\subsection{Lessons Learned}

\begin{table}[t]
\color{black}
\centering
\caption{Summary of Key Symbols in Section V}
\label{tab:key_symbol_summary_section_v}
\begin{tabular}{|m{0.2\linewidth}|m{0.65\linewidth}|}
\hline
\textbf{Symbol} & \textbf{Description} \\
\hline
$F(\cdot)$ & Socialized learning function \\
\hline
$\theta, \theta_t, \theta_i$ & Model parameters \\
\hline
$N$ & Number of devices or participants \\
\hline
$\mathcal{L}, \mathcal{L}_s, \mathcal{L}_t$ & Loss functions \\
\hline
$D_i$ & Local dataset for device $i$ \\
\hline
$\pi, \pi^G, \pi^g$ & Policies in reinforcement learning \\
\hline
$f_C, f_S$ & Device and server computation functions in split learning \\
\hline
$P_i$ & Personalized model for the $i$-th edge node \\
\hline
$w_G, w_i$ & Global and local model weights in FL \\
\hline
$Z_i, Z_{\text{combined}}$ & Feature representations and combined representation in multi-modal learning \\
\hline
$Q, K, V$ & Query, key, and value in attention mechanism \\
\hline
$X, Y$ & Treatment and outcome variables in causal learning \\
\hline
$\lambda$ & Regularization weight or trade-off parameter \\
\hline
$\mathcal{D}$ & Set of expert demonstrations in imitation learning \\
\hline
$H$ & Cross-entropy function in knowledge distillation \\
\hline
$\alpha$ & Weight for the distillation loss or learning rate \\
\hline
$P_{\text{ensemble}}, P_n$ & Ensemble and individual model predictions \\
\hline
$D, G$ & Discriminator and generator networks in adversarial learning \\
\hline
$M, m_i$ & Set of all modalities and individual modality in modality-agnostic learning \\
\hline
\end{tabular}
\end{table}

Integrating SL in EI systems accelerates distributed intelligence processing, necessitating robust encryption and advanced analytics. Key challenges include scalability to handle evolving learning algorithms, heightened security and privacy concerns, and ensuring efficient energy use. The variety of devices and the need for \textbf{seamless software-hardware integration} further complicate synchronized learning across platforms.
Reliable real-time processing and interoperability across diverse devices and models are crucial, particularly for critical SL applications. \textbf{This integration must strike a balance between centralized control and distributed resilience, optimizing resource allocation and making strategic processing decisions.}

%分段
Future advancements should concentrate on refining processes to enhance scalability, interoperability, and the development of robust networking strategies. This will support the increasing complexity and scalability needs of SL in EI, ensuring robust and efficient performance in varied operational environments. \textbf{Strategic leveraging of layer strengths and a proactive approach} to security and resource management will be paramount.

% \vspace{-0.5em}
\section{Socialized Training for EI}\label{section6}
In this section, we explore the intricacies of socialized training mechanisms within the broader context of EI. %Drawing analogies from human social interactions, we discuss various ML approaches that enable SL in EI. 
Our analysis is structured around three aspects of social interaction: cooperation, submissive and conflict interaction. {\color{black}A summary table of the key symbols is described in Table \ref{tab:key_symbol_summary_section_v}.}

\vspace{-1em}
\subsection{{Intra-layer Cooperation in Socialized Training}}
\subsubsection{\textbf{Mutual Assistance}}%合作：互助,平等关系
%In the struggle against external forces, such as nature, humans achieve self-preservation by compensating for individual limitations through mutual assistance. Drawing inspiration from the operational mechanisms of human society, when faced with large and complex training tasks, agents within a specific layer can also compensate for individual limitations through information and resource exchange. Additionally, by combining the intelligence of multiple agents in the training process, both model performance and robustness are significantly enhanced.
Humans overcome various challenges through mutual assistance. Hence, devices within specific layers compensate for individual limitations through information and resource exchange, enhancing performance and robustness.

\begin{figure}[pt]
  \centering
  \includegraphics[width=3.49in]{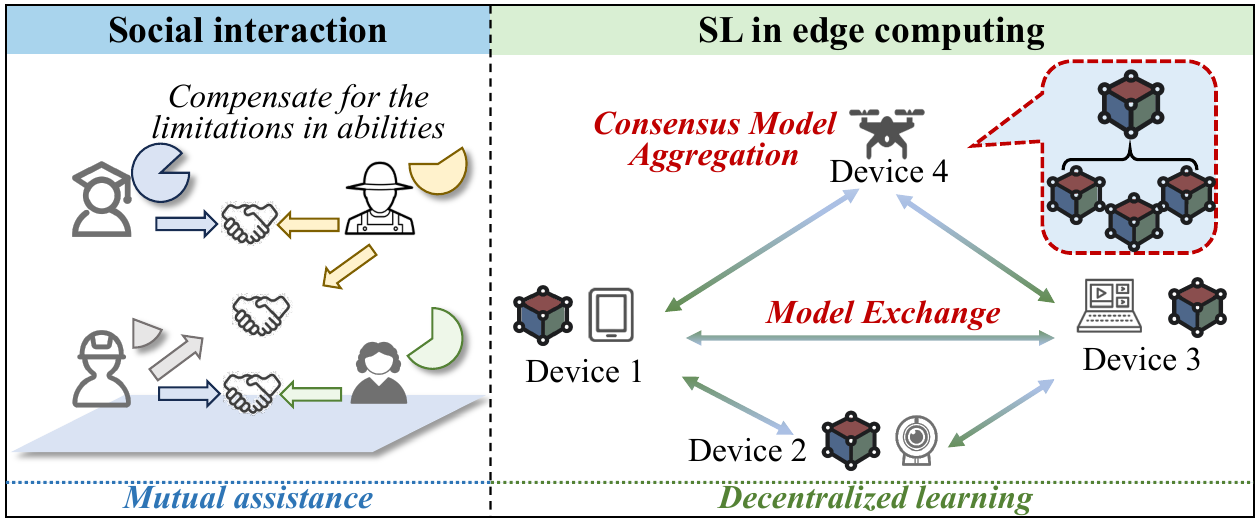}      
  \caption{Illustration of \textbf{decentralized learning} in socialized training.}\label{Decentralized learning}
  % \vspace{-1em}
\end{figure}

\subsubsubsection{Decentralized Learning} 
%Decentralized learning serves as an embodiment of mutual assistance within machine societies, akin to how individuals in human societies collaborate to accomplish common objectives. Unlike hierarchical structures where a central authority governs, decentralized learning enables peer-to-peer cooperation among agents in a given layer of the EI framework
Decentralized learning, much like human collaboration, allows devices in the EI framework to cooperate peer-to-peer, bypassing central authorities \cite{DBLP:conf/nips/LianZZHZL17}, as shown in Fig. \ref{Decentralized learning}. Devices share and combine local datasets or partial models, thus compensating for limitations in resources or information. 

{\color{black}The SL function Eq. \eqref{SL-function-define} for decentralized learning can be represented as $F([\text{devices}], [\text{mutual assistance}], \varnothing,$ $ \text{train}, \text{distributed} \text{ nodes})$. The optimization objective is:
\vspace{-0.5em}
\begin{equation}
\begin{aligned}
\min_{\theta} \frac{1}{N} \sum_{i=1}^{N} \mathcal{L}(D_i, \theta),
\end{aligned}
\vspace{-0.5em}
\end{equation}
where $\theta$ represents the global model parameters, $N$ is the total number of devices, $\mathcal{L}$ is the loss function, and $D_i$ is the local dataset for device $i$.
}

%This horizontal cooperation enhances the collective learning experience, paralleling mutual assistance in human communities where individuals collaborate to overcome challenges that they couldn't face alone.

Decentralized learning offers computing efficiency, scalability, and data locality, with SL building on these advantages to enhance privacy and resilience by maintaining data on local devices. These methods are notably robust against challenges such as data poisoning and system bottlenecks, depending on effective decentralized averaging and network topologies \cite{Swarm2021, DBLP:conf/icml/LuS21}. Popular techniques for data aggregation in such environments include consensus-based algorithms and diffusion strategies \cite{DBLP:journals/spm/SayedTCZT13}.

{\color{black}
\textbf{Communication Optimization in Decentralized Learning:}
Decentralized learning heavily relies on efficient communication among devices to achieve convergence and optimal performance. Minimizing communication overhead while ensuring fast and reliable information exchange is a critical challenge. Researchers have proposed various techniques to address this issue, such as graph-based aggregation schemes, quantization, and sparsification of model updates. These methods aim to reduce the amount of data transmitted between devices while maintaining the quality of the learned models. Moreover, the design of communication-efficient decentralized learning algorithms often involves joint optimization of learning and communication objectives, taking into account factors such as bandwidth limitations and device heterogeneity \cite{wang2019adaptive}.}

\begin{figure}[pt]
  \centering
  \includegraphics[width=3.49in]{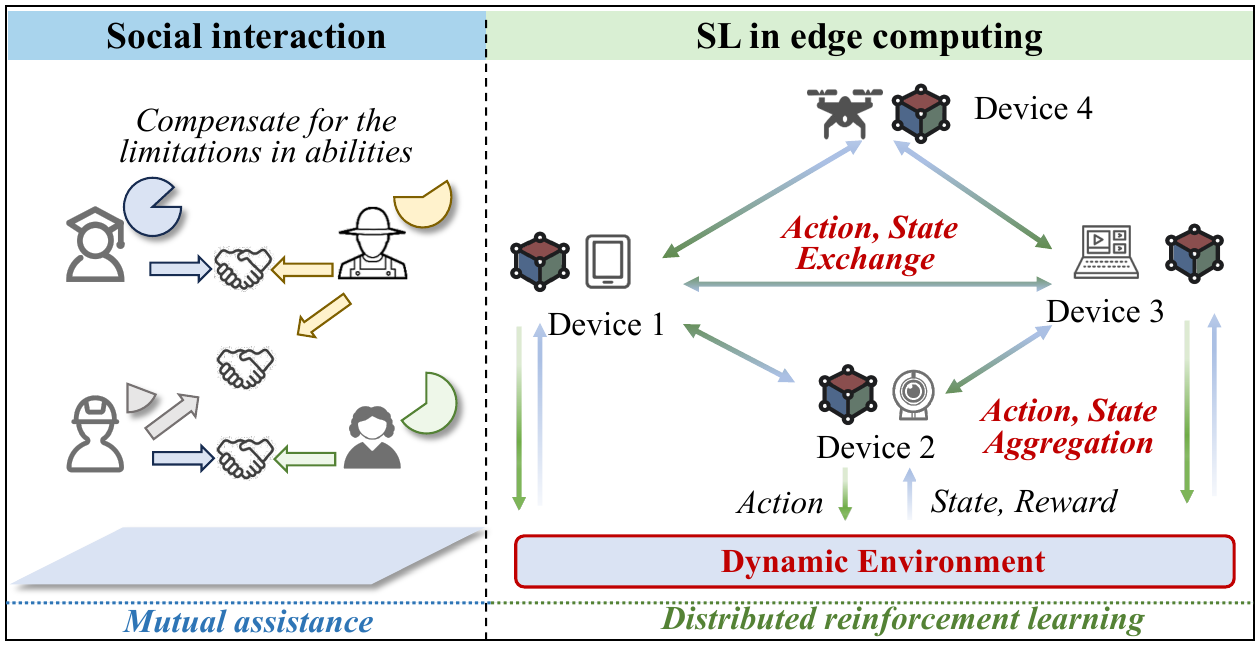} 
  \caption{Illustration of \textbf{distributed reinforcement learning} in socialized training.}\label{Distributed Reinforcement Learning}
  % \vspace{-0.5em}
\end{figure}

\subsubsubsection{Distributed Reinforcement Learning (DRL)} %分布式强化学习
%Distributed reinforcement learning (DRL) exemplifies mutual assistance in computational environments by facilitating coordinated decision-making in complex, dynamic settings \cite{DBLP:journals/spm/ArulkumaranDBB17}. This becomes particularly salient in multi-agent reinforcement learning (MARL), where agents collectively interact with their environment to achieve common objectives \cite{DBLP:conf/sigmetrics/ChenSX18}, 
DRL facilitates coordinated decision-making in dynamic settings \cite{DBLP:journals/spm/ArulkumaranDBB17}. Especially in multi-agent reinforcement learning (MARL), devices interact with their environment for shared objectives \cite{DBLP:conf/sigmetrics/ChenSX18},
as shown in Fig. \ref{Distributed Reinforcement Learning}.
%Rather than each agent selfishly maximizing its individual rewards, MARL promotes inter-agent cooperation.
%, akin to mutual assistance in human societies.
Devices share critical information, like state-action pairs or local policies, to improve collective decision-making. This collaborative behavior counters challenges of large state-action spaces, delayed rewards, and heterogeneous device behavior, leveraging mutual assistance to achieve higher performance than can be attained individually.

{\color{black}
The SL function Eq. \eqref{SL-function-define} for DRL is $F([\text{devices}], [\text{mutual} $ $\text{assistance}], \varnothing, \text{train}, \text{multi-agent})$. The optimization objective function can be shown as:
\vspace{-0.5em}
\begin{equation}
\vspace{-0.5em}
\begin{aligned}
\max_{\pi} \mathbb{E}{\tau \sim \pi} \left[\sum_{t=0}^{T} \gamma^t R(s_t, a_t)\right],
\end{aligned}
\end{equation}
where $\pi$ is the policy, $\tau$ represents the trajectories, $R(s_t, a_t)$ is the reward function, $\gamma$ is the discount factor, and $T$ is the time horizon.
}

In scenarios such as autonomous driving, where centralized control is not practical, devices collaborate using local communications on a network graph in MARL, ensuring peer-to-peer interaction for mutual assistance. Zhang \emph{et al.} advanced this field by introducing decentralized actor-critic algorithms with function approximation and an entropy-regularized policy gradient method that relies on neighborhood information sharing for multi-environment, multi-agent policy learning \cite{zeng2021decentralized}.

\vspace{-0.8em}
\subsection{{Inter-layer Cooperation in Socialized Training}}
\subsubsection{\textbf{Unidirectional Cooperation - Assistance}}
% In human society, assistance refers to the provision of social resources by actors to meet the needs of the assisted without the expectation of reciprocation. The actors will only undertake part of the work of the assisted or assist the assisted by completing the auxiliary task. Similarly, in machine society, the specific agents provide the assisted agents with the intermediate features needed for training, or they execute supplementary tasks to enhance the performance of the assisted agents, such as integrating knowledge of the assisted agents.
In human society, assistance involves providing resources without expecting reciprocation. Similarly, in machine society, devices offer intermediate features or perform supplementary tasks to enhance the performance of their counterparts.

\begin{figure}[pt]
  \centering
  \includegraphics[width=3.49in]{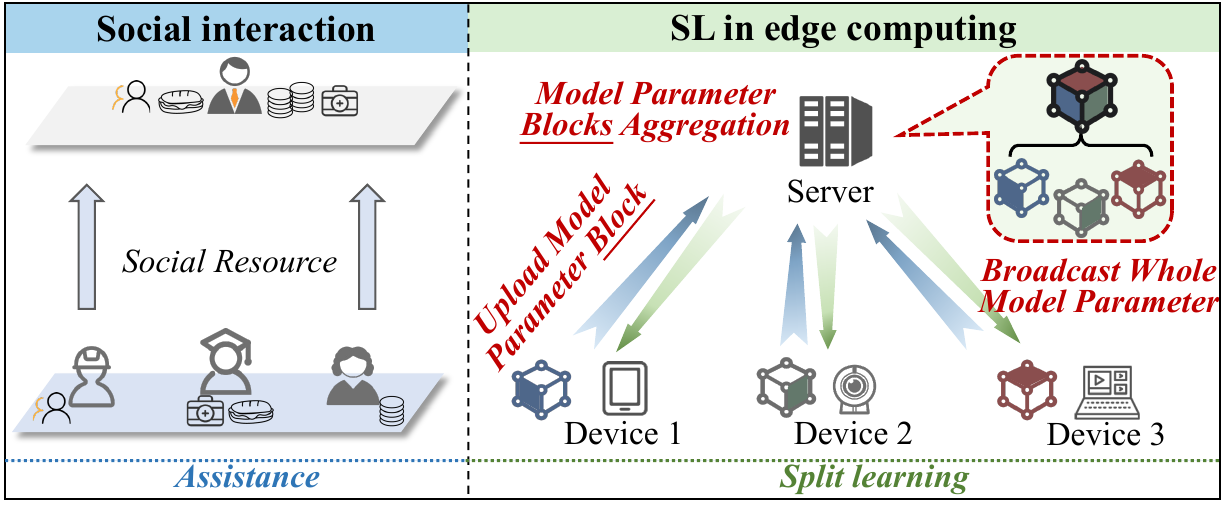}  
  \caption{Illustration of \textbf{split learning} in socialized training.}\label{Model split learning}
  % \vspace{-0.5em}
\end{figure}

\subsubsubsection{Split Learning} 
Split learning \cite{thapa2022splitfed} exemplifies unidirectional assistance in inter-layer cooperation. Fig. \ref{Model split learning} illustrates how the ML model is partitioned across multiple computational entities or `layers'. Devices process initial computation layers, forwarding outputs to ESs for further processing. Devices benefit from ESs' computational capabilities. Shared features let ESs process complex tasks while preserving data privacy and reducing communication overhead. Unlike approaches like FL, which update a global model across devices, it requires less data exchange.

{\color{black}
The SL function Eq. \eqref{SL-function-define} for split learning is $F([\text{server-}$ $\text{devices}],\varnothing, [\text{assistance}], \text{train}, \text{DNNs})$. The loss function is:
\vspace{-0.5em}
\begin{equation}
\begin{aligned}
\mathcal{L} = \text{Loss}\left( f_{S}\left( f_{C}(x; W_{C}); W_{S} \right), y \right),
\end{aligned}
\vspace{-0.5em}
\end{equation}
where $\mathcal{L}$ is the total loss, $f_{C}$ and $f_{S}$ represent the device and server computations with parameters $W_{C}$ and $W_{S}$, respectively, $x$ is the input data, and $y$ is the true output.}

Split learning is suitable for scenarios requiring data privacy, like medical diagnosis \cite{DBLP:journals/pieee/ParkSEKBKD21}, and in systems that require efficient use of computational resources. However, a challenge is managing the communication overhead from feature exchange in multi-device settings.
%By positioning split learning as a form of unidirectional assistance, we can better appreciate its role in fostering inter-layer cooperation. 
The devices unilaterally assist the ESs by undertaking a part of the overall computation, thereby allowing the assisted layer to perform more complex tasks or to do so more efficiently. 
%This one-sided contribution aligns closely with the social concept of assistance, making split learning a relevant and illustrative example of how unidirectional cooperation can manifest in machine societies.

{\color{black}\textbf{Communication Efficiency in Split Learning:}
Split learning is a distributed learning paradigm that enables collaborative training while preserving data privacy. However, the communication overhead incurred by exchanging intermediate activations and gradients between the client and server can be significant. To mitigate this issue, researchers have explored techniques such as activation compression, selective activation sharing, and model split optimization \cite{yansong2020end}. These approaches aim to reduce the amount of data communicated between the client and server while maintaining the performance of the split learning system. Additionally, the design of communication-efficient split learning architectures may involve optimizing the placement of the split layer and the trade-off between communication cost and model accuracy.}

\begin{figure}[pt]
  \centering
  \includegraphics[width=3.49in]{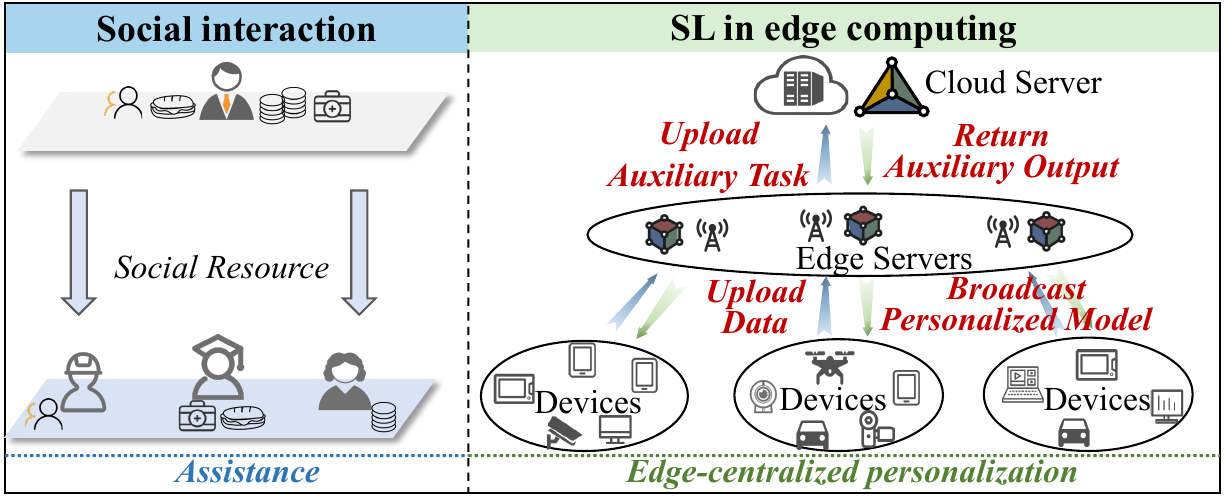}      
  \caption{Illustration of \textbf{edge-centralized personalization} in socialized training.}\label{Edge-centralized Personalization}
  % \vspace{-0.5em}
\end{figure}

\subsubsubsection{Edge-centralized Personalization} 
%服务器帮助设备训练，只有设备用
Edge-centralized personalization schemes feature unidirectional cooperation, where the cloud model enhances edge models for personalized tasks, as depicted in Fig. \ref{Edge-centralized Personalization}. The cloud model assists edges by aggregating knowledge, providing supervision, or recalibrating local models, fostering unidirectional cooperation. This approach fuses cloud computing power with edge personalization, underscoring unidirectional assistance's potential in inter-layer machine cooperation.

{\color{black}The SL function Eq. \eqref{SL-function-define} for edge-centralized personalization is $F([\text{cloud-edges}], \varnothing, [\text{assistance}], \text{train}, \text{DNNs})$. The personalized model for the $i$-th edge node is given by:
\vspace{-0.5em}
\begin{equation}
\begin{aligned}
P_i = f(C, E_i, D_i; \theta),
\end{aligned}
\vspace{-0.5em}
\end{equation}
where $P_i$ is the personalized model, $C$ represents the cloud model, $E_i$ is the local edge model, $D_i$ is the local data, and $\theta$ are the governing parameters.
}

In this context, the pioneering work by Lu \emph{et al.} \cite{DBLP:conf/edge/LuSTLZCP19} introduces COLLA, a collaborative learning framework designed for user location prediction. In COLLA, the cloud-based model acts as a global aggregator, distilling insights from multiple edge-based models to create personalized models for each edge. The cloud-based model effectively assists edges in generating more accurate predictive models, embodying the spirit of unidirectional cooperation.
%Similarly, Ding \emph{et al.} \cite{2020A} propose a framework where a cloudCNN provides ``soft supervision" to each local EdgeCNN. The CloudCNN generates higher-order features or learning signals that are transferred to the EdgeCNN, which then performs real-time inference on visual inputs. The cloud model essentially facilitates the edges in task-specific learning while expecting no direct benefits in return. 
Further extending this paradigm, Yao \emph{et al.} \cite{DBLP:conf/kdd/YaoWJHZY21} introduce a novel edge-cloud collaboration framework for recommendation systems that incorporates backbone-patch decomposition. This approach significantly reduces the computational burden on edges while maintaining high levels of personalization. The central model recalibrates edge backbone models to avoid local optimums through the MetaPatch mechanism.

\subsubsection{\textbf{Bidirectional Cooperation - Cooperation}}
% In human society, cooperation is defined as the collaborative effort of individuals within groups to achieve common goals. 
Cooperation is crucial in a machine society, where no single entity can achieve common goals independently. It depends on aligned objectives and an efficient division of labor, often through bidirectional collaboration at different layers. In training, his process entails both independent and collective modeling, characterized by continuous interaction through parameter exchanges and the potential sharing of representations. Such a process enables individuals to contribute to and benefit from the collective learning experience.

\begin{figure}[pt]
  \centering
  \includegraphics[width=3.49in]{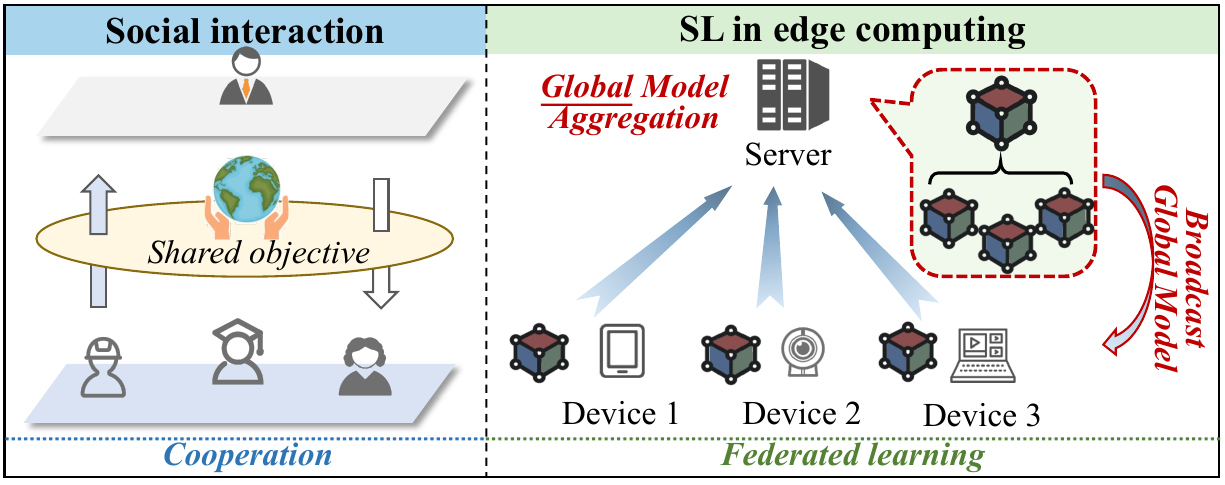}  
  \caption{Illustration of \textbf{federated learning} in socialized training.}\label{Federated learning}
  \vspace{-0.5em}
\end{figure}

\subsubsubsection{Federated Learning}
%FL represents an ideal case of inter-layer bidirectional cooperation because it inherently requires continuous, two-way interactions between participants for mutual benefit, as shown in Fig. \ref{Federated learning}. In this paradigm, all participating entities contribute to the collective intelligence and, in turn, benefit from it.
%, thus embodying the social essence of cooperation. 
FL is a bidirectional cooperation method where participants mutually benefit from shared collective intelligence, as illustrated in Fig. \ref{Federated learning}. It allows multiple participants to collaboratively train a global model without sharing private data. In FL, a coordinating server initializes the global model and distributes it to participants. These participants then update the model using their local data and send the updated models back to the server \cite{tan2022towards}. The server aggregates these updates to refine the global model. 

{\color{black} The SL function Eq. \eqref{SL-function-define} for FL is $F([\text{server-devices}], \varnothing,$ $[\text{cooperation}], \text{train}, \text{DNNs})$. The global model update in FL can be defined as:
\vspace{-0.5em}
\begin{equation}
\begin{aligned}
w_G^{t+1} = \frac{1}{N} \sum_{i=1}^{N} w_i^t,
\end{aligned}
\vspace{-0.5em}
\end{equation}
where $w_G^{t+1}$ is the updated global model at iteration $t+1$, $N$ is the total number of participants, and $w_i^t$ is the local model of participant $i$ at iteration $t$.}
Aggregation algorithms like FedAvg \cite{DBLP:journals/tpds/ZhouYL22} are integral to this process, combining the local models in a way that minimizes the global loss function.
%The bidirectional nature of this interaction is evident as each layer-both the server and edges-not only contributes to the global model but also benefits from the aggregated intelligence. 
This duality aligns closely with the social concept of cooperation, whereby mutual benefits are realized through collective efforts.

{\color{black}
\textbf{Communication Optimization in FL:}
FL has emerged as a promising approach for collaborative learning while preserving data privacy. However, the communication overhead associated with exchanging model updates between clients and the central server can be a significant bottleneck, especially in resource-constrained environments. To address this challenge, researchers have proposed various communication-efficient FL techniques, such as FetchSGD \cite{rothchild2020fetchsgd}, which combines sketching and quantization to reduce communication costs, and FedMix, which approximates mixup regularization under mean augmented FL to improve model generalization and reduce communication rounds. These methods aim to strike a balance between communication efficiency and model performance in FL systems, enabling their deployment in practical settings with limited bandwidth and computational resources.
}

\begin{figure}[pt]
  \centering
  \includegraphics[width=2.9in]{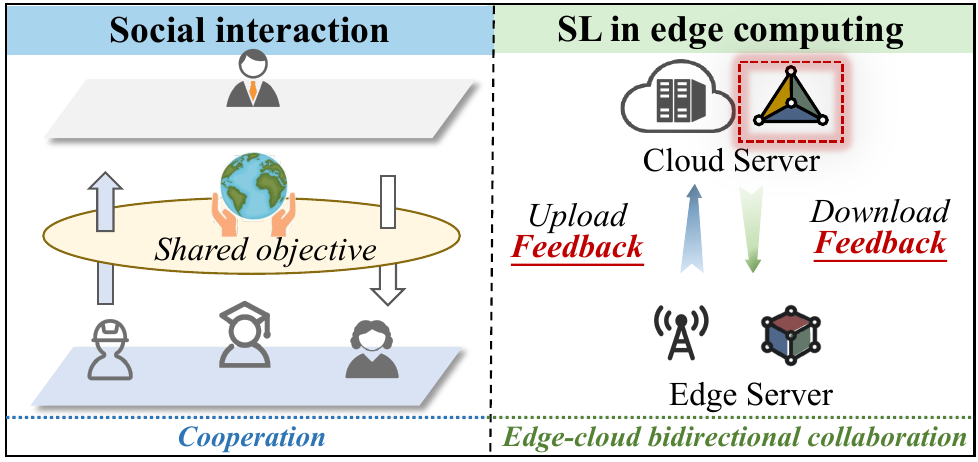}   
  \caption{Illustration of \textbf{edge-cloud bidirectional collaboration} in socialized training.}\label{Edge-cloud Bidirectional Collaboration}
  \vspace{-0.5em}
\end{figure}

\subsubsubsection{Edge-cloud Bidirectional Collaboration} 
Edge-cloud Bidirectional Collaboration, as depicted in Fig. \ref{Edge-cloud Bidirectional Collaboration}, capitalizes on the unique strengths of edge and cloud systems to address complex tasks such as real-time recommendations. This approach is exemplified by the Slow-Fast Learning mechanism on the Alibaba Luoxi Platform \cite{DBLP:journals/corr/abs-2109-12314}. In this model, the cloud and edge systems engage in a mutually beneficial exchange, where the cloud model aids the edge model in swift predictions, and the edge model provides real-time user feedback to the cloud. The interaction between edge and cloud is shown below, highlighting essence of cooperative intelligence.

%分段
{\color{black}The SL function Eq. \eqref{SL-function-define} for edge-cloud bidirectional collaboration is $F([\text{cloud-edges}], \varnothing, [\text{cooperation}], \text{train}, \text{DNNs})$. The collaborative outcome is:
\vspace{-0.5em}
\begin{equation}
\begin{aligned}
BC = L(E, C, D, R; \theta),
\end{aligned}
\vspace{-0.5em}
\end{equation}
where $BC$ is the collaborative outcome, $E$ and $C$ are the states of edge and cloud models, $D$ is the real-time edge data, $R$ is the cloud's response, and $\theta$ is the collaboration parameters.}

This bidirectional collaborative framework echoes the relationship between System I and II in human cognition, as described by Daniel Kahneman \emph{et al.} \cite{Kahneman2011}. System I is responsible for fast, intuitive reactions, while another undertakes slower, more deliberative reasoning \cite{DBLP:conf/iclr/MadanKGSB21}. The coexistence and interaction between these two systems enable them to exchange prior or privileged information in a timely manner, which in turn allows for better adaptive responses to environmental demands. This collaborative approach is increasingly relevant with the rise of advanced technologies like smartphones and IoT \cite{DBLP:journals/pieee/ZhouCLZLZ19}.

{\color{black}
\textbf{Communication-efficient Edge-cloud Collaboration:} Edge-cloud collaboration is a promising paradigm for leveraging complementary advantages of edge and cloud computing to enable intelligent and responsive services. However, the communication between edges and the cloud can introduce significant overhead and latency, hindering real-time performance of collaborative learning systems \cite{zhang2022communication}. 

Researchers have explored communication-efficient strategies for edge-cloud collaboration, such as hierarchical compression and selective update transmission. These techniques aim to reduce the volume of data exchanged between edge and cloud nodes while preserving the quality of the learned models and the timeliness of the collaborative inference process. Optimizing edge-cloud collaboration's communication efficiency allows SL systems to harness distributed computing resources while meeting real-world applications' stringent latency and bandwidth constraints.}

\subsubsection{\textbf{Bidirectional Cooperation - Exchange}}%合作：
% In human social interactions, exchange stands distinct from cooperation by emphasizing reciprocity. 
In human interactions, exchange requires reciprocity and is transactional, involving the giving and receiving of resources for mutual benefit. This contrasts with cooperation, which is oriented toward joint action and shared goals. In machine societies across cloud, edge, and device layers, this principle translates into a bidirectional flow of data and model parameters, optimizing system-wide performance despite each layer's autonomy.

\begin{figure}[pt]
  \centering
  \includegraphics[width=3.49in]{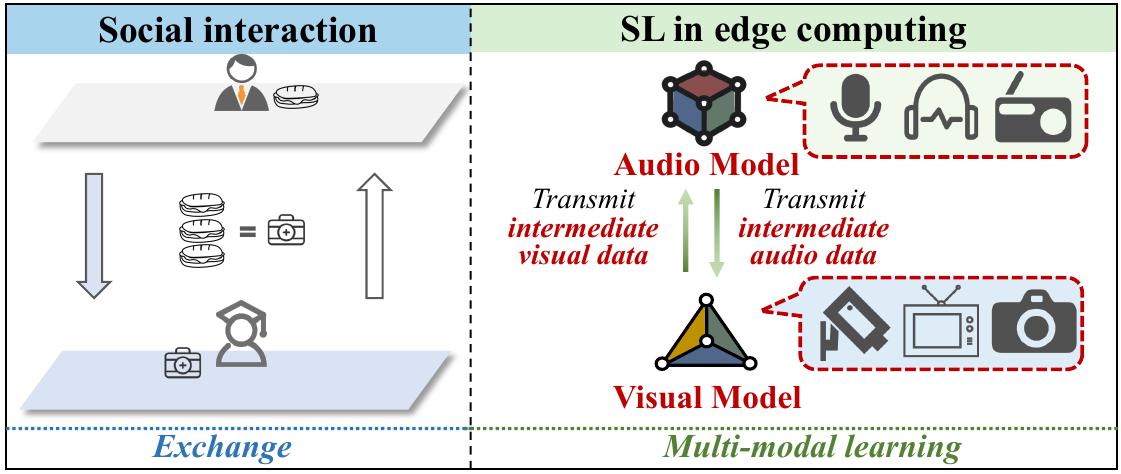}   
  \caption{Illustration of \textbf{multi-modal learning} in socialized training.}\label{Multi-modal learning}
  % \vspace{-0.5em}
\end{figure}

\subsubsubsection{Multi-modal Learning}
% Multi-modal learning exemplifies a highly nuanced form of inter-layer cooperation, grounded in the principle of exchange rather than mere cooperation. 
Multi-modal learning capitalizes on diverse data types, such as images, audio, and text, akin to human sensory processing \cite{DBLP:journals/inffus/UppalBHMPZZ22}. In systems integrating cloud, edge, and device layers, each layer processes different data modalities, as depicted in Fig. \ref{Multi-modal learning}. 

{\color{black} The SL function Eq. \eqref{SL-function-define} for multi-modal learning is $F([\text{cloud-edge-}$ $\text{devices}], \varnothing, [\text{exchange}], \text{train}, \text{modalities})$. The feature representations and combined representation are:
\vspace{-0.5em}
\begin{equation}
\begin{aligned}
Z_i = f_i(X_i; \theta_i), \quad Z_{\text{combined}} = L(Z_1, Z_2, \ldots, Z_n; \Theta_L),
\end{aligned}
\vspace{-0.5em}
\end{equation}
where $Z_i$ is the feature representation for modality $i$, $f_i$ is the transformation function with parameters $\theta_i$, $X_i$ is the input data of modality $i$, $Z_{\text{combined}}$ is the combined representation, and $L$ is the fusion function with parameters $\Theta_L$.
}

One noteworthy application that captures this essence of exchange is the multi-modal federated collaborative training framework by Qayyum \emph{et al.} \cite{9891834}. The framework builds a composite model for diagnosing COVID-19 by harmoniously integrating two modalities of data-X-ray and ultrasound images, across different centers in a cloud-edge environment. Here, the interaction is not merely cooperative but fundamentally reciprocal, as each layer significantly contributes to the shared model and benefits from it for its own objectives. 

We provide a comprehensive exploration of cooperation dynamics within ML layers, drawing parallels to human social interactions. A comparative overview is presented in Table \ref{table:Comparison of cooperation}, highlighting their distinct characteristics.

\begin{table*}[t]
\color{black}
\caption{\color{black}Comparative Analysis of Technical Approaches Reflecting Cooperation Relationships}
\centering
\begin{tabular}{|c|c|c|c|c|c|c|c|}
\hline
\textbf{\begin{tabular}[c]{@{}c@{}}Cooperation \\ Type\end{tabular}}           & \textbf{Technique}                                                                               & \textbf{\begin{tabular}[c]{@{}c@{}}Social \\ Relationship\end{tabular}}       & \textbf{\begin{tabular}[c]{@{}c@{}}Social \\ Characteristics\end{tabular}}                  & \textbf{\begin{tabular}[c]{@{}c@{}}Model \\ Interactivity\end{tabular}} & \textbf{\begin{tabular}[c]{@{}c@{}}Privacy \\ Maintenance\end{tabular}} & \textbf{Scalability} & \textbf{Resilience} \\ \hline
\multirow{3}{*}{Intra-layer}                                                   & \begin{tabular}[c]{@{}c@{}}Decentralized \\ Learning \cite{DBLP:conf/nips/LianZZHZL17,Swarm2021, DBLP:conf/icml/LuS21, DBLP:journals/spm/SayedTCZT13, wang2019adaptive}\end{tabular}                                & \multirow{3}{*}{\begin{tabular}[c]{@{}c@{}}Mutual \\ Assistance\end{tabular}} & \begin{tabular}[c]{@{}c@{}}Peer-to-peer \\ cooperation\end{tabular}                  & High                                                                    & High                                                                    & High                 & High                \\ \cline{2-2} \cline{4-8} 
                                                                               & \begin{tabular}[c]{@{}c@{}}Distributed \\ Reinforcement Learning \cite{zeng2021decentralized}\end{tabular}                    &                                                                               & \begin{tabular}[c]{@{}c@{}}Coordinated \\ decision-making\end{tabular}               & High                                                                    & Medium                                                                  & High                 & Medium              \\ \hline
\multirow{5}{*}{Inter-layer}                                                   & \begin{tabular}[c]{@{}c@{}}Model Split\\  Learning \cite{thapa2022splitfed, yansong2020end}\end{tabular}                                  & \multirow{2}{*}{Assistance}                                                   & \begin{tabular}[c]{@{}c@{}}Unilateral \\ assistance\end{tabular}                     & Medium                                                                  & High                                                                    & Medium               & Medium              \\ \cline{2-2} \cline{4-8} 
                                                                               & \begin{tabular}[c]{@{}c@{}}Edge-centralized\\ Personalization \cite{DBLP:conf/edge/LuSTLZCP19, DBLP:conf/kdd/YaoWJHZY21}\end{tabular}                       &                                                                               & Cloud aiding edges                                                                   & Medium                                                                  & High                                                                    & High                 & Medium              \\ \cline{2-8} 
                                                                               & \begin{tabular}[c]{@{}c@{}}Federated\\ Learning \cite{tan2022towards, DBLP:journals/tpds/ZhouYL22, rothchild2020fetchsgd}\end{tabular}                                     & \multirow{2}{*}{Cooperation}                                                  & \begin{tabular}[c]{@{}c@{}}Reciprocal benefit and \\ contribute equally\end{tabular} & High                                                                    & High                                                                    & High                 & High                \\ \cline{2-2} \cline{4-8} 
                                                                               & \begin{tabular}[c]{@{}c@{}}Edge-cloud Bidirectional\\ Collaboration \cite{DBLP:journals/corr/abs-2109-12314, Kahneman2011, DBLP:conf/iclr/MadanKGSB21, DBLP:journals/pieee/ZhouCLZLZ19, zhang2022communication}\end{tabular}                 &                                                                               & \begin{tabular}[c]{@{}c@{}}Interdependent and \\ reciprocal\end{tabular}             & High                                                                    & Moderate                                                                & High                 & High                \\ \cline{2-8} 
                                                                               & \begin{tabular}[c]{@{}c@{}}Multi-modal \\ Learning \cite{DBLP:journals/inffus/UppalBHMPZZ22, 9891834}\end{tabular}                                  & Exchange                                                                      & \begin{tabular}[c]{@{}c@{}}Exchanging diverse \\ data modalities\end{tabular}        & High                                                                    & Moderate                                                                & High                 & Moderate            \\ \hline
\end{tabular}
\label{table:Comparison of cooperation}
\end{table*}

\vspace{-0.8em}
\subsection{{Intra-layer Submissive Interaction in Socialized Training}}
\subsubsection{\textbf{Conformity}} % 可信学习___顺应
% In social interactions, individuals often conform to the norms and behaviors of the majority, either due to explicit pressures or implicit desires to fit in. 
In society, individuals often conform to majority norms due to various pressures or desires.
% Conformity is the process of actors acting in the way of others under certain pressure, that is, acting in the way of the majority of the group. 
In machine society, individuals mine separable discriminant features or similar parameter relationships among individuals in the training process and reduce model bias by increasing the impact of discriminant features on models or maximizing the retention of similar parameter relationships between individuals.

\begin{figure}[pt]
  \centering
  \includegraphics[width=3.49in]{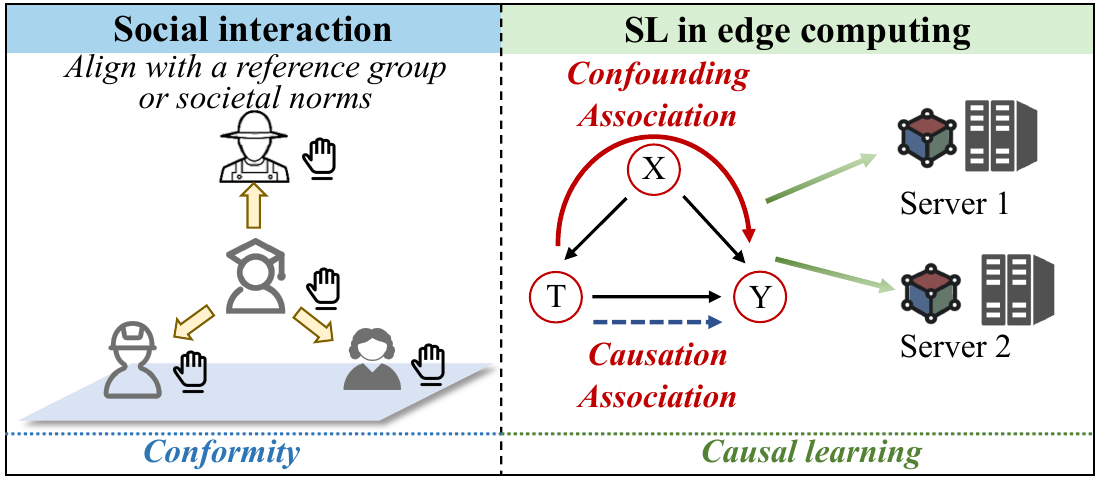}      
  \caption{Illustration of \textbf{causal learning} in socialized training.}\label{Causal Learning}
  % \vspace{-1em}
\end{figure}

 \subsubsubsection{Causal Learning}  
% Rather than merely mirroring observed behaviors, causality aims to unearth the fundamental reasons behind these behaviors, offering a more profound and genuine form of conformity. 
% Causality seeks the underlying reasons for behaviors, providing a deeper form of conformity. 
Causal learning in EI is crucial for maintaining fairness \cite{2020Causal} and reducing bias, particularly by facilitating universal model applicability through on-cloud training and mitigating overfitting in edge training. This approach is illustrated in Fig. \ref{Causal Learning}. A fundamental concept in causal learning is estimating the average causal effect (ACE) of one variable on another. Given a set of variables, we often are interested in the causal effect of one variable $X$ (treatment) on another $Y$ (outcome).
{\color{black}
The SL function Eq. \eqref{SL-function-define} for causal learning is $F([\text{servers}], [\text{conformity}], \varnothing, \text{train}, \text{variables})$. The average causal effect (ACE) is:
\vspace{-0.5em}
\begin{equation}
\begin{aligned}
ACE = \mathbb{E}[Y_1 - Y_0] = \mathbb{E}[Y_1] - \mathbb{E}[Y_0],
\end{aligned}
\vspace{-0.5em}
\end{equation}
where $Y_1$ and $Y_0$ are the potential outcomes under treatment ($X=1$) and control conditions ($X=0$), respectively, and $\mathbb{E}[\cdot]$ denotes the expectation.
}

Integrating model compression with causal learning for enhanced generalization is demonstrated in \cite{DBLP:journals/tacl/RotmanFR21}. Several other research studies have extensively explored the convergence of causality and out-of-domain generalization \cite{DBLP:conf/icml/TeshimaSS20, DBLP:conf/sigir/0007S0L20, DBLP:conf/iccv/YueS0Z21, DBLP:conf/kdd/KuangCAXL18, 9566788, DBLP:conf/mm/ZhangJWKZZYYW20}. For instance, \cite{DBLP:conf/iccv/YueS0Z21} underscores the importance of preserving semantics intrinsic to the target domain by leveraging disentangled causal mechanisms. The study in \cite{9566788} posits the enduring relationship between causal features and classifications across domains and champions the Markov Blanket \cite{DBLP:journals/csur/YuGLLWLW20} for strategic causal feature selection. The proposal by \cite{DBLP:conf/kdd/KuangCAXL18} introduces a causal regularizer, aiming to resurrect the causation between predictors and outcome variables when faced with unpredictable distributions. 
%In a similar vein, \cite{DBLP:journals/corr/abs-2110-01438} presents instrumental variable-based techniques, laying emphasis on sustaining invariant relationships between predictors and outcome variables across diverse domains.

\begin{figure}[pt]
    \centering
    \includegraphics[width=3.49in]{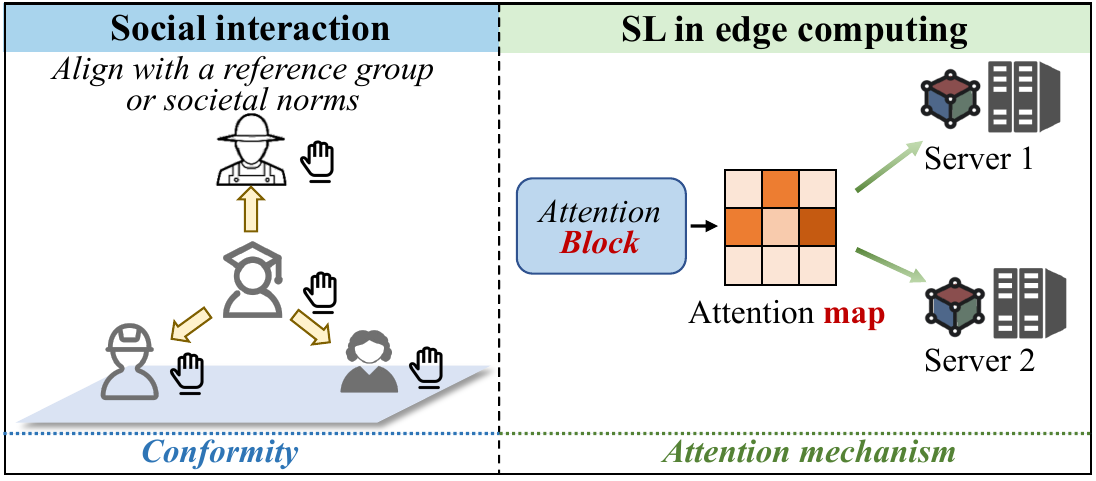}   
    \caption{Illustration of \textbf{attention mechanism} in socialized training.}\label{Attention mechanism}
    \vspace{-0.5em}
\end{figure}

\subsubsubsection{Attention Mechanism} %The attention mechanism in ML has drawn parallels to the social concept of conformity, where entities gravitate towards and focus on dominant or salient features, much like individuals conforming to the majority or prevalent norms. Such a mechanism, by selectively amplifying certain portions of the input while downplaying others, inherently embodies the principle of conforming to pivotal information \cite{vaswani2017attention}. For instance, in the realm of natural language processing, just as social groups give precedence to popular opinions, the attention mechanism prioritizes keywords in the source language to facilitate superior translations to the target language. Modern architectures like the Transformer, employing self-attention, underscore this conformity by weighting the importance of each word in relation to others. This evolution in design philosophy has ushered in models such as BERT\cite{kenton2019bert} and GPT\cite{brown2020language}, which further epitomize this 'conformity' by foregrounding contextually salient words in expansive text corpuses.
% The attention mechanism is where entities gravitate towards and focus on dominant or salient features, much like individuals conforming to the majority or prevalent norms. 
% The attention mechanism focuses on dominant features, similar to individuals conforming to prevalent norms.
Such a mechanism, by selectively amplifying certain portions of the input while downplaying others, embodies the principle of conforming to pivotal information \cite{vaswani2017attention}. %For instance, in the realm of natural language processing (NLP), the attention mechanism prioritizes keywords in the source language to facilitate superior translations to the target language.
Architectures like the Transformer, employing self-attention, underscore this conformity by weighting the importance of each word in relation to others. 
This evolution in design philosophy has ushered in models such as BERT \cite{kenton2019bert} and GPT \cite{brown2020language}, which further epitomize `conformity' by foregrounding contextually salient words in expansive text corpora. This mechanism is illustrated in Fig. \ref{Attention mechanism}.
{\color{black}The SL function Eq. \eqref{SL-function-define} for the attention mechanism is $F([\text{servers}], [\text{conformity}], \varnothing, \text{train}, \text{data})$. The attention mechanism is defined as follows:
\vspace{-0.5em}
\begin{equation}
\begin{aligned}
\text{Attention}(Q, K, V) = \text{softmax}\left(\frac{QK^T}{\sqrt{d_k}}\right)V,
\end{aligned}
% \vspace{-0.5em}
\end{equation}
where $Q$, $K$, and $V$ are the query, key, and value matrices, respectively, and $d_k$ is the dimension of the key vectors.}
$QK^T$ calculates attention scores, which, after normalization and conversion by the $\text{Softmax}$ function, help the model determine focus areas in the input data.

%In the visual realm, the attention mechanism reflects social conformity by emphasizing prominent and familiar regions of an image, much like society accentuates mainstream norms or trends. The diverse types of attention mechanisms, from channel to spatial attention, echo the varied nuances of social conformity. Channel attention, for example, mirrors our focus on prevailing trends, distinguishing vital channels akin to distinctive social figures \cite{chen2017sca}. Similarly, spatial attention resonates with the social spotlight on specific locations or events \cite{jaderberg2015spatial}. This selective focus of attention mechanisms, especially evident in models like BERT, serves as an embodiment of social conformity \cite{li2020does}. However, akin to social behaviors, it's essential to recognize that while attention provides significant insights, it's not an exhaustive explanation of model behavior, much like adhering to a norm doesn't fully capture society's complexity \cite{wiegreffe2020attention}.

The attention mechanism reflects social conformity by emphasizing prominent familiar regions of an image. Both channel and spatial attention mechanisms echo the varied nuances of social conformity. Channel attention mirrors our focus on prevailing trends, distinguishing vital channels akin to distinctive social figures \cite{chen2017sca}. Similarly, spatial attention resonates with the social spotlight on specific locations or events \cite{jaderberg2015spatial}. This selective focus of attention mechanisms, evident in models like BERT, serves as an embodiment of social conformity \cite{li2020does}. However, it's essential to recognize that while attention provides significant insights, it's not an exhaustive explanation of model behavior, much like adhering to a norm doesn't fully capture society's complexity \cite{wiegreffe2020attention}.

\vspace{-0.5em}
\subsection{{Inter-layer Submissive Interaction in Socialized Training}}
\subsubsection{\textbf{Unidirectional Submissive Interaction - Suggestion\&imitation}}
In the same way that suggestion and imitation play roles in human interactions, with individuals adapting behaviors observed in others, models also rely on similar principles during training. Complex models act as authoritative guides, transferring knowledge to newer models through intermediate features or outcomes. New models then aim to closely replicate these features and outcomes in an attempt to `imitate' the performance of more sophisticated models.

\begin{figure}[pt]
  \centering
  \includegraphics[width=3.42in]{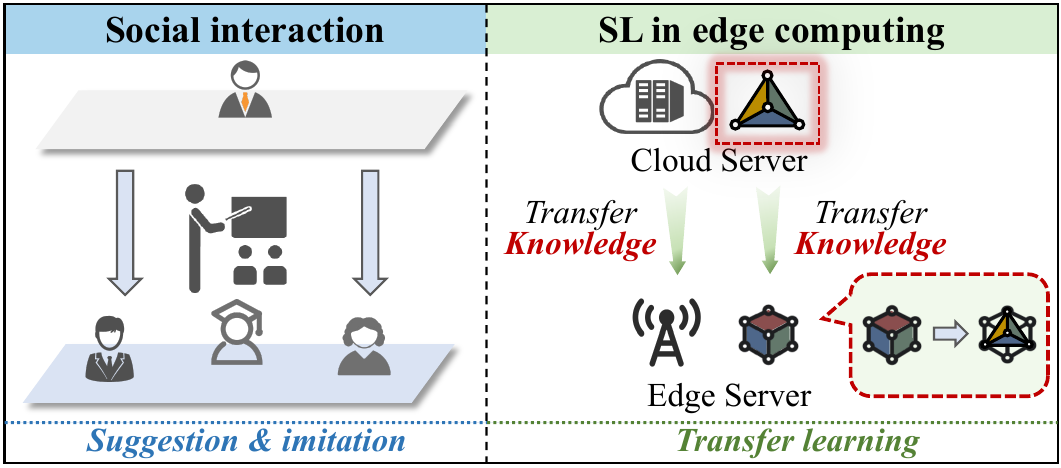}      
  \caption{Illustration of \textbf{transfer learning} in socialized training.}\label{Transfer learning}
  \vspace{-0.5em}
\end{figure}

\subsubsubsection{Transfer Learning} 
% Transfer learning, at its core, serves as a prime exemplar of the ``Inter-layer Submissive Interaction" paradigm, particularly resonating with the ``Unidirectional Submissive Interaction" dynamic. 
Transfer learning exemplifies the inter-layer submissive interaction concept, especially the ``Unidirectional Submissive Interaction" aspect.
% As observed in social structures, a novice or less-informed entity (the imitator) often seeks guidance and direction from a more experienced or authoritative figure (the suggester). 
Newer models can draw from the knowledge of pre-trained, more sophisticated models, as shown in Fig. \ref{Transfer learning}.
%Thus, these new models willingly submit to the insights of their predecessors, imitating their patterns and learning from the suggestions encapsulated within the transferred features.
% The process, as formulated below, encompasses knowledge transfer from a source domain to adapt to a target domain: % This holistic representation underscores the continuity and interdependence intrinsic to effective transfer learning.
% \vspace{-0.5em}
{\color{black}The SL function Eq. \eqref{SL-function-define} for transfer learning is $F([\text{server-devices}],\varnothing, $ $[\text{suggestion\&imitation}], \text{train}, \text{DNNs})$. The optimization objectives for the source and target domains are:
\vspace{-0.5em}
\begin{equation}
\begin{aligned}
&\underset{\theta}{\text{min}} \; \mathcal{L}_s(f(X_s; \theta), Y_s) + \lambda \mathcal{R}(\theta), 
    \quad \theta_t = \text{Transfer}(\theta^*), \\
&\quad\quad \underset{\theta_t}{\text{min}} \; \mathcal{L}_t(f(X_t; \theta_t), Y_t) + \lambda \mathcal{R}(\theta_t),
\end{aligned}
\vspace{-0.5em}
\end{equation}
where $\mathcal{L}_s$ and $\mathcal{L}_t$ are the loss functions for the source and target domains, $f$ is the model with parameters $\theta$ and $\theta_t$, $X_s$ and $X_t$ are the source and target data, $Y_s$ and $Y_t$ are the corresponding labels, $\mathcal{R}$ is the regularization term, and $\lambda$ is the regularization weight. $\text{Transfer}(\cdot)$ denotes the knowledge transfer function.}

%Given the intrinsic heterogeneity in data distribution across edges and between edge and cloud in collaborative systems, the need for such submissive interactions is accentuated. Models, recognizing the disparities, turn to the pre-trained knowledge hierarchies to bridge these gaps. Heterogeneous transfer learning, as explored in works such as \cite{DBLP:conf/eccv/ZhangQYPWT20}, acts as a conduit for this suggestive imitation. While symmetric transformations advocate for domain-agnostic representations \cite{DBLP:journals/tip/YehHW14}, echoing social norms of universal standards, asymmetric transformations \cite{DBLP:journals/tcyb/DengZKHPL22} emphasize domain alignment. This mirrors the social dynamic where one domain or group modulates its behavior to resemble another, often seen as superior. Such submissive interactions are further crystallized in methods like Semi-Supervised Kernel Matching Domain Adaptation \cite{DBLP:journals/pami/XiaoG15}, highlighting the model's endeavor to assimilate and conform to the insights of a more informed predecessor.

The diversity in data distribution across edges and the cloud highlights the significance of heterogeneous transfer learning, as demonstrated in \cite{DBLP:conf/eccv/ZhangQYPWT20}. This learning approach adapts to varying data characteristics, employing symmetric transformations for domain-agnostic representations, as discussed in \cite{DBLP:journals/tip/YehHW14}, and asymmetric transformations for domain-specific alignment, highlighted in \cite{DBLP:journals/tcyb/DengZKHPL22}. 
This mirrors the social dynamic where one domain or group modulates its behavior to resemble another, often seen as superior. This concept is further exemplified by semi-supervised kernel matching domain adaptation \cite{DBLP:journals/pami/XiaoG15}, showcasing models' capacity to assimilate and integrate more advanced insights.

\subsubsubsection{Imitation Learning}
In social constructs, people often follow the guidance of authoritative figures. Similarly, in imitation learning, a novice model observes and imitates the expert model's behavior, akin to learning through observation in social interactions. This method involves recording the expert's actions and corresponding states, then using this information to predict actions in similar situations \cite{zheng2022imitation}.

Consider the process of imitation learning encapsulated within the framework of behavior cloning, articulated as a system of equations representing the sequential stages from expert demonstration to action prediction. {\color{black}The SL function Eq. \eqref{SL-function-define} for imitation learning is $F([\text{server-devices}], \varnothing, $ $[\text{suggestion\&imitation}], \text{train}, \text{policies})$. The imitation learning process involves:
\vspace{-0.5em}
\begin{align}
&\begin{cases}
  \text{Expert Demonstrations}: \mathcal{D} = \{(s_i, a_i)\}_{i=1}^N, \\
  \text{Policy Learning}:  \theta^* = \underset{\theta}{\text{argmin}} \; \mathcal{L}(\pi_\theta(s), a), \; \forall (s, a) \in \mathcal{D}, \\
  \text{Action Prediction}: a_{\text{pred}} = \pi_{\theta^*}(s_{\text{new}}),
\end{cases}
\vspace{-1.5em}
\end{align}
where $\mathcal{D}$ is the set of expert demonstrations, $\pi_\theta$ is the learned policy with parameters $\theta$, $\mathcal{L}$ is the loss function, and $a_{\text{pred}}$ is the predicted action for a new state $s_{\text{new}}$.}

Imitation learning enables quick knowledge acquisition and application to various domains, effectively skipping learning steps. However, it depends on access to an expert model and can lead to overfitting issues \cite{zheng2022imitation}. Despite these challenges, imitation learning is a key technique in AI, particularly useful in fields such as autonomous driving and game AI.

\begin{figure}[pt]
  \centering
  \includegraphics[width=3.49in]{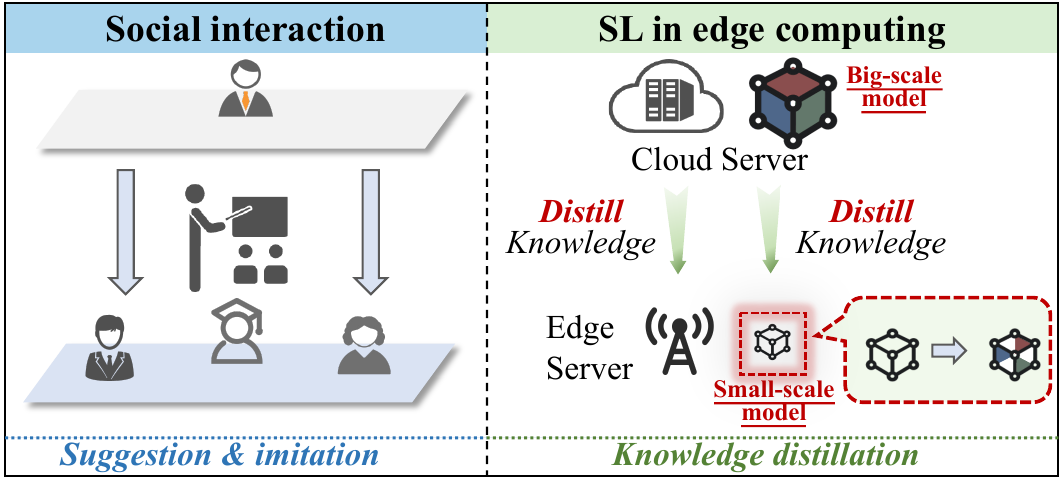}      
  \caption{\color{black}Illustration of \textbf{knowledge distillation} in socialized training.}\label{KD-major}
  % \vspace{-1em}
\end{figure}

\subsubsubsection{Knowledge Distillation (KD)} 
% In social constructs, the process of a novice imbibing wisdom and skills from an adept figure parallels a machine's emulation in the realm of KD. 
% KD is emblematic of ``Inter-layer Submissive Interaction", wherein a lightweight model (akin to the apprentice) passively inherits the wisdom from a more complex model or ensemble (akin to the mentor) \cite{DBLP:journals/tnn/LiLWWTSJ23}.
KD represents `inter-layer interaction' where a lightweight model inherits from a more complex one \cite{DBLP:journals/tnn/LiLWWTSJ23}, as shown in Fig. \ref{KD-major}.
A prevalent technique involves the student model learning from the softened output (logits) of the teacher model. 
{\color{black}The SL function Eq. \eqref{SL-function-define} for KD is $F([\text{server-devices}], \varnothing, [\text{suggestion\&}$ $\text{imitation}], \text{train}, \text{features})$. The knowledge distillation loss is:
\vspace{-0.5em}
\begin{equation}
\begin{aligned}
\mathcal{L}_{KD} = H(y, q^{\text{student}}) + \alpha H(q^{\text{teacher}}, q^{\text{student}}),
\end{aligned}
\vspace{-0.5em}
\end{equation}
where $H$ is the cross-entropy function, $y$ is the true label, $q^{\text{student}}$ and $q^{\text{teacher}}$ are the softened outputs of the student and teacher models, and $\alpha$ is the weight for the distillation loss.}

Submissive interaction in ML prominently manifests in three KD paradigms, each drawing parallels to social interactions. \textbf{Response-based KD} imitates the direct responses of expert models, using these soft targets to guide student models. \textbf{Feature-based KD} goes further, teaching student models to mimic intermediate features of teacher models, with FitNet as a prime example \cite{DBLP:journals/corr/RomeroBKCGB14}. It aligns student and teacher feature maps for deeper knowledge transfer. Lastly, \textbf{Relation-based KD} explores the relationships between layers, akin to the interplay of skills and concepts in social settings. The feature space projection matrix, capturing feature map pair relationships, is a notable method in this context \cite{DBLP:conf/cvpr/YimJBK17}. Together, these KD methods reflect the ML trends from novice to expert, mirroring human learning processes.

{\color{black}
\textbf{Communication-aware Knowledge Distillation:}
KD is a technique for transferring knowledge from a large, complex teacher model to a smaller, simpler student model, which can be deployed on resource-constrained devices for efficient inference. However, in a distributed setting, the communication cost of transferring the teacher model's knowledge to multiple student models can be prohibitive. 
Researchers have proposed communication-aware KD techniques \cite{wu2022communication}, which combine FL with KD to enable collaborative and communication-efficient knowledge transfer across multiple clients. These approaches aim to minimize the communication overhead of KD while ensuring that the student models can effectively absorb the knowledge of the teacher model and achieve comparable performance on the target task. By incorporating communication-awareness into the KD process, SL systems can facilitate the efficient deployment of powerful AI models on resource-constrained devices, enabling intelligent and responsive services at the edge.}

\begin{figure}[pt]
  \centering
  \includegraphics[width=3.49in]{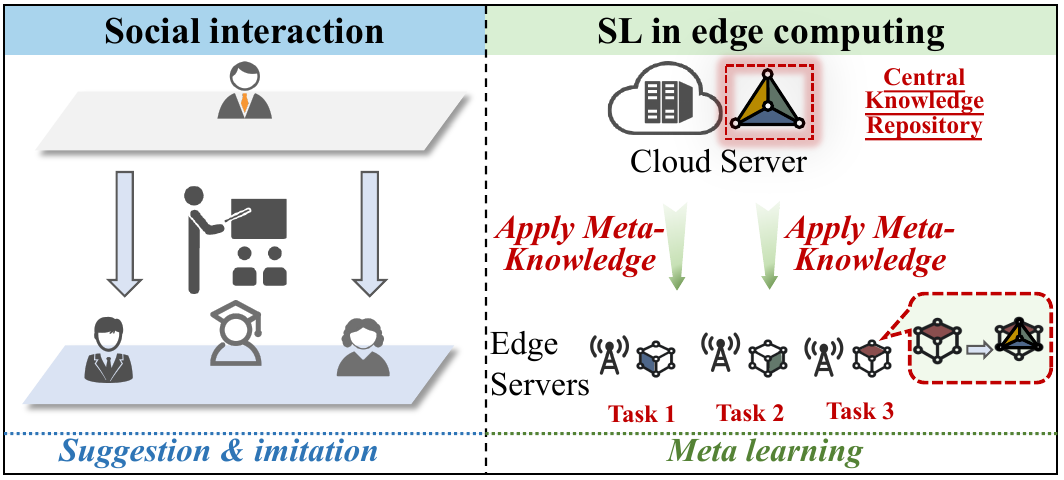}      
  \caption{\color{black}Illustration of \textbf{meta learning} in socialized training.}\label{meta learning-major}
  % \vspace{-1em}
\end{figure}

\subsubsubsection{Meta Learning} %Meta learning, in its essence, is reminiscent of social interactions where younger or less-experienced individuals, much like apprentice artisans, learn the nuances of a craft not just by mimicking the overt actions, but by internalizing the intricacies of the craft from seasoned experts.
% Unlike the broad strokes of transfer learning that focus on the direct application of knowledge, meta learning, much like a keen apprentice, emphasizes understanding the `how' behind learning from past tasks to enhance its capability in mastering new challenges swiftly. 
Meta learning, which aims to improve learning processes by utilizing experiences from past tasks, differs from transfer learning’s knowledge application, as shown in Fig. \ref{meta learning-major}. Advances such as MELO \cite{DBLP:journals/icl/HuangZYQW21}, which extends model-agnostic meta-Learning (MAML) and applications involving spiking neural networks, demonstrate the adaptation of meta learning in EI. This evolution, when combined with model compression, equips models for the unique demands of EC \cite{DBLP:conf/mm/YeZ021}.

Furthermore, the challenges posed by limited datasets in traditional ML methods echo the real-world scenarios where apprentices may only have a handful of opportunities to observe and learn from a master. Just as humans intuitively learn to learn, meta learning thrives on encapsulating patterns, problem properties, and past performances to accelerate the learning of new tasks, even with sparse data. In doing so, it not only epitomizes the spirit of human adaptability but also reinforces the social essence of learning through suggestion and imitation, achieving performance levels that potentially outstrip conventional methods.

{\color{black}The SL function Eq. \eqref{SL-function-define} for meta learning is $F([\text{server-}$ $\text{devices}], \varnothing,[\text{suggestion\&imitation}], \text{train}, \text{strategies})$. The optimization objective is:
\vspace{-0.5em}
\begin{equation}
\begin{aligned}
& \theta^* = \arg\min_{\theta} \mathcal{L}_{\text{meta-train}}(\theta),  \quad \theta_i' = \theta - \alpha \nabla_{\theta} \mathcal{L}_{\text{train}}(f_{\theta}(x_i), y_i), \\
& \mathcal{L}_{\text{meta-test}}(\theta') = \mathcal{L}_{\text{test}}(f_{\theta'}(x'), y'),
\end{aligned}
\vspace{-0.5em}
\end{equation}
where $\theta^*$ are the optimal parameters learned during meta-training, $\mathcal{L}_{\text{meta-train}}$ and $\mathcal{L}_{\text{meta-test}}$ are the meta-training and meta-testing loss functions, $f_{\theta}$ is the model with parameters $\theta$, $(x_i, y_i)$ and $(x', y')$ are the training and testing data pairs, $\alpha$ is the learning rate, and $\mathcal{L}_{\text{train}}$ and $\mathcal{L}_{\text{test}}$ are the loss functions for the training and testing tasks.}

\subsubsection{\textbf{Bidirectional Submissive Interaction}}
In machine societies, entities participate in bidirectional submissive interactions, meaning they both acquire and contribute knowledge. This dynamic allows them to switch roles between learner and mentor, ensuring that individual insights lead to collective system improvement and optimize overall efficiency.

\begin{table*}[t]
\color{black}
\centering
\caption{\color{black}Comparative Analysis of Technical Approaches Reflecting Submissive  Interactions}
\begin{tabular}{|c|c|c|c|c|c|c|c|c|}
\hline
\textbf{\begin{tabular}[c]{@{}c@{}}Submissive \\ Type\end{tabular}} & \textbf{Technique}                                            & \textbf{\begin{tabular}[c]{@{}c@{}}Social \\ Relationship\end{tabular}}                & \textbf{\begin{tabular}[c]{@{}c@{}}Social \\ Characteristics\end{tabular}}                  & \textbf{\begin{tabular}[c]{@{}c@{}}Model \\ Interactivity\end{tabular}} & \textbf{Complexity} & \textbf{Flexibility} & \textbf{Scalability} & \textbf{Resilience} \\ \hline
\multirow{2}{*}{Intra-layer}                                        & \begin{tabular}[c]{@{}c@{}}Causal \\ Inference \\ \cite{2020Causal, DBLP:journals/tacl/RotmanFR21, DBLP:conf/icml/TeshimaSS20, DBLP:conf/sigir/0007S0L20, DBLP:conf/iccv/YueS0Z21, DBLP:conf/kdd/KuangCAXL18, 9566788, DBLP:conf/mm/ZhangJWKZZYYW20, DBLP:journals/csur/YuGLLWLW20}\end{tabular}   & \multirow{2}{*}{Conformity}                                                            & \begin{tabular}[c]{@{}c@{}}Conforming to reasons\\  or norms\end{tabular}                   & Low                                                                     & High                & Moderate             & Moderate             & High                \\ \cline{2-2} \cline{4-9} 
                                                                    & \begin{tabular}[c]{@{}c@{}}Attention\\ Mechanism\\ \cite{vaswani2017attention, kenton2019bert, brown2020language, chen2017sca, jaderberg2015spatial, li2020does, wiegreffe2020attention}\end{tabular} &                                                                                        & \begin{tabular}[c]{@{}c@{}}Focusing on dominant\\  features\end{tabular}                    & Low                                                                     & Moderate            & High                 & High                 & Moderate            \\ \hline
\multirow{6}{*}{Inter-layer}                                        & \begin{tabular}[c]{@{}c@{}}Transfer\\ Learning \\ \cite{DBLP:conf/eccv/ZhangQYPWT20, DBLP:journals/tip/YehHW14, DBLP:journals/tcyb/DengZKHPL22, DBLP:journals/pami/XiaoG15}\end{tabular}   & \multirow{4}{*}{\begin{tabular}[c]{@{}c@{}}Suggestion \\ \& \\ Imitation\end{tabular}} & \begin{tabular}[c]{@{}c@{}}Drawing guidance from\\  experienced figures\end{tabular}        & Moderate                                                                & Moderate            & Moderate             & High                 & Moderate            \\ \cline{2-2} \cline{4-9} 
                                                                    & \begin{tabular}[c]{@{}c@{}}Imitation\\ Learning \cite{zheng2022imitation}\end{tabular}  &                                                                                        & \begin{tabular}[c]{@{}c@{}}Directly mimicking \\ the actions of an expert\end{tabular}      & Moderate                                                                & Low                 & Low                  & Moderate             & Low                 \\ \cline{2-2} \cline{4-9} 
                                                                    & \begin{tabular}[c]{@{}c@{}}KD\\  \cite{DBLP:journals/tnn/LiLWWTSJ23, DBLP:journals/corr/RomeroBKCGB14, DBLP:conf/cvpr/YimJBK17, wu2022communication} \end{tabular}                                                       &                                                                                        & \begin{tabular}[c]{@{}c@{}}Learning from an expert,\\ inheriting wisdom\end{tabular}        & Low                                                                     & Moderate            & High                 & High                 & High                \\ \cline{2-2} \cline{4-9} 
                                                                    & \begin{tabular}[c]{@{}c@{}}Meta-\\ Learning\\ \cite{DBLP:journals/icl/HuangZYQW21, DBLP:conf/mm/YeZ021}\end{tabular}      &                                                                                        & \begin{tabular}[c]{@{}c@{}}Learning the nuances of \\ a craft from experts\end{tabular}     & High                                                                    & High                & Very High            & High                 & High                \\ \cline{2-9} 
                                                                    & \begin{tabular}[c]{@{}c@{}}DA\\  \cite{DBLP:journals/tnn/ZhangG24, bruzzone2009domain, chu2013selective, gong2013connecting, gheisari2015unsupervised, ganin2015unsupervised, liu2021cycle, yang2020mobileda}  \end{tabular}                                                        & \multirow{2}{*}{\begin{tabular}[c]{@{}c@{}}Bidirectional \\ Submissive\end{tabular}}   & \begin{tabular}[c]{@{}c@{}}Mutual adaptation \\ between entities\end{tabular}               & High                                                                    & High                & High                 & Moderate             & High                \\ \cline{2-2} \cline{4-9} 
                                                                    & \begin{tabular}[c]{@{}c@{}}MTL \\ \cite{caruana1997multitask, he2022metabalance, xin2019multi, maurer2013sparse}  \end{tabular}                                                        &                                                                                        & \begin{tabular}[c]{@{}c@{}}Collaborative learning \\ with interdependent tasks\end{tabular} & Very High                                                               & High                & Very High            & Very High            & Moderate            \\ \hline
\end{tabular}
\label{table:Comparison of submissive}
\end{table*}

    \begin{figure}[pt]
  \centering
  \includegraphics[width=3.49in]{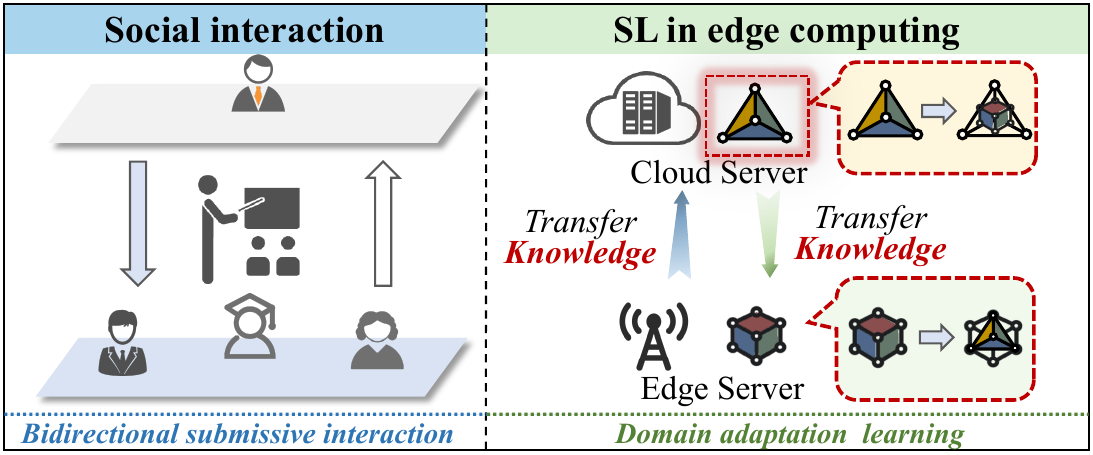}      
  \caption{Illustration of \textbf{domain adaptation learning} in socialized training.}\label{Domain Adaptation}
  \vspace{-0.5em}
\end{figure}

\subsubsubsection{Domain Adaptation (DA) Learning} %Domain Adaptation (DA) mirrors the intricate dynamics of bidirectional submissive interactions observed in social frameworks, aligning well with the conceptual depths of Inter-layer Submissive Interaction. 
% Drawing analogies from society, just as individuals not only influence but also evolve from mutual exchanges in diverse interactions, DA functions in a bidirectional or even multi-directional realm of transfer learning 
DA functions in a bidirectional or even multi-directional realm of transfer learning \cite{DBLP:journals/tnn/ZhangG24}. It leverages knowledge from a source domain to adapt to the target domain's data distribution. This exchange is depicted in Fig. \ref{Domain Adaptation}. Instance-based DA \cite{bruzzone2009domain,chu2013selective} reweights source instances to bridge disparities between source and target samples. Feature-based DA \cite{gong2013connecting,gheisari2015unsupervised,ganin2015unsupervised} creates a unified space for the cohesion of source and target features. Gani \emph{et al.}'s approach \cite{ganin2015unsupervised} utilizes generative adversarial networks (GANs) to create consistent features across domains.

In DA, a fundamental objective is to reduce the difference in distribution between the source and target domains. A prevalent strategy involves utilizing the maximum mean discrepancy (MMD) to measure and minimize the distance between distinct domain distributions. 
{\color{black}The SL function Eq. \eqref{SL-function-define} for DA learning is $F([\text{server-devices}], \varnothing,$ $[\text{bi-submissive interaction}], \text{train}, \text{features})$. The optimization objective is:
\vspace{-0.5em}
\begin{equation}
\begin{aligned}
\min_{\theta} \mathcal{L}(\theta) = \mathcal{L}_s(\theta) + \lambda \cdot \text{MMD}^2(H_s, H_t),
\end{aligned}
\vspace{-0.5em}
\end{equation}
where $\mathcal{L}_s$ is the source domain loss, $\text{MMD}$ is the maximum mean discrepancy, $H_s$ and $H_t$ are feature representations for the source and target domains, and $\lambda$ is the trade-off parameter.}

\begin{figure}[pt]
  \centering
  \includegraphics[width=3.49in]{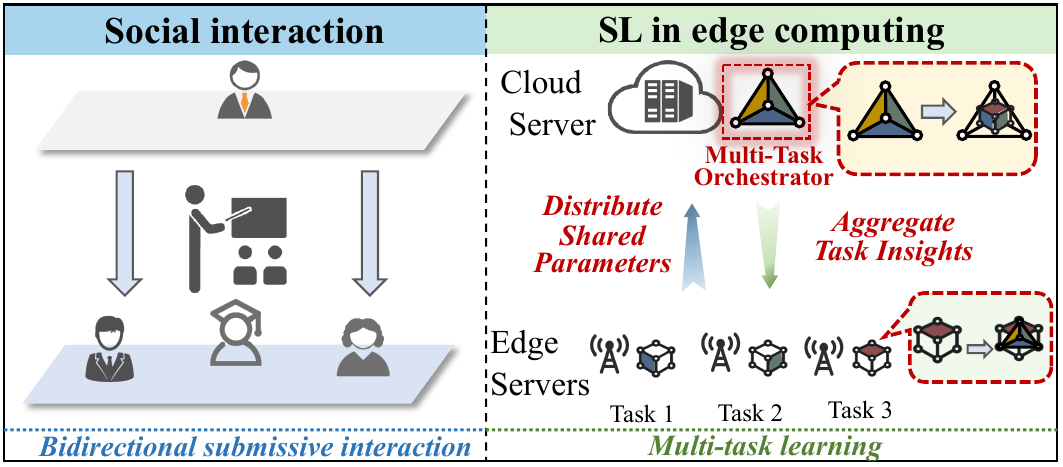}      
  \caption{\color{black}Illustration of \textbf{multi-task learning} in socialized training.}\label{MTL-major}
  % \vspace{-1em}
\end{figure}

% However, the brilliance of DA doesn't stop there. %Much like in social dynamics where interactions yield mutual growth, 
Insights from the target domain enhance the source's knowledge. Liu \emph{et al.} \cite{liu2021cycle} describe this with their cycle self-training (CST) model, which facilitates knowledge exchange between domains. Such intricate dynamics of DA become indispensable in realms like EI where data distributions diversify across devices. %This recalls the social analogy where knowledge-sharing bridges diverse communities. 
Representative of this is the MobileDA framework by Yang \emph{et al.} \cite{yang2020mobileda}, which not only combats domain shifts in EI devices but also facilitates seamless knowledge exchange between centralized servers and edge entities.

\subsubsubsection{Multi-task Learning (MTL)} 
%Multi-Task Learning (MTL) \cite{caruana1997multitask} serves as a quintessential model that orchestrates bidirectional submissive interactions among diverse tasks, much like interdependent social entities collaboratively evolving while preserving their distinctive identities. Drawing from social parallels, just as diverse domains in society - such as the realms of culture, economy, and governance - simultaneously influence and adapt from each other, MTL undertakes the harmonious act of sharing model parameters and feature representations.
MTL \cite{caruana1997multitask} orchestrates bidirectional submissive interactions among diverse tasks, as shown in Fig. \ref{MTL-major}. Just as diverse domains in society, such as the realms of culture, economy, and governance, simultaneously influence and adapt to each other, MTL undertakes the collaborative act of sharing model parameters and feature representations.

Assuming the presence of $T$ tasks, each with its distinctive loss function $\mathcal{L}_t$, the overarching loss function in MTL can be articulated as the weighted sum of these individual losses. 
{\color{black}The SL function Eq. \eqref{SL-function-define} for multi-task learning is $F([\text{server-devices}], \varnothing, [\text{bi-submissive interaction}], \text{train}, \text{tasks})$. The multi-task learning objective is:
\vspace{-0.5em}
\begin{equation}
\begin{aligned}
\mathcal{L}_{MTL} = \sum_{t=1}^{T} \alpha_t \mathcal{L}_t,
\end{aligned}
\vspace{-0.5em}
\end{equation}
where $\mathcal{L}_{MTL}$ is the total multi-task loss, $\mathcal{L}_t$ is the loss for task $t$, $\alpha_t$ is weight for task $t$, and $T$ is total number of tasks.}

Within the MTL, the categorization into hard sharing \cite{he2022metabalance} and soft sharing \cite{xin2019multi} mirrors social dynamics where certain norms are universally rigid (hard sharing) while others permit regional variances (soft sharing). Maurer \emph{et al.}'s exploration illustrates how shared features, envisioned as transformations through multi-layered networks \cite{maurer2013sparse}, parallel the way societies adapt universal principles into specific regional customs. In the larger picture, MTL champions a symbiotic feature interaction across multiple tiers, harmonizing model structures and fostering task interdependencies. 
%Much like societies that flourish by embracing and integrating the strengths of their myriad components, MTL capitalizes on this inter-task collaboration. 
The result is enhanced model robustness, adaptability, and commendable prowess to tackle a plethora of complex tasks.
Bidirectional submissive interactions, illustrated by DA and MTL, underscore the importance of knowledge exchange. For clarity, Table \ref{table:Comparison of submissive} provides a comparative analysis of these methods.

\vspace{-0.5em}
\subsection{{Intra-layer Conflict Interaction in Socialized Training}}
\subsubsection{\textbf{Competition}}
%In human societies, competition is an inherent drive, propelling individuals and groups to strive for supremacy in their endeavors, be it resources, recognition, or influence. Competition often arises when multiple entities, operating at similar levels or hierarchies, vie for the same limited resource or objective. The intrinsic tension and rivalry can either lead to innovation and progress or engender strife and discord. Within the machine society, this competitive spirit is mirrored in algorithms and models that aim to outperform their peers by constantly fine-tuning, adapting, and evolving. They are pitted against one another on the same datasets, under similar constraints, pushing the envelope to achieve unparalleled precision and efficiency. In this theater of machine rivalry, each model's ambition is to emerge as the front-runner, setting benchmarks that others aspire to eclipse.

% In human societies, competition is a driving force, spurring individuals to strive for supremacy, whether in resources, recognition, or influence. 
Competition is a driving force in many systems, pushing entities to strive for supremacy in various domains.
% Competition occurs when entities at similar levels vie for limited resources or objectives and leads to innovation or discord. 
In the machine society, algorithms and models mirror this competitiveness, aiming to outperform peers through constant fine-tuning, adaptation, and evolution. 
% They compete on similar datasets, pushing boundaries for precision and efficiency. 
Each model desires to set benchmarks for others to surpass.

\begin{figure}[pt]
  \centering
  \includegraphics[width=3.49in]{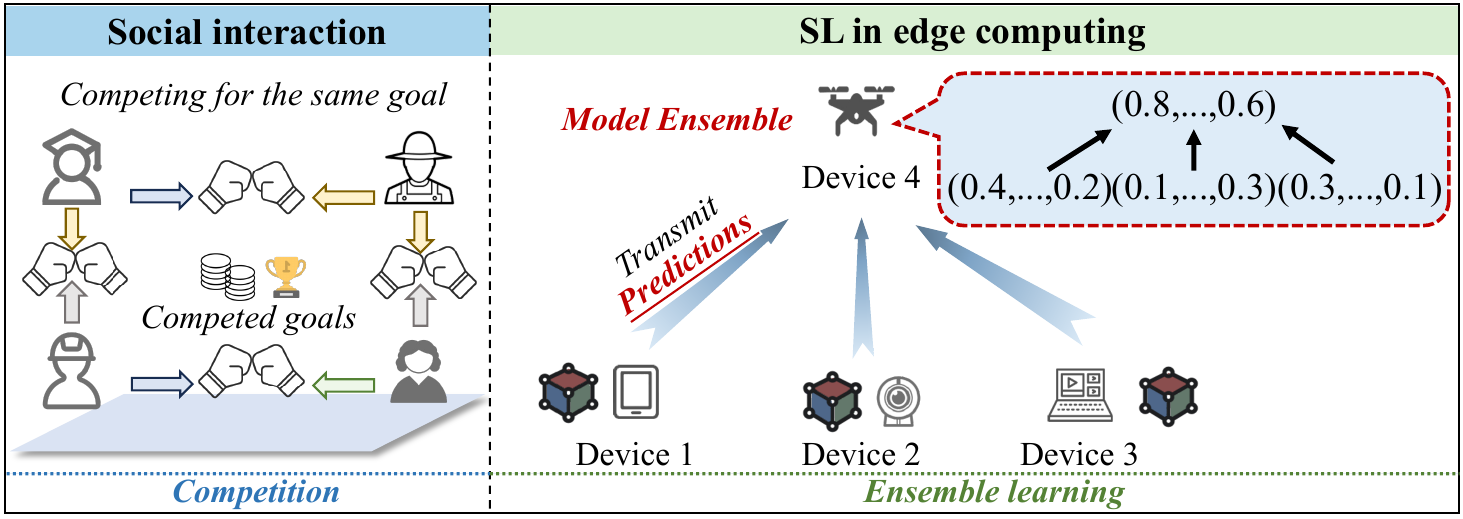} 
  \caption{Illustration of \textbf{ensemble learning} in socialized training.}\label{Ensemble Learning}
  % \vspace{-1em}
\end{figure}

\subsubsubsection{Ensemble Learning} 
%Intra-layer competition, a hallmark of human social dynamics, emphasizes the interactions and rivalries among entities of the same hierarchical tier. Analogously, Ensemble Learning epitomizes this essence of competition within the realm of ML. At the heart of Ensemble Learning is a cadre of models, each vying to showcase its predictive prowess on a common dataset, embodying the intra-layer competition in a machine society. While individual models are complete in themselves, their competitive edge is sharpened when they're juxtaposed with peers, pushing each model to accentuate its strengths and compensate for its weaknesses.

Ensemble learning epitomizes intra-layer competition where models vie to showcase their predictive prowess on a common dataset. While individual models are complete in themselves, their competitive advantages become more pronounced when juxtaposed with peers, pushing each model to highlight its strengths and compensate for its weaknesses. This concept is illustrated in Fig. \ref{Ensemble Learning}. 

A concept is the aggregation of decisions from multiple models to formulate a final prediction. 
{\color{black}The SL function Eq. \eqref{SL-function-define} for ensemble learning is $F([\text{devices}], [\text{competition}], \varnothing,$ $\text{train}, \text{predictions})$. The ensemble prediction is:
\vspace{-0.6em}
\begin{equation}
\begin{aligned}
P_{\text{ensemble}} = \sum_{n=1}^{N} w_n P_n,
\end{aligned}
\vspace{-0.5em}
\end{equation}
where $P_{\text{ensemble}}$ is the ensemble prediction, $P_n$ is the prediction of the $n$-th model, $w_n$ is the weight for the $n$-th model, and $N$ is the total number of models.}

Ensemble learning employs various strategies to harness this competition. 
% The diversity of individual model predictions, derived from different learning algorithms or hyperparameters, converges to form a more robust and reliable decision. 
In particular, approaches like voting \cite{DBLP:journals/air/YangLC23} by pooling predictions, whereas weighted voting \cite{onan2016multiobjective} confers preferential weights based on a model's relative performance. Further, the model averaging method \cite{ganaie2022ensemble} seeks a consensus by averaging individual model outputs. Bai \emph{et al.} \cite{bai2021automated} proposed an Automated DNN ensemble (AES), which selects optimal DNN models and tunes the ensemble according to nuances of the input data. Such intricate competitive dynamics have found success across many applications, from fraud detection \cite{DBLP:journals/asc/ForoughM21} and medical diagnosis \cite{suk2017deep} to image recognition \cite{DBLP:journals/air/ZhongDW23}, emphasizing its effectiveness in harnessing intra-layer competition for superior predictive outcomes.

\begin{figure}[t]
  \centering
  \includegraphics[width=3.4in]{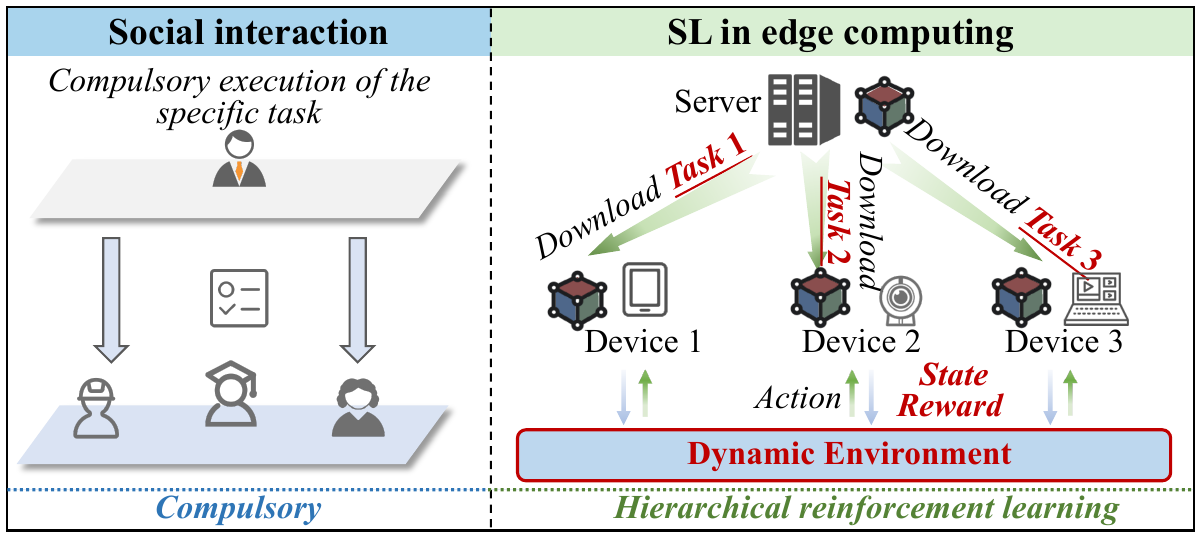}  
  \caption{Illustration of \textbf{hierarchical reinforcement learning} in socialized training.}\label{Hierarchical reinforcement learning}
  \vspace{-0.5em}
\end{figure}

\vspace{-1em}
\subsection{{Inter-layer Conflict Interaction in Socialized Training}}
\subsubsection{\textbf{Unidirectional Competition - Compulsory}}
%In social hierarchies, it's common for one group or layer to exert influence or command over another, leading to a unidirectional flow of authority. This compulsory interaction is characterized by the dominance of one group and the subsequent compliance of another. Often, this stems from structural hierarchies, such as in workplaces where managers direct employees, or in social constructs where certain groups hold sway over others. This dynamic is emblematic of a compulsory nature, with the higher group mandating actions or guidelines and the lower group compelled to follow. In the realm of ML, this interaction can be observed in hierarchical structures where higher-level algorithms or protocols dictate the functioning or constraints of those at lower levels. By understanding this unidirectional competition, researchers can harness the efficiency and directionality it offers, while being mindful of the inherent constraints and potential lack of flexibility it introduces.
Compulsory interaction is characterized by the dominance of one group and the subsequent compliance of another. Often, this stems from structural hierarchies, such as in workplaces where managers direct employees or in social constructs where certain groups have influence over others.
%This dynamic is emblematic of a compulsory nature, with the higher group mandating actions or guidelines and the lower group compelled to follow. 
In ML, this interaction can be observed in hierarchical structures where higher-level algorithms or protocols dictate the functioning or constraints of those at lower levels. By understanding this unidirectional competition, researchers can harness the efficiency and directionality it offers while being mindful of the inherent constraints and potential lack of flexibility it introduces.

 \subsubsubsection{Hierarchical Reinforcement Learning (HRL)} 
HRL mirrors unidirectional compulsory interactions where a dominant group or layer prescribes rules or guidelines for another group, mandating compliance.
% HRL decomposes intricate tasks into a stratified structure, each tier embodying a specific abstraction level. 
This structure is visually represented in Fig. \ref{Hierarchical reinforcement learning}. The essence of compulsory is exhibited as the superiors set overarching goals, while the subordinate layers are obligated to adhere. Dayan \emph{et al.} \cite{dayan1992feudal} offer a quintessential example with a system partitioned into managers and sub-managers. Governed by principles like reward hiding and information hiding, this structure encapsulates unidirectional competition, with each layer, while autonomous to some extent, invariably remaining in contact with its superiors.

In HRL, decision-making is layered, requiring representations that capture the complexity of its dynamics.
{\color{black}The SL function Eq. \eqref{SL-function-define} for HRL is $F([\text{server-devices}], \varnothing, [\text{competition}], $ $\text{train}, \text{DRLs})$. The HRL objective is:
\vspace{-0.5em}
\begin{equation}
\begin{aligned}
H(s, \pi^G, \pi^g, \text{Transit}, R) = [\text{Transit}(s, \pi^g(s, \pi^G(s))), \\
R(s, \pi^g(s, \pi^G(s)), \text{Transit}(s, \pi^g(s, \pi^G(s))))],
\end{aligned}
\vspace{-0.5em}
\end{equation}
where $H$ is the main function, $s$ is the state, $\pi^G$ and $\pi^g$ are the high-level and low-level policies, $\text{Transit}(\cdot)$ is the transition function, and $R$ is the reward function.}
Such hierarchical frameworks not only streamline learning but also afford the assimilation of prior knowledge.
As in social structures, HRL embodies the directives of one group to the behaviour of another. The directives from higher layers, while potentially restrictive, shape the trajectory of the entire system, ensuring alignment with overarching objectives \cite{DBLP:journals/ral/GieselmannP21}.
 
\subsubsection{\textbf{Bidirectional Competition - Struggle}}
%The concept of struggle is intrinsic to social dynamics when entities are in conflict over shared resources, power, or influence. This struggle is bidirectional in nature, with both entities exerting influence over each other in an ongoing battle for dominance. The ML realm mirrors this phenomenon through adversarial learning techniques, where two components continuously challenge each other. This paradigm is not merely a static competition, but an evolving duel, where each party continually adapts based on the strategies of the opponent. It is through this relentless back-and-forth that models evolve, mature, and ultimately produce better results. Drawing a parallel to social dynamics, this bidirectional struggle mimics the intricate phenomena of competition seen in nature, economies, and societies at large, where constant challenges lead to growth, innovation, and resilience.

% The concept of struggle is intrinsic to social dynamics when entities are in conflict over shared resources, power or influence. Struggle is bidirectional in nature. Adversarial learning techniques mirror this phenomenon, where two components continuously challenge each other. 
% % This paradigm is not merely a static competition, but an evolving duel, where each party continually adapts based on the strategies of the opponent. 
% It is through relentless back-and-forth that models evolve, mature, and ultimately produce better results.

The concept of struggle in social dynamics, when entities compete for shared resources, power, or influence, is reflected in adversarial learning techniques. In these techniques, two components challenge each other, and it's through this continuous push and pull that the model evolves, yielding improved outcomes.

\begin{table*}[t]
\color{black}
\centering
\caption{\color{black}Comparative Analysis of Technical Approaches Reflecting Conflict Interactions}
\begin{tabular}{|c|c|c|c|c|c|c|c|c|}
\hline
\textbf{\begin{tabular}[c]{@{}c@{}}Conflict \\ Type\end{tabular}} & \textbf{Technique}                                                    & \textbf{\begin{tabular}[c]{@{}c@{}}Social \\ Relationship\end{tabular}} & \textbf{\begin{tabular}[c]{@{}c@{}}Social \\ Characteristics\end{tabular}}         & \textbf{\begin{tabular}[c]{@{}c@{}}Model \\ Interactivity\end{tabular}} & \textbf{Complexity} & \textbf{Scalability} & \textbf{Adaptability}                               & \textbf{\begin{tabular}[c]{@{}c@{}}Outcome\\ Diversity\end{tabular}} \\ \hline
Intra-layer                                                       & \begin{tabular}[c]{@{}c@{}}Ensemble\\ Learning \\ \cite{DBLP:journals/air/YangLC23,onan2016multiobjective,ganaie2022ensemble,bai2021automated,DBLP:journals/asc/ForoughM21,suk2017deep,DBLP:journals/air/ZhongDW23}\end{tabular}           & Competition                                                             & \begin{tabular}[c]{@{}c@{}}Collaborative rivalry\\ among peers\end{tabular}        & Moderate                                                                & Moderate            & High                 & Moderate                                            & High                                                                 \\ \hline
\multirow{3}{*}{Inter-layer}                                      & HRL  \cite{dayan1992feudal,DBLP:journals/ral/GieselmannP21}                                                                & Compulsory                                                              & \begin{tabular}[c]{@{}c@{}}Top-down directive\\ from superiors\end{tabular}        & Low                                                                     & High                & Moderate             & Low                                                 & Moderate                                                             \\ \cline{2-9} 
                                                              & \begin{tabular}[c]{@{}c@{}}Adversarial\\ Learning\\ \cite{DBLP:journals/csur/BrophyWSW23,mirza2014conditional,radford2015unsupervised,adler2018banach,ganin2018synthesizing,yin2019utilizing,DBLP:conf/nips/SunCGY19}\end{tabular}        & \multirow{2}{*}{Struggle}                                               & \begin{tabular}[c]{@{}c@{}}Mutual power \\ struggle between \\ opponents\end{tabular}  & High                                                                    & High                & Moderate             & High                                                & High                                                                 \\ \cline{2-2} \cline{4-9} 
                                                              & \begin{tabular}[c]{@{}c@{}}Modality Agnostic\\  Learning \\ \cite{jaegle2021perceiver,vaswani2017attention}\end{tabular} &                                                                         & \begin{tabular}[c]{@{}c@{}}Adaptability across\\ diverse environments\end{tabular} & Low                                                                     & High                & High                 & \begin{tabular}[c]{@{}c@{}}Very\\ High\end{tabular} & \begin{tabular}[c]{@{}c@{}}Very\\ High\end{tabular}                  \\ \hline
\end{tabular}
\label{table:Comparison of conflict}
\end{table*}

\begin{figure}[pt]
    \centering
    \includegraphics[width=3.49in]{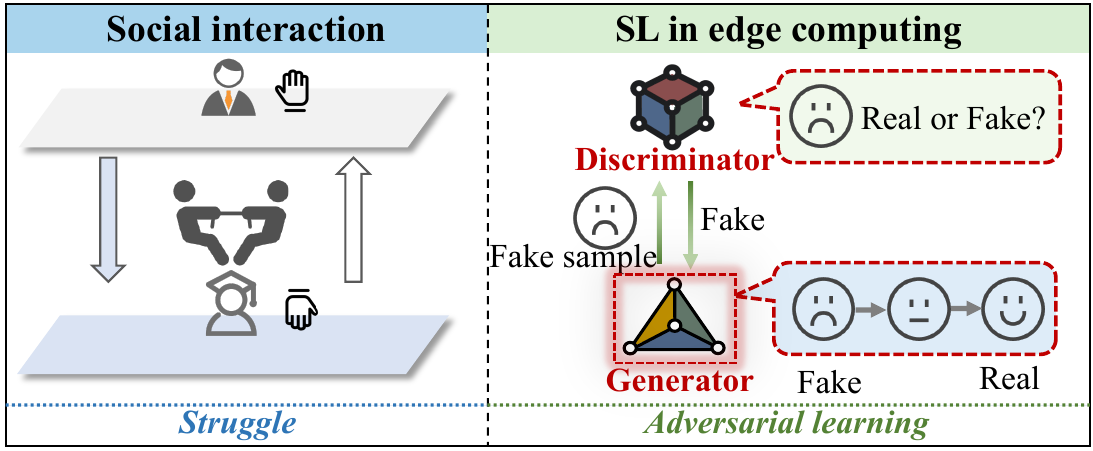}      
    \caption{Illustration of \textbf{adversarial learning} in socialized training.}\label{Adversarial learning}
    \vspace{-0.5em}
\end{figure}

\subsubsubsection{Adversarial Learning} 
%In the context of social dynamics, this bidirectional competition is reminiscent of the power plays witnessed between rival entities, each continuously adapting and recalibrating strategies based on the countermeasures of the opponent. 
% Rooted in the principles of generative adversarial networks (GANs) \cite{DBLP:journals/csur/BrophyWSW23}, adversarial learning epitomizes this struggle between two diametrically opposing components: the generator and the discriminator. 
In adversarial learning, particularly notable in GANs \cite{DBLP:journals/csur/BrophyWSW23}, there is a struggle between two components: the generator and the discriminator, as illustrated in Fig. \ref{Adversarial learning}. The generator crafts authentic samples to deceive the discriminator. Conversely, the discriminator endeavors to discern real data from the counterfeits created by the generator. However, challenges like instability in the training process and homogeneous sample generation persist. Original GANs often produced uncontrollable image variations from random noise.

{\color{black}The SL function Eq. \eqref{SL-function-define} for adversarial learning is $F([\text{server-devices}], \varnothing, [\text{struggle}], \text{train}, \text{DNNs})$. The adversarial learning objective is:
\vspace{-0.5em}
\begin{equation}
\begin{aligned}
&\min_G \max_D V(D, G) = \mathbb{E}_{x \sim p_{\text{data}}(x)}[\log D(x)] + \\
&\mathbb{E}_{z \sim p_z(z)}[\log (1 - D(G(z)))],
\end{aligned}
\vspace{-0.5em}
\end{equation}
where $G$ and $D$ are the generator and discriminator networks, $p_{\text{data}}$ is the real data distribution, $p_z$ is the noise distribution, and $V(D, G)$ is the value function.
}

Solutions emerged in the form of conditional GANs \cite{mirza2014conditional}, introducing conditional data and Deep Convolution GCNs (DCGANs) \cite{radford2015unsupervised}, swapping fully connected layers for convolutional ones. The quest for stability birthed the Wasserstein GAN \cite{adler2018banach} with its novel loss function. Drawing parallels to social struggle, adversarial learning represents inter-layer conflicts.
This dynamic framework extends beyond image generation, spanning areas like RL \cite{ganin2018synthesizing} and natural language processing (NLP) \cite{yin2019utilizing}.

{\color{black}
\textbf{Communication-aware Adversarial Learning:}
Adversarial learning techniques, such as GANs, have shown remarkable success in various applications. However, in a distributed setting, the communication between the generator and discriminator networks can be costly and may impact the overall performance. Communication-aware adversarial learning techniques have been proposed to address this issue by reducing communication overhead while maintaining the quality of the generated samples. These methods may involve strategies such as gradient compression, selective parameter sharing, and decentralized architectures for adversarial learning. The design of communication-efficient adversarial learning systems often requires considering the trade-off between communication cost and the stability and convergence of the training process \cite{DBLP:conf/nips/SunCGY19}.}

\subsubsubsection{Modality Agnostic Learning}
%In social systems, entities often find themselves entrenched in a bidirectional struggle as they strive to adapt, survive, and thrive in varied environments. This adaptability to diverse conditions is reminiscent of the modality agnostic learning paradigm emerging in the ML arena. Traditionally, ML models have been tailored to specific input data types, necessitating dedicated architectures that often operate in silos. Just as societies evolve through interactions, grappling with diverse challenges and adapting to multifaceted environments, ML models too are now moving toward a more flexible and modality-independent approach. The dawn of B5G/6G networks heralds the proliferation of varied dataset modalities, including vision, audio, and point cloud. In such a vast and diverse landscape, the requirement for a universally adaptable model becomes palpable.
% In social systems, entities engage in bidirectional struggles to adapt and thrive, similar to the emerging modality-agnostic learning in ML.
Modality-agnostic learning signifies a shift in ML toward models that can process various data types, akin to a society that adapts to diverse challenges. Traditionally, models were specialized for specific modalities, but the emerging B5G/6G networks, with their wide array of data ranging from vision to audio and point cloud, demand models that are versatile across these modalities.
We perform modality-agnostic learning by considering a unified learning model that operates on an ensemble of different data modalities.
{\color{black}
The SL function Eq. \eqref{SL-function-define} for modality-agnostic learning is $F([\text{server-devices}], \varnothing, [\text{struggle}], \text{train}, \text{modalities})$. The learning objective is:
\vspace{-0.5em}
\begin{align}
&\begin{cases}
  x \in M = \{m_1, m_2, \ldots, m_n\}, \\
  y = f(x; \theta), 
\end{cases}
\vspace{-0.5em}
\end{align}
where $M$ is the set of all modalities, $x$ is the input data from any modality in $M$, $f$ is the modality-agnostic model with parameters $\theta$, and $y$ is the model output.}

\begin{table*}[p]
\color{black}
\centering
\caption{{\color{black}Comparative Analysis of ML Methods in Socialized Training for SL}}
\label{tab:ml_methods_comparison}
\begin{tabular}{|m{0.12\linewidth}|m{0.22\linewidth}|m{0.28\linewidth}|m{0.28\linewidth}|}
\hline
\textbf{ML Method} & \textbf{Scope of Application} & \textbf{Advantages} & \textbf{Disadvantages} \\
\hline
Decentralized Learning \cite{DBLP:conf/nips/LianZZHZL17,Swarm2021,DBLP:conf/icml/LuS21,DBLP:journals/spm/SayedTCZT13,wang2019adaptive} & Peer-to-peer collaboration among devices with limited resources. & Efficient resource utilization, improved privacy, and robustness against single points of failure. & Increased communication overhead, the potential for slower convergence. \\
\hline
DRL \cite{DBLP:journals/spm/ArulkumaranDBB17,DBLP:conf/sigmetrics/ChenSX18,zeng2021decentralized} & Multi-agent systems with shared objectives and dynamic environments. & Enables coordinated decision-making and adaptation to complex environments. & Challenges in managing large state-action spaces, delayed rewards, and heterogeneous agents. \\
\hline
Split Learning \cite{thapa2022splitfed,DBLP:journals/pieee/ParkSEKBKD21,yansong2020end} & Systems with privacy constraints and heterogeneous computational capabilities. & Preserves data privacy, reduces communication overhead, and enables collaboration among devices with varying resources. & Potential for increased computational complexity and challenges in managing multiple local models. \\
\hline
Edge-centralized Personalization \cite{DBLP:conf/edge/LuSTLZCP19,DBLP:conf/kdd/YaoWJHZY21} & Applications requiring personalized models at the edge with centralized control. & Enables personalized edge models while leveraging cloud resources and global knowledge. & Increased dependency on the cloud, potential for higher latency, and privacy concerns. \\
\hline
FL \cite{tan2022towards,DBLP:journals/tpds/ZhouYL22,rothchild2020fetchsgd}& Scenarios with distributed data and privacy requirements. & Enables collaborative learning without sharing raw data, reduces communication costs, and preserves privacy. & Challenges in managing non-IID data, model divergence, and potential security vulnerabilities. \\
\hline
Edge-cloud Bidirectional Collaboration \cite{DBLP:journals/corr/abs-2109-12314,Kahneman2011,DBLP:conf/iclr/MadanKGSB21,DBLP:journals/pieee/ZhouCLZLZ19,zhang2022communication}
& Applications requiring real-time insights and global knowledge. & Leverages the strengths of both edge and cloud computing, enables fast local decisions and global model updates. & Increased complexity in managing the collaboration, potential for higher communication costs. \\
\hline
Multi-modal Learning \cite{DBLP:journals/inffus/UppalBHMPZZ22,9891834}& Systems with heterogeneous data types and modalities. & Enables learning from diverse data sources, improves model generalization and robustness. & Challenges in data alignment, fusion, and increased computational complexity. \\
\hline
Causal Learning \cite{2020Causal,DBLP:journals/tacl/RotmanFR21,DBLP:conf/icml/TeshimaSS20,DBLP:conf/sigir/0007S0L20,DBLP:conf/iccv/YueS0Z21,DBLP:conf/kdd/KuangCAXL18,9566788,DBLP:conf/mm/ZhangJWKZZYYW20,DBLP:journals/csur/YuGLLWLW20}& Scenarios requiring unbiased and interpretable models. & Enables the discovery of causal relationships, improves model interpretability and fairness. & Requires domain knowledge for causal graph construction, potential for increased computational complexity. \\
\hline
Attention Mechanism \cite{vaswani2017attention,kenton2019bert,brown2020language,chen2017sca,jaderberg2015spatial,li2020does,wiegreffe2020attention}& Applications with sequential or spatial dependencies. & Enables focusing on relevant information, improves model interpretability and performance. & Increased computational complexity, potential for overemphasizing certain features. \\
\hline
Transfer Learning \cite{DBLP:conf/eccv/ZhangQYPWT20,DBLP:journals/tip/YehHW14,DBLP:journals/tcyb/DengZKHPL22,DBLP:journals/pami/XiaoG15}& Scenarios with limited labeled data or related source domains. & Enables knowledge transfer from related tasks, reduces the need for labeled data, and improves model performance. & Challenges in identifying suitable source domains, potential for negative transfer. \\
\hline
Imitation Learning \cite{zheng2022imitation}& Applications with expert demonstrations or optimal behaviors. & Enables learning from expert demonstrations, reduces the need for explicit reward functions. & Requires access to expert demonstrations, potential for suboptimal policies and lack of generalization. \\
\hline
KD \cite{DBLP:journals/tnn/LiLWWTSJ23,DBLP:journals/corr/RomeroBKCGB14,DBLP:conf/cvpr/YimJBK17,wu2022communication}& Scenarios with resource-constrained devices and pre-trained large models. & Enables transfer of knowledge from large to small models, reduces computational requirements. & Potential for loss of information during distillation, challenges in managing the trade-off between model size and performance. \\
\hline
Meta Learning \cite{DBLP:journals/icl/HuangZYQW21,DBLP:conf/mm/YeZ021}& Applications requiring fast adaptation to new tasks or environments. & Enables rapid learning and adaptation to new tasks, improves model generalization. & Increased computational complexity, challenges in defining appropriate meta learning objectives. \\
\hline
DA Learning \cite{DBLP:journals/tnn/ZhangG24,bruzzone2009domain,chu2013selective,gong2013connecting,gheisari2015unsupervised,ganin2015unsupervised,liu2021cycle,yang2020mobileda}& Scenarios with domain shift or varying data distributions. & Enables knowledge transfer across different domains, improves model performance and robustness. & Challenges in identifying and aligning relevant domain features, potential for negative transfer. \\
\hline
MTL \cite{caruana1997multitask,he2022metabalance,xin2019multi,maurer2013sparse}& Applications with related tasks or shared representations. & Enables learning from multiple tasks simultaneously, improves model generalization and efficiency. & Challenges in balancing task-specific and shared representations, potential for task interference. \\
\hline
Ensemble Learning \cite{DBLP:journals/air/YangLC23,onan2016multiobjective,ganaie2022ensemble,bai2021automated,DBLP:journals/asc/ForoughM21,suk2017deep,DBLP:journals/air/ZhongDW23,breiman1996bagging}& Scenarios requiring robust and accurate predictions. & Improves model performance and robustness, reduces overfitting, and enables uncertainty quantification. & Increased computational complexity, potential for increased model size and inference time. \\
\hline
HRL \cite{dayan1992feudal,DBLP:journals/ral/GieselmannP21}& Applications with hierarchical task structures or long-term goals. & Enables learning at different levels of abstraction, improves sample efficiency and transfer learning. & Challenges in defining appropriate hierarchies, potential for suboptimal policies due to hierarchical constraints. \\
\hline
Adversarial Learning \cite{DBLP:journals/csur/BrophyWSW23,mirza2014conditional,radford2015unsupervised,adler2018banach,ganin2018synthesizing,yin2019utilizing,DBLP:conf/nips/SunCGY19}& Scenarios requiring realistic data generation or robust models. & Enables the generation of realistic data, improves model robustness and performance. & Challenges in training stability, mode collapse, and potential for adversarial attacks. \\
\hline
Modality Agnostic Learning \cite{jaegle2021perceiver,vaswani2017attention}& Applications with diverse and evolving data modalities. & Enables learning from various data types, improves model flexibility and adaptability. & Increased computational complexity, challenges in handling the heterogeneity of data modalities. \\
\hline
\end{tabular}
\end{table*}

Central to this evolution is the perceptron model \cite{jaegle2021perceiver}, which drives the development of modality-agnostic learning. This model finds its roots in the Transformers networks \cite{vaswani2017attention}, which contrary to their traditional domain of sequence transduction, have demonstrated prowess in computer vision tasks, all while consuming lesser computational resources. 
% The beauty of the Perceiver lies in its ability to handle varied data configurations, akin to how societies seamlessly transition between diverse challenges. 
This dynamicity not only facilitates broader participation in EI but also sets the stage for more robust and reliable network optimization tasks. %Just as social entities continually refine their strategies in the face of bidirectional challenges, the domain of modality-agnostic learning stands as a testament to the ever-evolving nature of ML.

In summary, we explore the dynamics of conflict interactions within and between layers of ML techniques. %Ensemble Learning showcases the power of collective effort, HRL demonstrates structured, top-down decision making, adversarial learning highlights the competitive drive for dominance, and modality agnostic learning shows the importance of versatility across various data types. 
Each approach contributes uniquely to our understanding of interactivity, complexity, scalability, adaptability, and outcome diversity in ML models, as outlined in Table \ref{table:Comparison of conflict}.

{\color{black}
To provide a comprehensive overview of the ML methods discussed in this section, we present a comparative analysis of their scope of application, advantages, and disadvantages in Table \ref{tab:ml_methods_comparison}. This table enables readers to quickly grasp the key characteristics and potential use cases of each method in the context of socialized  training and EI.}

\vspace{-1em}
\subsection{Lessons Learned}

Our examination within the context of SL and EI underscores the merging of complex human interactions and ML training. This fusion points to an emerging trend of AI models that increasingly incorporate human social behavior. The trend towards adaptable AI, influenced by frameworks like the perceiver model, mirrors social adaptability during transitions. ML's pursuit of devising versatile algorithms reflects the need for \textbf{a balance between efficiency and adaptability}. 

%分段
Additionally, our study delves into the dichotomy of \textbf{individual versus collective efficiency}, as reflected in ensemble learning's portrayal of social individualism and collectivism. This also relates to the HRL's \textbf{directive versus autonomy} debate, reflected in the balance of control and independence seen in society.
Developing ML models that \textbf{balance resilience and adaptability} while retaining specialization is challenging. The apparent synergy between ML and social dynamics signals a future maturity for \textbf{cross-disciplinary cooperation}. 

\begin{figure}[pt]
  \centering
  \includegraphics[width=3.35in]{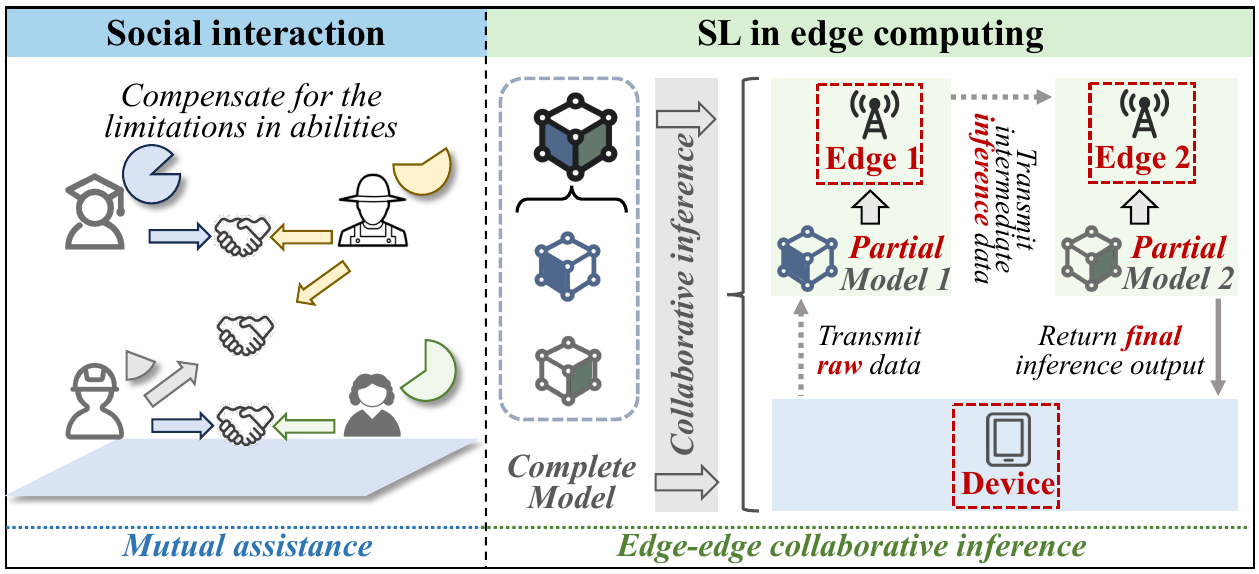}      
  \caption{Illustration of \textbf{edge-edge collaborative inference}.}\label{Edge-edge distributed Inference}
  \vspace{-0.5em}
\end{figure}

% \vspace{0.5em}
\section{Socialized Inference for EI}\label{section7}
% This section explores the integration of socialized inference in the EI framework, akin to social operations through interconnected interactions. 
This section delves into the integration of socialized inference in EI. It examines how social dynamics can improve decision-making, highlighting social interactions and their effects on inference.

\vspace{-1em}
\subsection{{Intra-layer Cooperation in Socialized Inference}}
\subsubsection{\textbf{Mutual Assistance}} 
% Within EI, mutual assistance is crucial for collaborative inference, reminiscent of the interdependence observed in tight-knit human societies.
In EI, mutual assistance is essential for collaborative inference.
Despite limitations, each device contributes unique abilities for the group's benefit.

\subsubsubsection{Edge-edge Collaborative Inference} 
As shown in Fig. \ref{Edge-edge distributed Inference}, EI employs edge cooperation in distributed inference, leveraging collective resources for better inference. This approach negates central oversight as edges converge on shared goals.
Optimization goals categorize this collaboration into:
\begin{itemize}[leftmargin=*]
\item \textbf{Total Inference Latency Minimization:} Edges accelerate inferences collaboratively, avoiding central delays. Just as members of society possess varied abilities, edge resources also are different. Addressing this involves task allocation and workload balancing. Studies such as Wan \emph{et al.} \cite{DBLP:journals/tvt/WanJGZY23} and Mohammed \emph{et al.} \cite{mohammed2020distributed} delve into this, proposing DL-based strategies that maximize resource utilization in EI.

\item \textbf{Total Cost Minimization:} Beyond speed, holistic costs, such as energy and communication overheads, reflect sustainable social cooperation goals. The focus here is on achieving a balance and distributing tasks in a manner that minimizes collective costs. Studies like those proposed by Fan \emph{et al.} \cite{DBLP:journals/tvt/FanHZSLTWL23} have ventured into this area, optimizing edge collaborations for minimum costs.
\end{itemize}   

A view of edge-edge collaborative inference encapsulates resource constraints, computational costs, communication overheads, and latency considerations. 
{\color{black}The SL function Eq. \eqref{SL-function-define} for edge-edge collaborative inference is $F([\text{edges}],[\text{mutual}$ $\text{assistance}], \varnothing, \text{inference}, \text{tasks})$. The optimization objective is:
\vspace{-0.5em}
\begin{equation}
\begin{aligned}
\underset{E}{\text{min}} & \sum_{j=1}^{m} \sum_{i=1}^{n} L(t_j, e_i) + \lambda \sum_{i<k} D(e_i, e_k), \\
\text{s.t.} & \quad C(t_j, e_i) \leq R(e_i), \quad \sum_{j=1}^{m} C(t_j, e_i) \leq \omega, \quad \forall i,j,
\end{aligned}
\vspace{-0.5em}
\end{equation}
where $E = \{e_1, e_2, \ldots, e_n\}$ is the set of edges, $T = \{t_1, t_2, \ldots, t_m\}$ is the set of tasks, $R(e_i)$ is the available resources on edge $e_i$, $C(t_j, e_i)$ is the computational cost of task $t_j$ on edge $e_i$, $D(e_i, e_k)$ is the communication cost between edges $e_i$ and $e_k$, $L(t_j, e_i)$ is the latency of task $t_j$ on edge $e_i$, $\lambda$ is the trade-off parameter, and $\omega$ is the maximum workload allowed on each edge.
}

\begin{figure}[t]
  \centering
  \includegraphics[width=\linewidth]{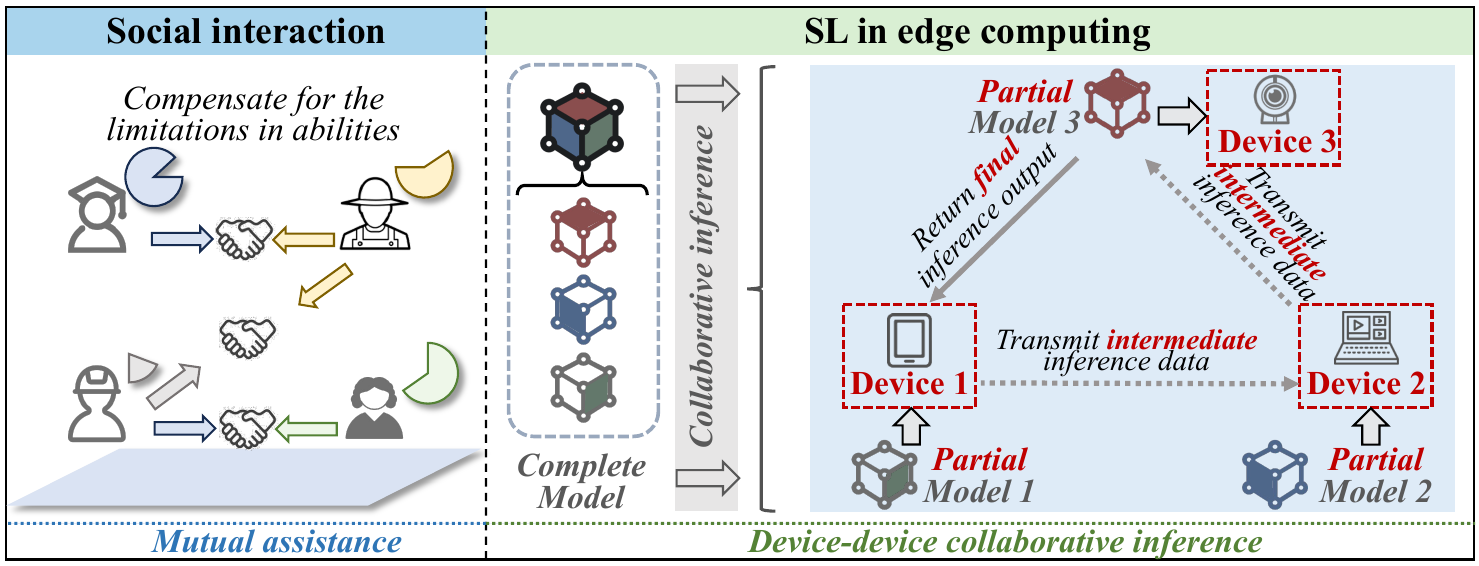}      
  \caption{Illustration of \textbf{device-device collaborative inference}.}\label{device-device collaborative DNN inference}
  \vspace{-0.5em}
\end{figure}

Edge-edge collaborative inference, seen through the lens of mutual assistance, boasts several merits:
1) \textbf{Privacy and Security.} The communal edge approach safeguards sensitive data, echoing the social emphasis on individual privacy.
2) \textbf{Scalability and Flexibility.} Reflecting the adaptability of social cooperation, this method is inherently scalable and accommodative of diverse edges.
3) \textbf{Low Latency.} Emphasizing timely responses, the method caters to real-time applications, paralleling societies' emphasis on immediacy in collaboration.

%分段
While the virtues are evident, the challenges in edge-edge cooperation also mirror those in social mutual assistance:
1) \textbf{Communication Overhead.} Collaboration invariably involves communication, which can be burdensome, more so with increasing collaborators.
2) \textbf{Heterogeneity.} Just as individuals in a society vary, so do edges in capability, posing optimization challenges.
3) \textbf{Fault Tolerance.} Societies must contend with unpredictable challenges; similarly, edge collaboration needs to account for device failures and network anomalies.

{\color{black}
\textbf{Communication Optimization in Edge-edge Collaborative Inference:}
Edge-edge collaborative inference heavily relies on efficient communication among edges to achieve optimal performance and minimize latency. One key challenge is to optimize the communication network topology and routing strategies to ensure fast and reliable information exchange. This can be formulated as a joint inference and communication optimization problem, aiming to minimize total inference latency while satisfying communication bandwidth and reliability constraints. Techniques such as network encoding and cooperative communication can be applied to improve the communication efficiency and robustness in edge-edge collaborative inference scenarios \cite{shao2023task}.}

\subsubsubsection{Device-device Collaborative Inference} 
% The concept of mutual assistance emphasizes the collective efforts of entities working in harmony to achieve a common objective, as illustrated in Fig. \ref{device-device collaborative DNN inference}. This idea is mirrored in device-device collaborative inference in EI. 
Device-device collaborative inference in EI highlights the combined efforts of devices, as illustrated in Fig. \ref{device-device collaborative DNN inference}.
Just as individuals in a society pool their strengths and resources to overcome challenges that are beyond their capability alone, devices, particularly those with limited computational capacities. They can collaborate, exchange information, and process data jointly.

\begin{itemize}[leftmargin=*]
\item \textbf{Total Inference Latency Minimization:} This aspect involves the strategic sharing of computational tasks, akin to public resource allocation, aimed at reducing overall inference delays. Especially critical for UAVs and other latency-sensitive applications, this strategy involves optimizing task allocation across devices, even when faced with inherent resource limitations \cite{ DBLP:journals/jsac/LiuLH23}. The objective remains to prevent resource overloading on a single device, ensuring harmonious operations \cite{DBLP:journals/tc/DisabatoRA21}.

\item \textbf{Total Cost Minimization:} 
The collaborative EI also extends to encompass cost-effectiveness, where devices, particularly those limited by battery life and storage, operate under a paradigm of energy efficiency. This approach necessitates a balance between energy consumption for inference operations and communication, as highlighted in studies by Yang \emph{et al.} \cite{DBLP:journals/tpds/YangZSX23} and Zeng \emph{et al.} \cite{9296560}.
\end{itemize}

The objective is to optimally allocate tasks to devices to minimize both inference latency and energy consumption.
{\color{black}The SL function Eq. \eqref{SL-function-define} for device-device collaborative inference is $F([\text{devices}], [\text{mutual assistance}], \varnothing, \text{inference}, \text{tasks})$. The optimization objective is:
\vspace{-0.7em}
\begin{equation}
\begin{aligned}
\underset{x_{ij}}{\text{min}} & \sum_{i=1}^{n} \sum_{j=1}^{m} (L_{ij} x_{ij} + E_{ij} x_{ij}), \\
\text{s.t.} & \quad \sum_{i=1}^{n} x_{ij} = 1, \quad \sum_{j=1}^{m} R_{ij} x_{ij} \leq R_i, \  x_{ij} \in \{0,1\}, \ \forall i,j,
\end{aligned}
\vspace{-0.2em}
\end{equation}
where $L_{ij}$ is the inference latency of task $t_j$ on device $d_i$, $E_{ij}$ is the energy consumption of task $t_j$ on device $d_i$, $x_{ij}$ is the assignment variable indicating whether task $t_j$ is assigned to device $d_i$, $R_{ij}$ is the resource requirement of task $t_j$ on device $d_i$, and $R_i$ is the available resources on device $d_i$.
}

\begin{figure}[pt]
  \centering
  \includegraphics[width=3.49in]{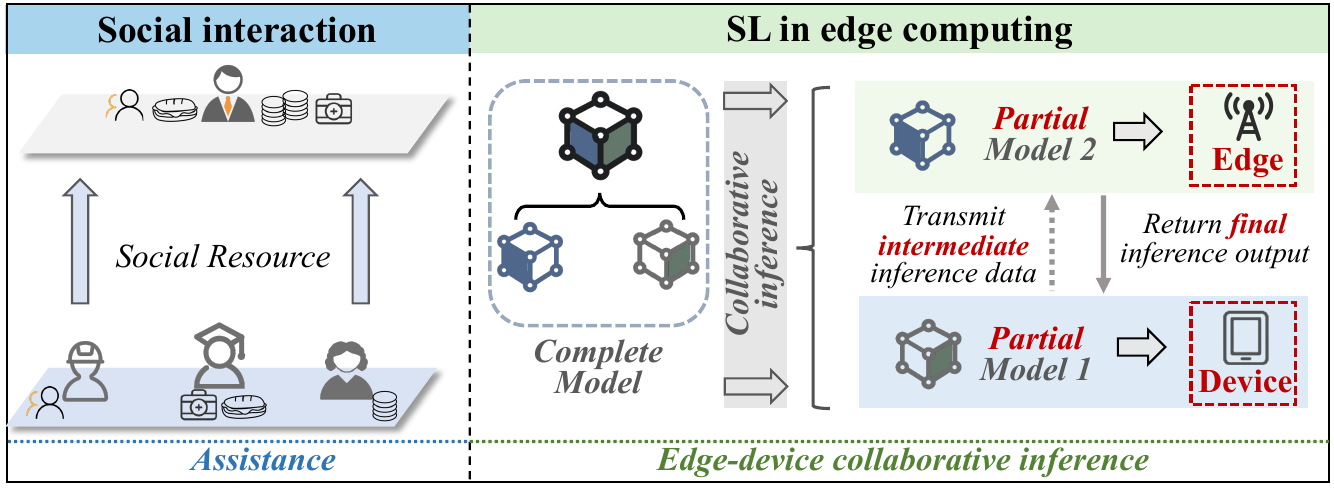}    \caption{Illustration of \textbf{edge-device collaborative inference}.}\label{edge-device collaborative DNN inference}
  \vspace{-0.5em}
\end{figure}

The advantages of this collaborative inference model include:
1) \textbf{System independence:} It promotes operational autonomy of devices, mitigating reliance on central infrastructures, and thereby reducing associated costs.
2) \textbf{Reduced latency:} Collaborative efforts enhance decision-making speeds, a feature reminiscent of quick social problem-solving dynamics.
3) \textbf{Adaptability:} The system exhibits resilience against disruptions, reflective of social robustness.

%分段
However, it also introduces challenges:
1) \textbf{Privacy concerns:} Increased inter-device communication potentially elevates the risk of data leakage, a dilemma paralleling privacy concerns in human societies.
2) \textbf{Synchronization among heterogeneous devices:} Variances in devices may complicate synchronization, an issue analogous to aligning individuals with diverse abilities in a society.
3) \textbf{Task adaption:} The necessity for rapid adaptability in task management, considering variables like energy and computation, reflects the dynamism inherent in social collaborations.

\vspace{-1.5em}
\subsection{{Inter-layer Cooperation in Socialized Inference}}%合作：合作
\subsubsection{\textbf{Unidirectional Cooperation - Assistance}}%合作：互助,平等关系
% In social contexts, assistance typically involves one party offering capabilities or resources to another, with the benefits flowing unidirectionally. 
In EI during inference, certain computing layers provide insights or intermediary results to speed up others' processes.
The aided entities gain the advantages while the helpers play a supporting role. 
% Translating this to EI during inference, certain computing layers or nodes provide insights or intermediary results to accelerate others' inference processes. 
Unlike training, where knowledge transfer is paramount, in EI's inference phase, the focus shifts to utilizing specific layers or devices to minimize latency and expedite speed up decision-making, mirroring how society leverages resources for swift outcomes.
    
\subsubsubsection{Edge-device Collaborative Inference} 
% As depicted in Fig. \ref{edge-device collaborative DNN inference}, edge-device collaborative inference in EI reflects the social principle of unidirectional assistance, where entities with greater resources and capabilities offer support to those in need. 
As depicted in Fig. \ref{edge-device collaborative DNN inference}, in edge-device collaborative inference, entities with greater resources offer support to those in need.
Different computing layers collaborate to optimize DNN inference processes. The categorization of assistant efforts can be understood through distinct performance optimization objectives:

\begin{itemize}[leftmargin=*]
\item \textbf{Total Inference Latency Minimization:} 
In EI, minimizing inference latency is analogous to the rapid social response to challenges. Methods like the DINA framework \cite{mohammed2020distributed} segment DNNs for optimal task allocation, a strategy inspired by social cooperation and matching theory \cite{roth1992two}. This emphasis on collaborative problem-solving for reduced response times is also proposed in studies by Chen \emph{et al.} \cite{10330751} and Li \emph{et al.} \cite{10413648}.

\item \textbf{Inference Accuracy Maximization:}
Beyond speed, the quality of assistance is vital, reflected in EI's pursuit of enhanced inference accuracy. Hanyao \emph{et al.} \cite{DBLP:conf/infocom/Hanyao0Q0L21} highlight this with a system that leverages collaborative inference to maintain accuracy under challenging conditions. Similar studies are explored by Yun \emph{et al.} \cite{DBLP:journals/tvt/YunKCK21} and Kong \emph{et al.} \cite{DBLP:conf/mm/KongYC23}, emphasizing how shared resources and knowledge can lead to better performance.

\item \textbf{Total Cost Minimization:} 
Reflecting the social requirement for efficient resource use, EI strives for cost-effective operations. Fang \emph{et al.} \cite{DBLP:conf/mobicom/FangZ018} demonstrate this through DNN filter pruning, optimizing resource utilization. This balance is also the focus of research by Du \emph{et al.} \cite{DBLP:journals/tcom/DuZFC18} and Dai \emph{et al.} \cite{DBLP:journals/tsc/DaiWWYZX23}, which mirror social efforts in optimal allocation.
\end{itemize}

{\color{black}The SL function Eq. \eqref{SL-function-define} for edge-device collaborative inference is $F([\text{edge-devices}], \varnothing, [\text{assistance}], \text{inference}, \text{tasks})$. The optimization objective is:
\vspace{-1em}
\begin{equation}
\begin{aligned}
\underset{x_{ij}^d, x_{ij}^e}{\text{min}} & \sum_{i=1}^{n} \sum_{j=1}^{m} (L_{ij}^d + E_{ij}^d) x_{ij}^d + \sum_{i=1}^{k} \sum_{j=1}^{m} (L_{ij}^e + E_{ij}^e) x_{ij}^e, \\
\text{s.t.} & \quad \sum_{i=1}^{n} x_{ij}^d + \sum_{i=1}^{k} x_{ij}^e = 1, \quad \sum_{j=1}^{m} R_{ij}^d x_{ij}^d \leq R_{di}, \quad \forall i,j, \\
& \quad \sum_{j=1}^{m} R_{ij}^e x_{ij}^e \leq R_{ej}, \quad x_{ij}^d, x_{ij}^e \in \{0,1\}, \quad \forall i, j,
\end{aligned}
\vspace{-0.5em}
\end{equation}
where $L_{ij}^d$ and $L_{ij}^e$ are the inference latency of task $t_j$ on device $d_i$ and edge $e_i$, respectively, $E_{ij}^d$ and $E_{ij}^e$ are the energy consumption of task $t_j$ on device $d_i$ and edge $e_i$, respectively, $x_{ij}^d$ and $x_{ij}^e$ are the assignment variables indicating whether task $t_j$ is assigned to device $d_i$ or edge $e_i$, $R_{ij}^d$ and $R_{ij}^e$ are the resource requirements of task $t_j$ on device $d_i$ and edge $e_i$, respectively, and $R_{di}$ and $R_{ej}$ are the available resources on device $d_i$ and edge $e_j$, respectively.
}

Edge-device collaborative DNN inference brings about:
1) \textbf{Lower latency}, akin to rapid social assistance.
2) \textbf{Lower cost}, analogous to social efforts aimed at optimizing resource utilization.
Despite these advantages, challenges analogous to those in social assistance persist. Limitations in real-time performance reflect social constraints, underscoring that while it introduces a novel paradigm, its evolution is progressive, similar to the evolution of social structures over time.

\begin{figure}[pt]
  \centering
  \includegraphics[width=3.49in]{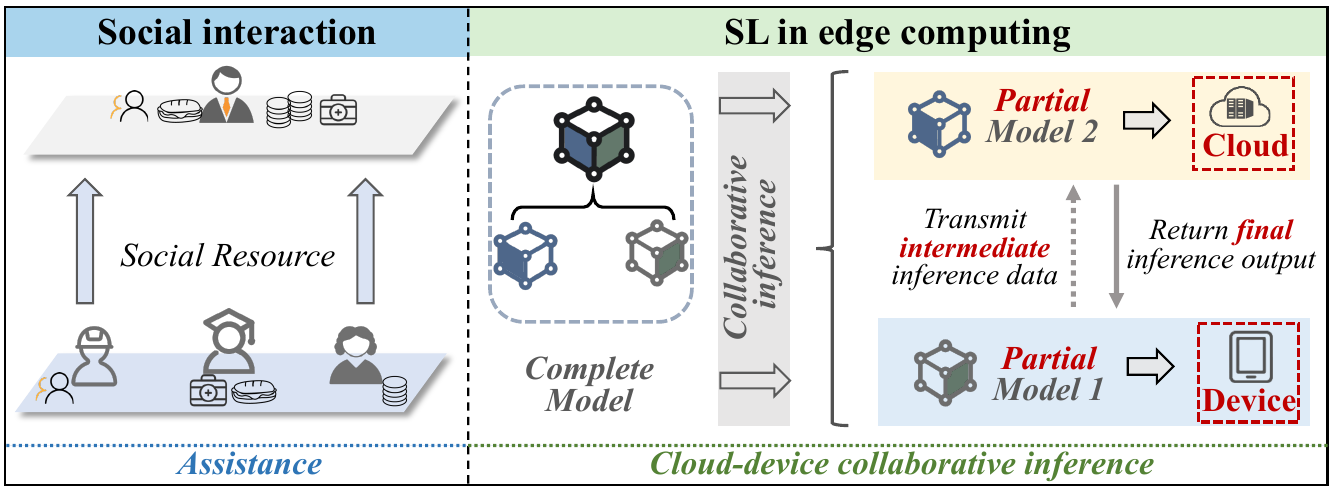}     
  \caption{Illustration of \textbf{cloud-device collaborative inference}.}\label{Cloud-device collaborative inference}
  \vspace{-0.5em}
\end{figure}

{\color{black}
\textbf{Communication-efficient Edge-device Collaborative Inference:}
Edge-device collaborative inference involves frequent communication between ESs and devices, which can be a bottleneck in resource-constrained environments. To address this challenge, various communication-efficient strategies have been proposed, such as model compression, quantization, and selective transmission of intermediate features \cite{shao2020normalization}. These techniques aim to reduce the amount of data exchanged between ESs and devices while maintaining inference accuracy. Additionally, the design of communication-efficient protocols and scheduling algorithms is crucial for optimizing the overall performance of edge-device collaborative inference systems.
}    
    
\subsubsubsection{Cloud-device Collaborative Inference} 
Fig. \ref{Cloud-device collaborative inference} illustrates the collaborative DNN inference within EI, showcasing the dynamic between the cloud and the device that mirrors the social norm of unidirectional assistance. Here, the cloud, rich in resources, empowers devices to enhance their operational efficiency. This interplay is discussed further through the lens of performance optimization objectives:
    
\begin{itemize}[leftmargin=*]
\item \textbf{Total Inference Latency Minimization:} 
Vital for real-time applications such as autonomous driving and video surveillance, the balance between accuracy and latency is non-negotiable. Emmons \emph{et al.} \cite{DBLP:conf/mobicom/EmmonsFAVSW19} highlight the merit of synergy between cloud and devices. The introduction of complex DNN structures, such as GoogleNet \cite{DBLP:conf/cvpr/SzegedyLJSRAEVR15} and ResNet \cite{DBLP:conf/cvpr/HeZRS16}, complicates the model partitioning. This challenge has been addressed by Hu \emph{et al.} \cite{DBLP:conf/infocom/HuBWL19} through the development of an adaptive scheme. Further research by Duan \emph{et al.} \cite{DBLP:journals/tmc/DuanW24} and Jeong \emph{et al.} \cite{DBLP:conf/cloud/JeongLSM18} emphasizes parallel DNN computation as a solution for latency.

\item \textbf{Energy Consumption Minimization:} 
While the cloud dominates in computational capacity, device energy limitations are a stark reality. Research by Kang \emph{et al.} \cite{DBLP:conf/asplos/KangHGRMMT17} and Hauswald \emph{et al.} \cite{DBLP:conf/icassp/HauswaldMZDCM14} provides strategies for computing partitioning, optimizing both latency and energy use. Collaborative frameworks, as suggested by Eshratifar \emph{et al.} \cite{DBLP:journals/tmc/EshratifarAP21} and Jiang \emph{et al.} \cite{DBLP:conf/icc/DengTF16}, further highlight the need for collaborative frameworks that balance performance with energy consumption.
\end{itemize}

We encapsulate the optimization task of balancing latency and energy consumption for a collaborative inference strategy between the cloud and devices. 
{\color{black}The SL function Eq. \eqref{SL-function-define} for cloud-device collaborative inference is $F([\text{cloud-devices}],\varnothing , [\text{assistance}], \text{inference}, \text{DNNs})$. The optimization objective is:
% \vspace{-0.5em}
\begin{equation}
\begin{aligned}
\underset{a}{\text{min}} & \quad \alpha L + \beta E, \
\text{s.t.} & \quad 0 \leq a \leq 1,
\end{aligned}
\vspace{-0.5em}
\end{equation}
where $L = aL_c + (1-a)L_d$ is the total latency, $E = aE_c + (1-a)E_d$ is the total energy consumption, $a$ is the partition point of DNN, $L_c$ and $L_d$ are the latency on the cloud and device, respectively, $E_c$ and $E_d$ are energy consumption on the cloud and device, respectively, and $\alpha$ and $\beta$ are weights for latency and energy consumption, respectively.
}

\begin{figure}[pt]
  \centering
  \includegraphics[width=3.49in]{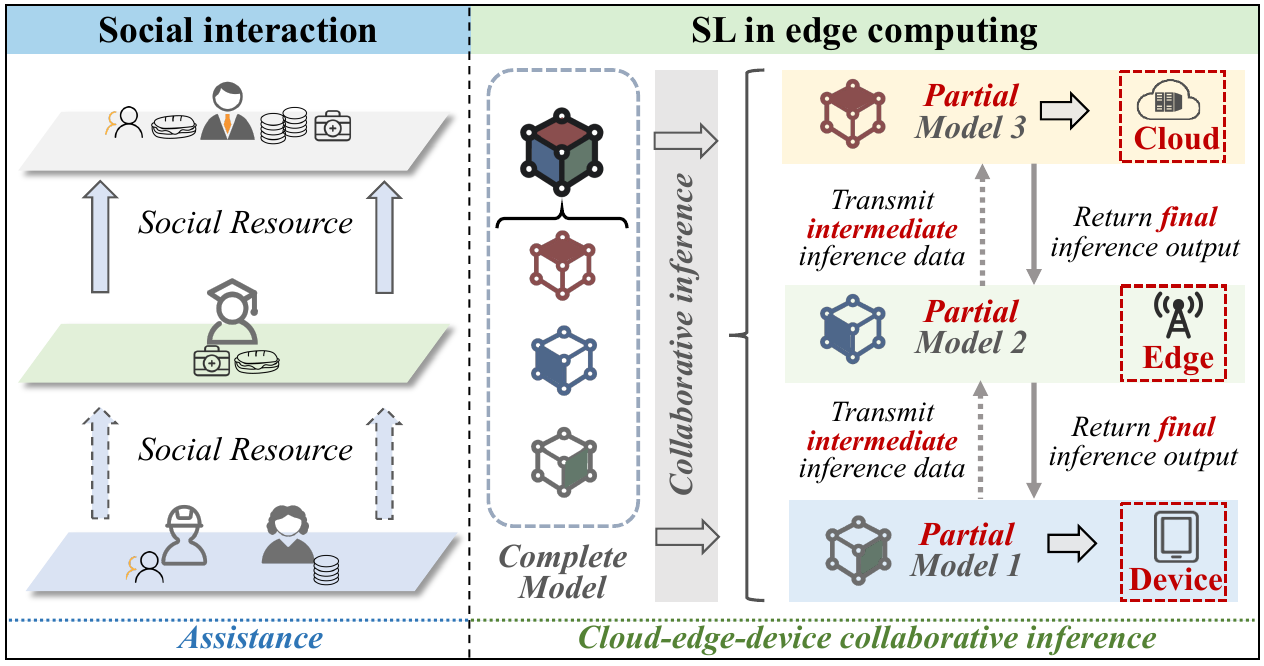}    
  \caption{Illustration of \textbf{cloud-edge-device collaborative inference}.}\label{cloud-edge-device collaborative DNN inference}
  % \vspace{-0.5em}
\end{figure}

The comparison between assistance and cloud-device collaboration in inference clearly demonstrates similar advantages, such as:
1) \textbf{Lowered inference latency}, achieved through strategic model partitioning and task allocation.
2) \textbf{Reduced energy consumption}, by optimizing the use of communication and computation resources.
However, this collaboration also faces distinct challenges:
1) \textbf{Maintaining performance in unstable conditions}, where factors like rapid mobility or communication disruptions can present significant hurdles.
2) \textbf{Navigating the physical disconnect} between cloud and devices, which can limit the scope of applications.
This comparison underscores the essential synergy required to achieve superior results in both human society and EI.

\subsubsubsection{Cloud-edge-device Collaborative Inference} 
% Drawing inspiration from unilateral aid in human societies, cloud-edge-device collaboration reflects a scenario where specific components (be it cloud, edge, or device) bolster another's performance without necessarily receiving an immediate reciprocal advantage. 
Cloud-edge-device collaboration involves specific components (cloud, edge, or device) assisting others without expecting immediate reciprocal benefits.
This collaboration branches into three distinct categories, each focusing on enhancing various performance aspects, as shown in Fig. \ref{cloud-edge-device collaborative DNN inference}.

\begin{itemize}[leftmargin=*]
% \item \textbf{Total Inference Latency Minimization:} Addressing the latency issue inherent in expansive collaborative DNN inference systems, Ren \emph{et al.} \cite{9495122} have formulated a distributed DNN computing partition. This setup, much like a supportive entity in social assistance, coordinates computations across heterogeneous devices to boost inference efficiency. 
\item \textbf{Total Inference Latency Minimization:}  To address latency issues in collaborative DNN inference systems, Ren \emph{et al.} \cite{9495122} formulate a distributed DNN partition. This setup coordinates computations across heterogeneous entities to boost inference efficiency.
Further contributions by Xue \emph{et al.} \cite{9534773} focus on offloading and distributed DNN deployment, providing beneficial support to the whole system in terms of latency reduction.

\item \textbf{Total Cost Minimization:} Within this collaboration, cost reduction is integral, akin to a form of aid that elevates system functionality. Lin \emph{et al.} \cite{lin2019cost} propose an adaptive optimization algorithm, reducing system offloading costs significantly. Similarly, strategies by Teerapittayanon \emph{et al.} \cite{teerapittayanon2017distributed} and Liu and Wang \emph{et al.} \cite{10472084} develop approaches that strategically map parts of a DNN across the computing hierarchy, optimizing overall costs associated with communication and inference processes.

\item \textbf{Failure-Resilient Distributed Model:} In situations of component failures, the concept of assistance becomes paramount to ensure system robustness. The importance of assistance is portrayed by Yousefpour \emph{et al.} \cite{yousefpour2019guardians}, where skipping hyperconnections in distributed DNNs provides a safeguarding mechanism, ensuring continuity in the face of physical node failures.
\end{itemize}

We encapsulate the optimization process to achieve the best collaborative inference while considering latency, cost, and failure resilience across cloud, edge, and device.
{\color{black}The SL function Eq. \eqref{SL-function-define} for cloud-edge-device collaborative inference is $F([\text{cloud-edge-devices}], [\text{assistance}], \varnothing, \text{inference}, \text{DNNs})$. The optimization objective is:
\vspace{-0.5em}
\begin{equation}
\begin{aligned}
\text{min} & \quad L + C - F, \\
\text{s.t.} & \quad M = M_c \cup M_e \cup M_d, \\
& \quad L = \alpha L(M_c) + \beta L(M_e) + \gamma L(M_d), \\
& \quad C = \delta C(M_c) + \xi C(M_e) + \pi C(M_d), \\
& \quad F = \epsilon F(M_c) + \zeta F(M_e) + \eta F(M_d),
\end{aligned}
\vspace{-0.5em}
\end{equation}
where $M$ is the entire DNN, $M_c$, $M_e$, and $M_d$ are the partitions of the model executed on the cloud, edge, and device, respectively. $L(M_x)$, $C(M_x)$, and $F(M_x)$ are the latency, cost, and failure resilience of executing partition $M_x$ on the corresponding platform, respectively. $\alpha, \beta, \gamma, \delta, \xi, \pi, \epsilon, \zeta, \eta$ are the weights for the respective terms.
}

\begin{figure}[pt]
  \centering
  \includegraphics[width=\linewidth]{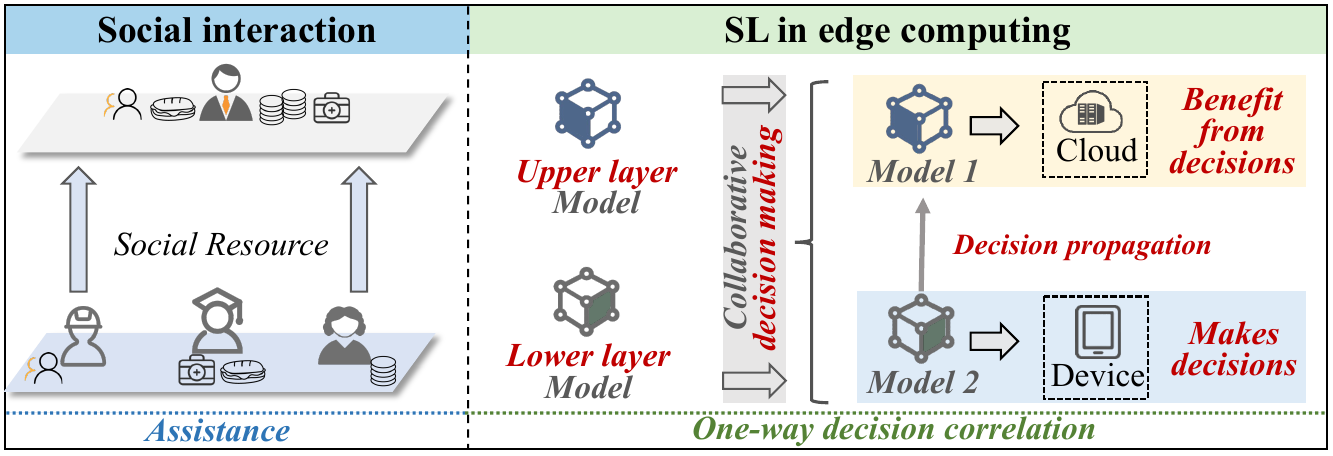} 
  \caption{\color{black}Illustration of \textbf{one-way decision correlation} in socialized inference.}\label{one-way decision-major}
  \vspace{-0.5em}
\end{figure}

% Compared with the previous two inference paradigms, the research on cloud-edge-device collaborative inference can also be considered as a supplement to cloud-device collaborative inference and edge-device collaborative inference, and it has serval advantages as follows: 
Compared to the prior two inference paradigms, cloud-edge-device collaborative inference serves as a supplement to both cloud-device and edge-device collaborative inferences, offering several advantages:
1) \textbf{Optimized resource utilization}, through detailed model segmentation that integrates the full potential of clouds, edges, and devices.
2) \textbf{Enhanced fault recovery capabilities}, by establishing a robust distributed system model that improves sensor fusion, system resilience, and data confidentiality.

{\color{black}
\textbf{Communication Optimization in Cloud-edge-device Collaborative Inference:}
Cloud-edge-device collaborative inference systems involve coordination and communication among multiple layers, which can introduce significant overhead and complexity. Optimizing the communication efficiency in such systems is crucial for achieving low-latency and high-performance inference. This can be formulated as a multi-objective optimization problem, where the goals are to minimize the communication cost, inference latency, and energy consumption while ensuring the required inference accuracy \cite{wang2022pcnncec}. Techniques such as hierarchical communication, data compression, and adaptive transmission strategies can be employed to optimize communication efficiency in cloud-edge-device collaborative inference scenarios.}

\subsubsubsection{One-way Decision Correlation} 
% Drawing inspiration from human social behaviors where assistance is often extended in a unidirectional manner, the dynamics of one-way decision correlation in the EI echo the principles of unidirectional cooperation. 
The dynamics of one-way decision correlation in EI focus on unidirectional cooperation, as shown in Fig. \ref{one-way decision-major}.
Decisions made within one hierarchical layer guide subsequent layers, assisting them without direct reciprocation. We explore the implications of unidirectional assistance in one-way decision correlation within EI.

\begin{itemize}[leftmargin=*]
% \item \textbf{Total Inference Latency Minimization:} Analogous to an expert offering guidance to a novice, one-way decision correlation streamlines processes by transferring decisions from one layer to another in a unidirectional manner, thus averting redundant back-and-forth communications. 
\item \textbf{Total Inference Latency Minimization:} One-way decision correlation streamlines processes by transferring decisions from one layer to another, reducing communication load.
Benefiting from upper decisions, lower layers can enhance decision speed and quality, streamlining decision-making and reducing inference latency. Cao \emph{et al.} \cite{cao2021edge} present a framework where the cloud `big model' assists edge `small models' by differentiating case complexity, and optimizing task allocation.

\item \textbf{Energy Consumption Minimization:} Similarly, in the energy consumption area, the lower layers align their operations efficiently by receiving and capitalizing on decisions from an upper layer, mitigating unnecessary computational costs. The layer that benefits from this transfer of decisions saves energy, mirroring the dynamic where one party assists another by offering resources or expertise, leading to optimized outcomes. Kosuru \emph{et al.} \cite{kosuru2023intelligent} showcase this principle within their methods, providing energy-efficient solutions in their respective contexts.
\end{itemize}

\begin{figure}[pt]
  \centering
  \includegraphics[width=3.49in]{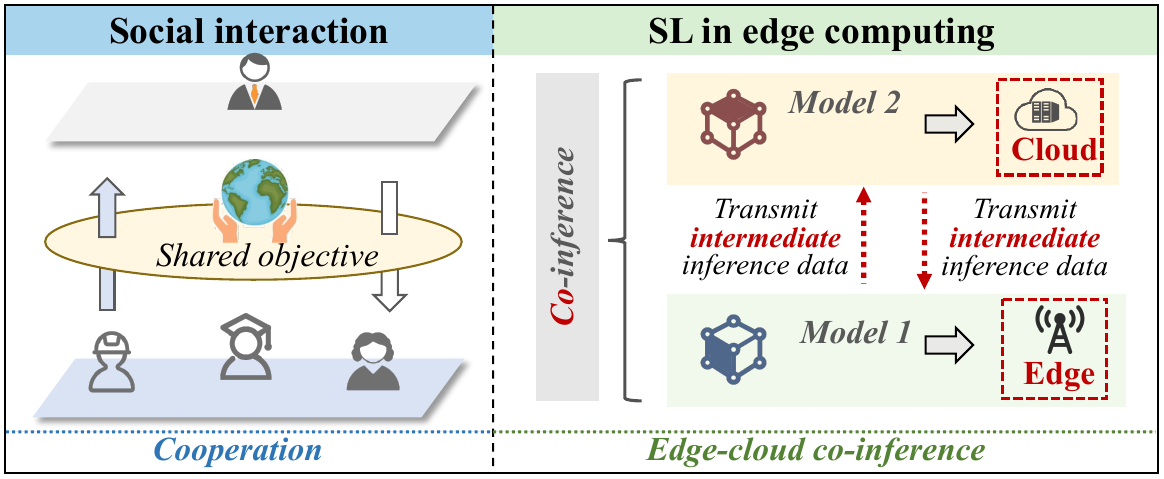}      
  \caption{Illustration of \textbf{edge-cloud co-inference} in socialized inference.}\label{Edge-cloud Co-inference}
  % \vspace{-0.5em}
\end{figure}

In the context of one-way decision correlation, we model layer interactions and decision propagation using the following equations. {\color{black}The SL function Eq. \eqref{SL-function-define} for one-way decision correlation is $F([\text{multi-layer}], \varnothing, [\text{assistance}], \text{inference}, \text{decisions})$. The decision propagation is modeled as:
\vspace{-0.5em}
\begin{equation}
\begin{aligned}
\text{min} & \quad \sum_{i} L_i(D_{i})+\sum_{i} E_i(D_{i}), \
D_{i+1} = f_{i+1}(D_{i}), 
\end{aligned}
\vspace{-0.5em}
\end{equation}
where $D_{i}$ is the decision made at layer $i$, $f_{i}$ is the decision-making function at layer $i$, $L_i(D_{i})$ and $E_i(D_{i})$ are the latency and energy consumption of making decision $D_{i}$ at layer $i$, respectively.}
%In reflection, the one-way decision correlation in EI mirrors the unidirectional cooperation observed in human interactions.
This non-reciprocal decision transfer aligns with altruism, streamlines processes, reduces inference latency, enhances real-time decision-making for time-sensitive tasks, and conserves energy by optimizing resource use and eliminating unnecessary computations.
 
\subsubsection{\textbf{Bidirectional Cooperation - Cooperation}}
% Bidirectional cooperation in EI, much like in human societies, signifies a form of mutualistic interaction where entities engage in reciprocal information and resource exchange to achieve common goals. 
Bidirectional cooperation in EI signifies mutual interaction where devices engage in reciprocal information and resource exchange.
% This collaborative spirit in EI extends beyond simple model parameter sharing, encompassing sharing intermediate results, and refining both individual and collective intelligence.
This interplay ensures a collaborative evolution where individuals learn and adapt based on the broader insights from the EI, leading to more accurate and efficient inferences.
    
\subsubsubsection{Edge-cloud Co-inference}
The edge-cloud co-inference, as shown in Fig. \ref{Edge-cloud Co-inference}, exemplifies bidirectional cooperation in EI. In this scheme, both cloud and edge devices actively participate, leveraging their unique strengths to enhance overall performance and efficiency.

\begin{itemize}[leftmargin=*]
\item \textbf{Total Inference Latency Minimization:}
This collaborative dynamic optimizes task allocation based on the inherent capabilities of cloud and edge resources. The cloud, with its superior computing power, handles more complex, resource-intensive tasks, while edges, being closer to the data source, address latency-sensitive tasks. This judicious task allocation offloads certain operations to the cloud, allowing the edge to focus on timely processing, thereby enhancing system adaption. Reinforcing this scheme, Li \emph{et al.} \cite{li2021appealnet} introduce AppealNet, an architecture aiming to synergize accuracy with low latency in inference. Similarly, Banitalebi Dehkordi \emph{et al.} \cite{banitalebi2021auto} advance the collaborative edge-cloud AI through Auto-Split, a mechanism tailoring performance optimization with minimum latency.

\item \textbf{Energy Consumption Minimization:}
Energy efficiency is another critical focus of the edge-cloud co-inference. By intelligently offloading tasks to the cloud, edges can operate within their energy constraints, thereby promoting sustainable energy use across the EI system. Further, optimizing resource allocation and power management across this collaborative framework encapsulates the principle of mutual resource conservation. Xiang \emph{et al.} \cite{xiang2021energy} delve into the cooperation of service provisioning in EC systems, advocating for an energy-efficient scheme. Concurrently, Zhang \emph{et al.} \cite{zhang2021energy} explore strategies to optimize energy consumption during task offloading in edge-cloud co-inference.
\end{itemize}

We can model the cooperation between the edge and cloud for latency and energy consumption with a set of equations.
{\color{black}The SL function Eq. \eqref{SL-function-define} for edge-cloud co-inference is $F([\text{cloud-edge}], \varnothing, [\text{cooperation}], \text{inference}, \text{tasks})$. The optimization objective is:
\vspace{-0.5em}
\begin{equation}
\begin{aligned}
\text{min} & \quad \alpha T_{\text{total}} + \beta E_{\text{total}}, \\
\text{s.t.} & \quad T_{\text{total}} = T_{\text{edge}}(D) + T_{\text{cloud}}(D) + T_{\text{comm}}(D), \\
& \quad E_{\text{total}} = E_{\text{edge}}(D) + E_{\text{cloud}}(D) + E_{\text{comm}}(D), \\
& \quad A_{\text{inference}} \geq A_{\text{threshold}},
\end{aligned}
% \vspace{-0.5em}
\end{equation}
where $T_{\text{total}}$ and $E_{\text{total}}$ are the total inference latency and energy consumption, respectively, $T_{\text{edge}}(D)$, $T_{\text{cloud}}(D)$, and $T_{\text{comm}}(D)$ are the latency of executing decisions $D$ on the edge, cloud, and communication, respectively. $E_{\text{edge}}(D)$, $E_{\text{cloud}}(D)$, and $E_{\text{comm}}(D)$ are the energy consumption of executing decisions $D$ on the edge, cloud, and communication, respectively. $A_{\text{inference}}$ is the inference accuracy, $A_{\text{threshold}}$ is the minimum required accuracy. $\alpha$ and $\beta$ are the weights for latency and energy, respectively.
}

% The cooperation between the cloud and edge encapsulates the spirit of mutual aid and symbiotic growth, reminiscent of social bidirectional collaborations. 
Harnessing both cloud and edge strengths ensures task processing with maximum efficiency, minimum latency and energy conservation. Due to challenges such as network latency, scalability and data privacy, ongoing refinement and adjustment is essential to harness the framework's potential.

\begin{figure}[pt]
  \centering
  \includegraphics[width=3.49in]{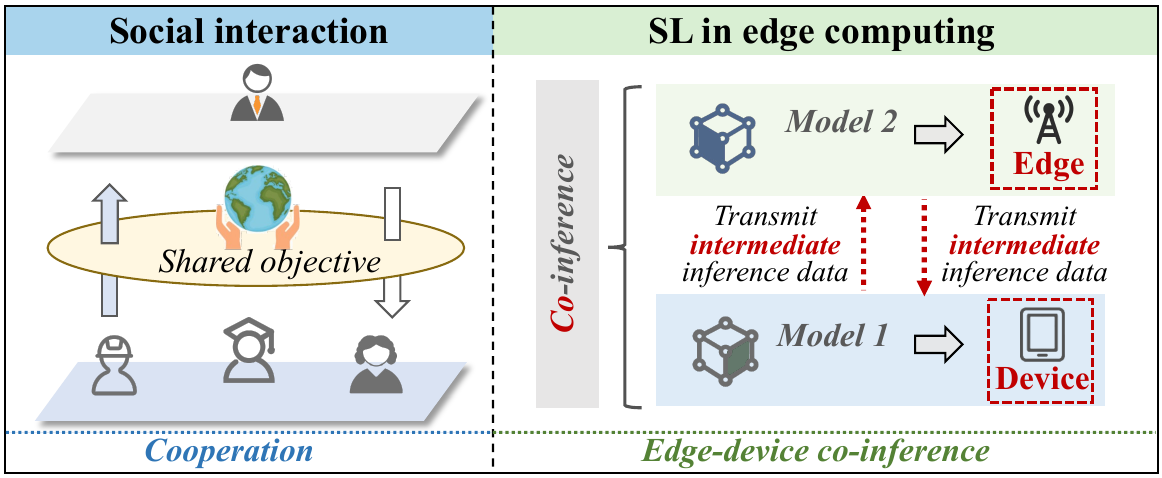}  
  \caption{Illustration of \textbf{edge-device co-inference} in socialized inference.}\label{Edge-device Co-inference}
  \vspace{-0.5em}
\end{figure}
    
\subsubsubsection{Edge-device Co-inference} 
% Expanding on the edge-cloud collaboration, Edge-device Co-inference represents a further evolution of bidirectional cooperation within EI, as illustrated in Fig. \ref{Edge-device Co-inference}. This paradigm mirrors social structures, wherein diverse actors work in concert for mutual advancement. 
Building on edge-cloud collaboration, edge-device co-inference, as depicted in Fig. \ref{Edge-device Co-inference}, represents an evolution of bidirectional cooperation in EI, highlighting the synergistic roles of edges and devices. The ensuing discourse delineates the framework of cooperation, underscoring the participatory roles of both edges and devices.

\begin{itemize}[leftmargin=*]
\item \textbf{Total Inference Latency Minimization:}
Analogous to social cooperation optimizing task offloading, the edge-device dynamic leverages the strengths of both tiers for latency reduction. Edges, equipped with robust computational resources, absorb significant processing loads, enabling devices to efficiently manage local tasks and circumvent excessive transmission delays. Such an interaction finds support in the work of Bai \emph{et al.} \cite{bai2020latency}, where the balance between benefits of computational offloading and challenges of communication latency is assessed, highlighting the utility of intelligent reflecting surfaces (IRSs) in enhancing efficiency.

\item \textbf{Energy Consumption Minimization:}
Drawing parallels with social emphasis on resource preservation, the edge-device collaboration prioritizes energy conservation. Distributed energy strategies, as seen in Krouka \emph{et al.} \cite{krouka2021energy}, implement model compression techniques for energy optimization at the edge without model performance compromise. Likewise, Xiao \emph{et al.} \cite{xiao2022reinforcement} propose a dynamic, RL-driven approach, aligning devices to real-time conditions for sustained energy efficiency.
\end{itemize}

This mechanism balances task allocation between edge and device to optimize latency and energy. 
{\color{black}The SL function Eq. \eqref{SL-function-define} for edge-device co-inference is $F([\text{edge-devices}], \varnothing,$ $[\text{cooperation}], \text{inference}, \text{tasks})$. The optimization objective is:
\vspace{-0.5em}
\begin{equation}
\begin{aligned}
\text{min} & \quad L = L_e(T_e) + L_d(T_d), \quad E = E_e(T_e) + E_d(T_d), \\
\text{s.t.} & \quad 0 \leq T_e \leq 1, \quad 0 \leq T_d \leq 1, \quad T_e + T_d = 1,
\end{aligned}
\end{equation}
where $L$ and $E$ are the total inference latency and energy consumption, respectively. $L_e(T_e)$ and $L_d(T_d)$ are the latency of executing tasks $T_e$ and $T_d$ on the edge and device, respectively. $E_e(T_e)$ and $E_d(T_d)$ are the energy consumption of executing tasks $T_e$ and $T_d$ on the edge and device, respectively. $T_e$ and $T_d$ are the proportions of tasks executed on the edge and device, respectively.
}
Building on edge-cloud co-inference, edge-device co-inference greater highlights the benefits of bidirectional cooperation. Emphasizing task offloading, adaptive management, and proximity benefits, this mechanism minimizes latency and ensures energy efficiency.

\begin{figure}[pt]
  \centering
  \includegraphics[width=\linewidth]{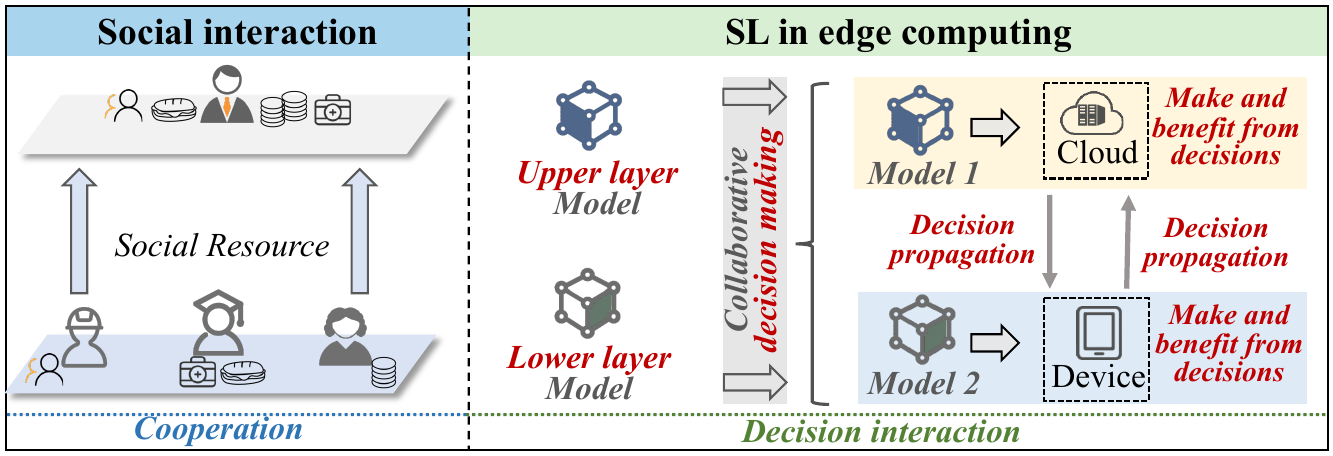} 
  \caption{\color{black}Illustration of \textbf{decision interaction} in socialized inference.}\label{decision interaction-major}
  \vspace{-0.5em}
\end{figure}
    
\subsubsubsection{Decision Interaction} 
% Interdependence is a cornerstone of social evolution, critical for both survival and advancement. In EI, bidirectional decision-making cooperation mirrors this social interdependence, emphasizing the reciprocal nature of decision processes between entities.
Bidirectional decision-making cooperation in EI emphasizes the importance of interdependence between entities, as shown in Fig. \ref{decision interaction-major}.
This cooperation transcends mere task offloading, focusing instead on collaborative exchange and coordination of decision-making processes and highlighting the interconnectedness of decision-making and the mutual exchange of decision-related insights.

\begin{itemize}[leftmargin=*]
\item \textbf{Total Inference Latency Minimization:}
% Much like communities that harmonize decisions by pooling knowledge to respond swiftly to environmental stimuli, bidirectional decision-making cooperation seeks to minimize total inference latency. 
Bidirectional decision-making cooperation aims to minimize inference latency through efficient knowledge pooling. This is achieved by fostering a cooperative environment where devices exchange decision-related insights or intermediate results, synergizing their expertise for timely decisions. An example is presented by Carroll \emph{et al.} \cite{carroll2022towards}, where the FlexiBiT framework illuminates the potential of masked modeling in sequential decision-making. Just as collective decision-making in societies often utilises diverse knowledge sources for better outcomes, this framework's capacity to be attuned to a range of tasks exemplifies the benefits of collaborative decision-making.

\item \textbf{Energy Consumption Minimization:} 
% Historical societies often operated under resource constraints, making collective decisions on rationing and distribution vital. Similarly, in the EI, energy remains a finite resource, making its judicious use a collaborative imperative. 
In the bidirectional decision-making domain, devices are keenly aware of energy saving. They engage in strategies like intelligent resource allocation or the selective sharing of decision data, all aimed at optimizing energy consumption without compromising decision accuracy. Guo \emph{et al.} \cite{tang2022effects} present an exploration into the optimization of energy consumption in distributed inference, consistent with the social principle of strategic resource division for the collective good.
\end{itemize}

{\color{black}The SL function Eq. \eqref{SL-function-define} for decision interaction is $F([\text{server-devices}],$ $\varnothing, [\text{cooperation}], \text{inference}, \text{decisions})$. The optimization objective is:
\vspace{-0.5em}
\begin{equation}
\begin{aligned}
\underset{D_1, D_2}{\text{min}} \quad L(D_1, D_2) + \lambda E(D_1, D_2), \
\text{s.t.} \quad I(D_1, D_2) \geq \theta,
\end{aligned}
\vspace{-0.4em}
\end{equation}
where $D_1$ and $D_2$ are the decision vectors of the two interacting entities, $L(D_1, D_2)$ and $E(D_1, D_2)$ are the latency and energy consumption of making decisions $D_1$ and $D_2$, respectively. $I(D_1, D_2)$ is the interaction quality between the two decision vectors. $\theta$ is the minimum required interaction quality, and $\lambda$ is the weight for energy consumption.
}

Decision-making in this cooperative environment focuses on both speed (lower latency) and sustainability (less energy consumption). Emphasizing mutual interaction and coordination underlines the well-known concept that collaboration, whether in society or networks, often exceeds individual efforts. However, to fully benefit from collaborative decision-making, challenges like decision synchronization and information exchange must be managed. 

\subsubsection{\textbf{Bidirectional Cooperation - Exchange}}
In social interactions, exchanging resources and knowledge is crucial for mutual benefit. This principle, when applied to EI, entails an interaction between hierarchical layers during the inference phase. The cloud, with its extensive data repositories, requires data from edges for context-aware inferences, while edge layers use the cloud’s computing power for deeper analysis. This interplay ensures inferences are prompt and relevant, and capitalize on each stratum's strengths, fostering an integrated and efficient system. We explore how clouds, edges, and devices while maintaining their distinct objectives, engage in this reciprocal exchange during task inference.

\begin{itemize}[leftmargin=*]
\item \textbf{Total Inference Latency Minimization:} 
Reciprocity in bidirectional exchanges aids in optimizing inference latency. While the cloud focuses on depth, the edge emphasizes speed and context. Devices, being closest to the source of data, prioritize immediate, low-latency inferences, drawing insights from the edge or cloud. The aim is a harmonious distribution of tasks that benefits from each layer's strengths. Li \emph{et al.} \cite{li2022multi} delve into reducing latency, particularly during simultaneous inference across multiple models with resource constraints. The research pivots to task scheduling in real-time, utilizing the ONNX runtime engine for inference. It highlights the challenges in lowering system latency and maintaining compatibility across different platforms, particularly during simultaneous model execution.

\begin{table*}[t]
\color{black}
\centering
\caption{Comparative Analysis of Inference Approaches Reflecting Cooperation Interactions}
\begin{tabular}{|c|c|c|c|c|c|c|c|}
\hline
\textbf{Type}                                                              & \textbf{\begin{tabular}[c]{@{}c@{}}Social \\ Relationship\end{tabular}}      & \textbf{Technique}                                                                       & \textbf{Efficiency}                                                                                     & \textbf{Scalability}                                                                          & \textbf{\begin{tabular}[c]{@{}c@{}}Communication\\ Overhead\end{tabular}}             & \textbf{\begin{tabular}[c]{@{}c@{}}Heterogeneity\\ Handling\end{tabular}}     & \textbf{\begin{tabular}[c]{@{}c@{}}Cost\\ Efficiency\end{tabular}}                                                     \\ \hline
\multirow{2}{*}{\begin{tabular}[c]{@{}c@{}}Intra\\ -\\ layer\end{tabular}} & \multirow{2}{*}{\begin{tabular}[c]{@{}c@{}}Mutual\\ Assistance\end{tabular}} & \begin{tabular}[c]{@{}c@{}}Edge-edge\\ Collaborative\\ Inference \\ \cite{DBLP:journals/tvt/WanJGZY23,mohammed2020distributed,DBLP:journals/tvt/FanHZSLTWL23}\end{tabular}              & \begin{tabular}[c]{@{}c@{}}Optimized for\\ minimum latency\end{tabular}                                 & \begin{tabular}[c]{@{}c@{}}Highly scalable\\  due to \\ decentralization\end{tabular}         & \begin{tabular}[c]{@{}c@{}}Moderate due to\\ inter-edge\\ communication\end{tabular}  & \begin{tabular}[c]{@{}c@{}}Challenges due\\ to diverse edge\end{tabular}      & \begin{tabular}[c]{@{}c@{}}Costs saved by \\ reduce server\\ but increase with\\ communication\\ overhead\end{tabular} \\ \cline{3-8} 
                                                                           &                                                                              & \begin{tabular}[c]{@{}c@{}}Device-device\\ Collaborative\\ Inference\\ \cite{DBLP:journals/jsac/LiuLH23,DBLP:journals/tc/DisabatoRA21,DBLP:journals/tpds/YangZSX23,9296560}\end{tabular}        & \begin{tabular}[c]{@{}c@{}}Balances between\\ transmission and\\ computational\\ latency\end{tabular} & \begin{tabular}[c]{@{}c@{}}Scalable but \\ dependent on \\ device\\ capabilities\end{tabular} & \begin{tabular}[c]{@{}c@{}}High due to \\ frequent \\ communication\end{tabular}      & \begin{tabular}[c]{@{}c@{}}Sync delays\\ due to \\ varied device\end{tabular} & \begin{tabular}[c]{@{}c@{}}Costs saved on\\ centralized\\ processing\end{tabular}                                      \\ \hline
\multirow{8}{*}{\begin{tabular}[c]{@{}c@{}}Inter\\ -\\ layer\end{tabular}} & \multirow{4}{*}{Assistance}                                                  & \begin{tabular}[c]{@{}c@{}}Edge-device\\ Collaborative\\ Inference \cite{mohammed2020distributed}, \\ \cite{roth1992two,10330751,10413648,DBLP:conf/infocom/Hanyao0Q0L21,DBLP:journals/tvt/YunKCK21,DBLP:conf/mm/KongYC23,DBLP:conf/mobicom/FangZ018,DBLP:journals/tcom/DuZFC18,DBLP:journals/tsc/DaiWWYZX23}\end{tabular}            & \begin{tabular}[c]{@{}c@{}}Optimized by\\ task allocation\end{tabular}                                & \begin{tabular}[c]{@{}c@{}}High with edge\\ scalability\end{tabular}                          & \begin{tabular}[c]{@{}c@{}}Moderate due to\\ edge-device\\ communication\end{tabular} & \begin{tabular}[c]{@{}c@{}}High due to\\ diverse device\end{tabular}          & \begin{tabular}[c]{@{}c@{}}High with \\ optimized\\ resource use\end{tabular}                                          \\ \cline{3-8} 
                                                                           &                                                                              & \begin{tabular}[c]{@{}c@{}}Cloud-device\\ Collaborative\\ Inference \\
                                                                        \cite{DBLP:conf/mobicom/EmmonsFAVSW19,DBLP:conf/cvpr/SzegedyLJSRAEVR15,DBLP:conf/cvpr/HeZRS16,DBLP:conf/infocom/HuBWL19,DBLP:journals/tmc/DuanW24,DBLP:conf/cloud/JeongLSM18,DBLP:conf/asplos/KangHGRMMT17,DBLP:conf/icassp/HauswaldMZDCM14,DBLP:journals/tmc/EshratifarAP21,DBLP:conf/icc/DengTF16}  \end{tabular}         & \begin{tabular}[c]{@{}c@{}}Optimized by cloud's\\ computing power\end{tabular}                      & \begin{tabular}[c]{@{}c@{}}High with cloud\\ scalability\end{tabular}                         & \begin{tabular}[c]{@{}c@{}}High due to\\ cloud-device\\ distance\end{tabular}         & Moderate                                                                      & \begin{tabular}[c]{@{}c@{}}Moderate due\\ to cloud costs\end{tabular}                                                  \\ \cline{3-8} 
                                                                           &                                                                              & \begin{tabular}[c]{@{}c@{}}Cloud-edge\\ -device\\ Collaborative\\ Inference \\ \cite{9495122,9534773,lin2019cost,teerapittayanon2017distributed,10472084,yousefpour2019guardians}
\end{tabular} & \begin{tabular}[c]{@{}c@{}}High with multi\\ -layer\\ optimization\end{tabular}                         & Very High                                                                                     & \begin{tabular}[c]{@{}c@{}}High with \\ multi-layer\\ communication\end{tabular}      & Very High                                                                     & \begin{tabular}[c]{@{}c@{}}High with \\ optimized\\ resource \\ distribution\end{tabular}                              \\ \cline{3-8} 
                                                                           &                                                                              & \begin{tabular}[c]{@{}c@{}}One-way\\ Decision\\ Correlation \\ \cite{cao2021edge,kosuru2023intelligent}\end{tabular}                 & \begin{tabular}[c]{@{}c@{}}Streamlined by\\ decision transfer\end{tabular}                              & Moderate                                                                                      & \begin{tabular}[c]{@{}c@{}}Low due to\\ one-way transfer\end{tabular}                 & Moderate                                                                      & \begin{tabular}[c]{@{}c@{}}High with\\ reduced\\ redundancy\end{tabular}                                               \\ \cline{2-8} 
                                                                           & \multirow{3}{*}{Cooperation}                                                 & \begin{tabular}[c]{@{}c@{}}Edge-cloud\\ Co-inference \\ \cite{li2021appealnet,banitalebi2021auto,xiang2021energy,zhang2021energy}\end{tabular}                        & \begin{tabular}[c]{@{}c@{}}High with shared\\ responsibilities\end{tabular}                             & Very High                                                                                     & \begin{tabular}[c]{@{}c@{}}Moderate due to\\ edge-cloud proximity\end{tabular}        & High                                                                          & \begin{tabular}[c]{@{}c@{}}High with\\ resource sharing\end{tabular}                                                   \\ \cline{3-8} 
                                                                           &                                                                              & \begin{tabular}[c]{@{}c@{}}Edge-device\\ Co-inference \\ \cite{bai2020latency,krouka2021energy,xiao2022reinforcement}\end{tabular}                       & \begin{tabular}[c]{@{}c@{}}High with\\ localized\\ processing\end{tabular}                              & \begin{tabular}[c]{@{}c@{}}High with\\ edge scalability\end{tabular}                          & \begin{tabular}[c]{@{}c@{}}Moderate due to\\ edge-device\\ proximity\end{tabular}     & Very High                                                                     & \begin{tabular}[c]{@{}c@{}}High with\\ localized\\ processing\end{tabular}                                             \\ \cline{3-8} 
                                                                           &                                                                              & \begin{tabular}[c]{@{}c@{}}Decision\\ Interaction \\ \cite{carroll2022towards,tang2022effects}\end{tabular}                           & \begin{tabular}[c]{@{}c@{}}High with mutual\\ decision insights\end{tabular}                            & High                                                                                          & \begin{tabular}[c]{@{}c@{}}High due to\\ continuous\\ interaction\end{tabular}        & High                                                                          & \begin{tabular}[c]{@{}c@{}}Moderate with\\ decision overheads\end{tabular}                                             \\ \cline{2-8} 
                                                                           & Exchange                                                                     & -                                                                                        & \begin{tabular}[c]{@{}c@{}}High with device-\\ level optimization\end{tabular}                          & \begin{tabular}[c]{@{}c@{}}Moderate with\\ device limits\end{tabular}                         & \begin{tabular}[c]{@{}c@{}}Moderate due to \\ device-device\\ proximity\end{tabular}  & \begin{tabular}[c]{@{}c@{}}High with\\ device\\ compatibility\end{tabular}    & \begin{tabular}[c]{@{}c@{}}High with\\ device-\\ level optimization\end{tabular}                                       \\ \hline
\end{tabular}
\label{table:Comparison inference of cooperation}
\end{table*}

\item \textbf{Energy Consumption Minimization:}
The principle of bidirectional exchange also applies to energy consumption. Each layer aims to use energy efficiently in making inferences. The cloud leverages its vast resources, offloading tasks to the edge to conserve energy. The edge, due to its proximity to data sources, minimizes energy consumption during data transmission and focuses on local inferences. Devices, inherently resource-constrained, advocate for energy-efficient inferences and selective data transmissions. Hu \emph{et al.} \cite{hu2022overview} spotlight energy conservation in mobile networks, illustrating how edge situated closer to mobile devices can optimize energy consumption. When a device's computational resources are insufficient, offloading tasks to the edge presents an energy-conservative solution.
\end{itemize}

We encapsulate the interactions between cloud, edge, and devices while preserving their distinct objectives in the inference process.
{\color{black}The SL function Eq. \eqref{SL-function-define} for bidirectional exchange is $F([\text{cloud-edge-devices}], \varnothing, [\text{exchange}], $ $\text{inference}, \text{tasks})$. The optimization objective is:
\begin{equation}
\begin{aligned}
{\text{max}} & \quad \theta_C O_C(X_C) + \theta_E O_E(X_E) + \theta_D O_D(X_D) \\
& - \lambda [L_c(X_C) + L_e(X_E) + L_d(X_D)] \\
& - \omega [E_c(X_C) + E_e(X_E) + E_d(X_D)], \\
\text{s.t.} & \quad X_C + X_E + X_D \leq T.
\end{aligned}
\vspace{-0.5em}
\end{equation}
Task and data exchanges between the cloud ($X_C$), edge ($X_E$), and devices ($X_D$) are optimized in our model to meet the objectives of each ($O_C$, $O_E$, $O_D$), while considering latency ($L_c$, $L_e$, $L_d$) and energy consumption ($E_c$, $E_e$, $E_d$). Weights ($\theta_C$, $\theta_E$, $\theta_D$, $\lambda$, $\omega$) prioritize these factors accordingly, ensuring that the total exchange ($T$) aligns with system capabilities and objectives. 
This inter-layer collaboration, mirroring social reciprocity, improves task allocation efficiency, reduces latency and conserves energy by leveraging participant's strengths.}

The discussion on Socialized Inference in EI draws an analogy between social cooperation and collaborative decision-making, such as edge-device and cloud-edge interactions. Detailed in Table \ref{table:Comparison inference of cooperation}, these mechanisms manage the trade-off between latency and energy efficiency, some focusing on real-time response and others on energy savings.

\vspace{-0.5em}
\subsection{{Intra-layer Submissive Interaction in Socialized Inference}}
\subsubsection{\textbf{Conformity}}
%Within society, conformity is a prevalent behavior where individuals, either willingly or due to social pressures, adopt the predominant behaviors and practices of their group. This tendency to align with the majority ensures harmony, reduces conflicts, and fosters a cohesive environment. Similarly, in EI, especially during the inference phase, entities within the same layer exhibit conformity dynamics. Conformity dynamics establish a shared mental model, facilitate information sharing and collaboration, enhance decision-making accuracy, and optimize resource utilization. However, it isn't mere mimicry. Instead, these entities harmonize their inference strategies to ensure a coherent and uniform approach, especially in response to real-time data. This is paramount in scenarios demanding rapid decisions, where divergence can lead to inefficiencies and misaligned actions.
% In societies, conformity, a phenomenon where individuals align their actions with group norms, fosters cohesion and collective efficiency.
% In societies, individuals often align their actions with group norms to promote cohesion.
% %This principle is reflected in EI, where entities synchronize their inference strategies, not merely for mimicry, but to ensure uniformity and efficiency in real-time decision-making.
% In EI, entities synchronize their inference strategies to ensure uniformity and real-time efficiency.

Similar to how individuals in society align their actions with group norms to promote cohesion, devices also adhere to established norms to enhance efficiency.

\subsubsubsection{Ultra-Reliable and Low-Latency Communication} 
Conformity in EI focuses on enabling ultra-reliable and low-latency communication (URLLC) by streamlining processes and eliminating superfluous exchanges to facilitate swift and reliable decision-making.
To encapsulate these ideas into a formula, we can represent the system's efficiency as a function of conformity, latency, and reliability. 
{\color{black}The SL function Eq. \eqref{SL-function-define} for URLLC is $F([\text{devices}], [\text{conformity}], \varnothing, \text{inference}, \text{tasks})$. The optimization objective is:
\begin{equation}
\begin{aligned}
\text{max} & \quad U(C, L, R) = C \times R - \alpha \times L, \\
\text{s.t.} & \quad 0 \leq C \leq 1, \quad L \geq 0, \quad 0 \leq R \leq 1,
\end{aligned}
\vspace{-0.5em}
\end{equation}
where $C$ is the conformity among devices, $L$ is the communication latency, $R$ is the reliability, $U(C, L, R)$ is the utility function. Constant $\alpha$ weights the significance of latency, balancing it against reliability.
}

Han \emph{et al.} \cite{han2022dynamics} present ``coded edge federation" (CEF), where edge infrastructure providers (EIPs) join forces to allocate resources for coded distributed computing (CDC) tasks. Their approach shows how conformity leads to a Nash equilibrium in resource allocation, with no benefit from unilateral strategy changes. Tran \emph{et al.} \cite{tran2017collaborative} highlight how conformity to protocols in 5G networks enhances EI system performance.

% In conclusion, conformity within EI serves as a catalyst for URLLC, simplifying communication complexities and ensuring robust, reliable inter-entity interactions. By embracing shared norms and strategies, EI entities can navigate decision-making landscapes more efficiently, mirroring the social benefits of conformity in group dynamics.

\begin{figure}[pt]
  \centering
  \includegraphics[width=3.49in]{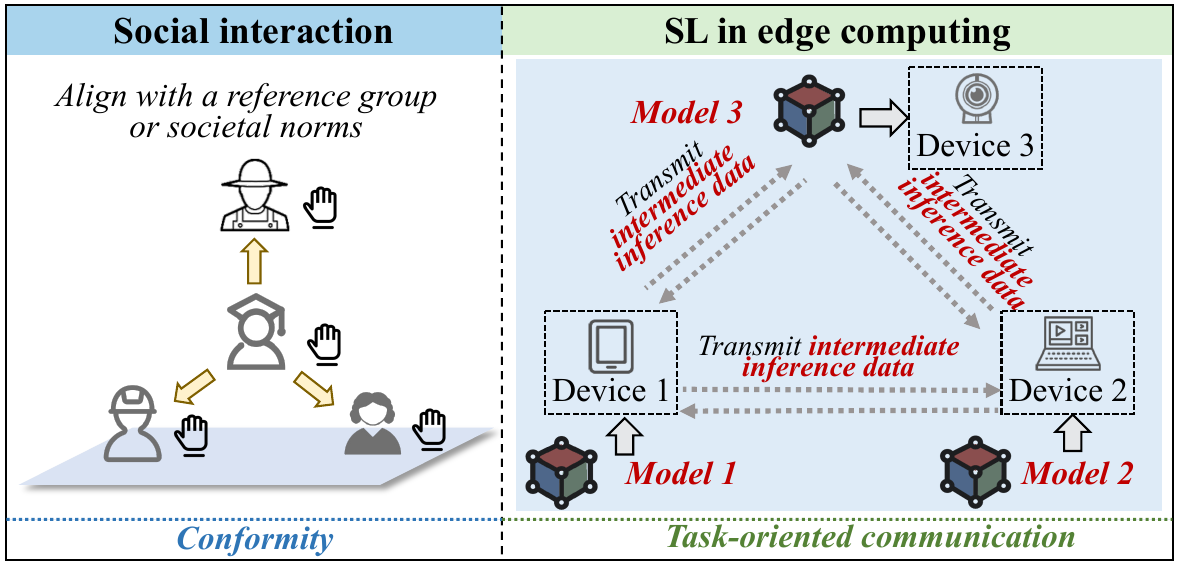}   
  \caption{Illustration of \textbf{task-oriented communication} in socialized inference.}\label{Fig-Task-oriented Communication}
  % \vspace{-0.5em}
\end{figure}

\subsubsubsection{Task-oriented Communication} 
% Conformity in human society extends beyond behavioral mimicry, significantly impacting cooperative task execution. 
In human societies, conformity impacts cooperative task execution.
Likewise, in EI, conformity dynamics serve as the backbone for task-oriented communication. Devices streamline communications to be task-centric. Conformity ensures interactions align with collective task completion goals. The shared understanding that arises from conformity forms the foundation for devices to synchronize their activities and share task-relevant information, as shown in Fig. \ref{Fig-Task-oriented Communication}. It facilitates collaborative decision-making and reduces potential for communication conflicts.
{\color{black}The SL function Eq. \eqref{SL-function-define} for task-oriented communication is $F([\text{devices}], [\text{conformity}], \varnothing, \text{inference}, \text{communications})$. The optimization objective is:
\vspace{-0.5em}
\begin{equation}
\begin{aligned}
{\text{max}}  \quad C \times f - \alpha \times L, \quad
\text{s.t.} \quad 0 \leq C, f \leq 1,  L \geq 0,
\end{aligned}
\vspace{-0.5em}
\end{equation}
where $C$ is the conformity among entities, $f$ is the relevance of features to the task, $L$ is the communication latency, and $\alpha$ is the weight for latency.
}

Shao \emph{et al.} \cite{shao2022task} delve into this paradigm by investigating task-oriented communication in multi-device cooperative edge inference. Distributed edge devices transmit local sample features to a more powerful edge for inference. While it enhances the collective sensing capability beyond that of a single device, it does introduce challenges, notably increased communication overheads and potential latency. However, Shao \emph{et al.} navigate this challenge by proposing a learning-based communication model that optimally extracts local features and encodes distributed features in a strictly task-oriented manner.

% In summary, by adhering to a common set of norms, EI entities facilitate more focused, task-oriented communication, increasing system efficiency in achieving collective objectives. Its emphasis on conformity not only streamlines interactions but elevates the effectiveness of task execution across the network.

\vspace{-0.5em}
\subsection{{Inter-layer Submissive Interaction in Socialized Inference}}
\subsubsection{\textbf{Unidirectional Submissive Interaction - Suggestion\&imitation}}
\begin{figure}[pt]
  \centering
  \includegraphics[width=3.49in]{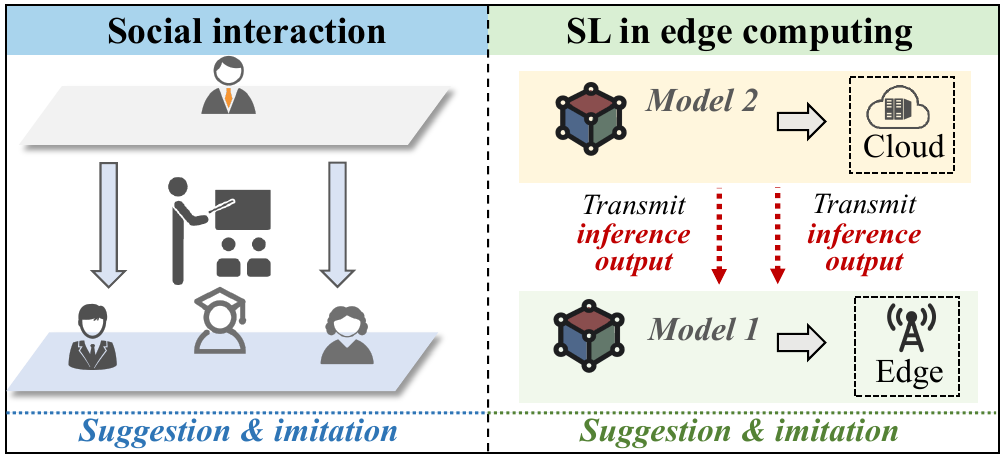}    
  \caption{Illustration of \textbf{suggestion \& imitation} in socialized inference.}\label{Suggestion & Imitation}
  % \vspace{-0.5em}
\end{figure}

% In traditional social frameworks, authoritative entities guide behaviors through suggestions, prompting others to act in an imitation mode. 
In traditional societies, authoritative figures guide behaviors. The suggestion\&imitation pattern, depicted in Fig. \ref{Suggestion & Imitation}, shows how leaders guide actions. In EI inference, layers with contextual or computational advantages provide `suggestions' as processed insights. Other layers then `imitate' by aligning their strategies with these insights. Unlike model parameter and feature transfer in Socialized Training, Socialized Inference focuses on real-time decision alignment, harmonizing actions with guiding layers. This ensures EI cohesion, reflecting how authoritative figures influence group decisions. We show how different layers' dynamics affect communication latency, energy usage, and computation.

\begin{itemize}[leftmargin=*]
\item  \textbf{Resource Allocation:} Resource allocation in EI, emulating social dynamics, involves the harnessing of resources and the imitation of efficient strategies. Authoritative layers guide resource allocation by suggesting optimal routes or efficient edge strategies, steering lower layers towards optimization and reducing communication delays. Cen \emph{et al.} \cite{DBLP:journals/candie/ChenZWW23} explore an RL approach to these challenges in cloud-edge environments. Similarly, Hu \emph{et al.} \cite{hu2023intelligent} investigate intelligent resource allocation in edge-cloud networks using a hybrid DDPG-D3QN model.

\item \textbf{Energy Management:} Authoritative layers' suggestions drive energy-efficient strategies, urging edges to mimic energy-saving techniques or adopt operational modes. Such imitation dynamics prolong device lifespans, reducing overall energy wastage. Ruan \emph{et al.} \cite{ruan2019priority} propose a three-tier architecture, emphasizing latency improvements and energy-efficient scheduling.

\item \textbf{Computational Offloading:} Task balancing between cloud and edge is central to EI. Authoritative entities steer lower layers' offloading decisions for balanced computational loads, reducing communication latency and enhancing system responsiveness. Wang \emph{et al.} \cite{wang2017computation} focus on computational offloading in wireless networks, balancing ES capacity and energy consumption. Zeng \emph{et al.} \cite{zeng2016joint} present a joint optimization for computational offloading, assessing its effects on energy usage and performance.
\end{itemize}

{\color{black}The SL function Eq. \eqref{SL-function-define} for suggestion\&imitation is $F([\text{server-devices}], \varnothing, [\text{suggestion\&imitation}],\text{inference}, \text{decis-}$ $\text{ions})$. The optimization objective is:
\vspace{-0.5em}
\begin{equation}
\begin{aligned}
{\text{max}} & \quad \theta_1 R(S, I) + \theta_2 E(S, I) + \theta_3 O(S, I), \\
\text{s.t.} & \quad 0 \leq S, I \leq 1,
\end{aligned}
\vspace{-0.5em}
\end{equation}
where $S$ and $I$ are the levels of suggestion and imitation, respectively, $R(S, I)$, $E(S, I)$, and $O(S, I)$ are the performance metrics for resource allocation, energy management, and computational offloading, respectively, and $\theta_1$, $\theta_2$, and $\theta_3$ are the weights for the respective metrics.
}

\begin{table*}[t]
\color{black}
\centering
\caption{\color{black}Comparative Analysis of Inference Approaches Reflecting Submissive Interactions}
\begin{tabular}{|c|c|c|c|c|c|c|c|}
\hline
\textbf{Type}                                                              & \textbf{\begin{tabular}[c]{@{}c@{}}Social \\ Relationship\end{tabular}} & \textbf{Technique}                                                    & \textbf{Efficiency}                                                                     & \textbf{Scalability}                                                                  & \textbf{\begin{tabular}[c]{@{}c@{}}Communication\\ Overhead\end{tabular}}              & \textbf{\begin{tabular}[c]{@{}c@{}}Heterogeneity\\ Handling\end{tabular}}       & \textbf{\begin{tabular}[c]{@{}c@{}}Cost\\ Efficiency\end{tabular}}                     \\ \hline
\multirow{2}{*}{\begin{tabular}[c]{@{}c@{}}Intra\\ -\\ layer\end{tabular}} & \multirow{2}{*}{Conformity}                                             & \multirow{2}{*}{\begin{tabular}[c]{@{}c@{}}URLLC \\ \cite{han2022dynamics,tran2017collaborative} \end{tabular}}                                                                & \begin{tabular}[c]{@{}c@{}}Optimized due to\\ streamlined \\ communication\end{tabular} & \begin{tabular}[c]{@{}c@{}}High due to\\ uniformity\end{tabular}                      & \begin{tabular}[c]{@{}c@{}}Reduced due to \\ streamlined \\ channels\end{tabular}      & \begin{tabular}[c]{@{}c@{}}Moderate due to\\ shared standards\end{tabular}      & \begin{tabular}[c]{@{}c@{}}Optimized due to\\ reduced redundancies\end{tabular}        \\ \cline{3-8} 
                                                                           &                                                                         & \begin{tabular}[c]{@{}c@{}}Task-oriented\\ Communication \\
                                                                           \cite{shao2022task}\end{tabular} & \begin{tabular}[c]{@{}c@{}}Enhanced due to \\ task-centric focus\end{tabular}           & \begin{tabular}[c]{@{}c@{}}High due to \\ shared task\\ objectives\end{tabular}       & \begin{tabular}[c]{@{}c@{}}Increased due to \\ task-specific \\ exchanges\end{tabular} & \begin{tabular}[c]{@{}c@{}}High due to \\ task-specific\\ handling\end{tabular} & \begin{tabular}[c]{@{}c@{}}Optimized due to\\ task-centric\\ focus\end{tabular}        \\ \hline
\begin{tabular}[c]{@{}c@{}}Inter\\ -\\ layer\end{tabular}                  & \begin{tabular}[c]{@{}c@{}}Suggestion \\ \&\\ Imitation\end{tabular}    & \begin{tabular}[c]{@{}c@{}}-  \\ \cite{DBLP:journals/candie/ChenZWW23,hu2023intelligent,ruan2019priority,wang2017computation,zeng2016joint}   \end{tabular}                                                                & \begin{tabular}[c]{@{}c@{}}Optimized by\\ task allocation\end{tabular}                & \begin{tabular}[c]{@{}c@{}}Enhanced due to\\ authoritative\\ suggestions\end{tabular} & \begin{tabular}[c]{@{}c@{}}Reduced due to\\ streamlined\\ suggestions\end{tabular}     & \begin{tabular}[c]{@{}c@{}}High due to\\ centralized\\ guidance\end{tabular}    & \begin{tabular}[c]{@{}c@{}}Optimized due to\\ authoritative\\ suggestions\end{tabular} \\ \hline
\end{tabular}
\label{table:Comparison inference of submissive}
\end{table*}

% In EI, intra-layer and inter-layer dynamics profoundly mirror social behaviors, notably in the realms of conformity, suggestion, and imitation. 
% EI's intra-layer and inter-layer dynamics reflect social behaviors, especially conformity, suggestion, and imitation.
% Intra-layer conformity fortifies URLLC and Task-oriented Communication, fostering system reliability and efficiency akin to social adherence to group norms. In contrast, inter-layer interactions characterized by unidirectional submissiveness reflect social hierarchies, facilitating streamlined decision-making within EI. 

{\color{black}
\textbf{Communication Efficiency in Suggestion\&Imitation Mechanisms:}
Communication efficiency plays a crucial role in ensuring effective information transmission from higher to lower layers in Suggestion\&Imitation mechanisms. To optimize end-to-end communication performance, various strategies can be employed, such as information compression, quantization, and selective transmission. These techniques aim to reduce data transferred between layers while maintaining inference accuracy. Moreover, designing efficient communication protocols and scheduling algorithms is essential for minimizing latency in the suggestion and imitation processes \cite{zeng2023copriv}. By optimizing communication efficiency in Suggestion\&Imitation mechanisms, socialized inference systems can achieve faster convergence and lower resource consumption.}

% The essence of unidirectional submissive interactions, particularly suggestion and imitation dynamics, isn't merely confined to human societies but is intrinsically woven into the fabric of EI during inference. 
Unidirectional submissive interactions, like suggestion and imitation dynamics, are evident in both human societies and EI inference, emphasizing the potential benefits of such collaborative frameworks. Embracing these dynamics enables a harmonized, efficient, and responsive system, optimizing resource allocation, energy management, and computational tasks. This hierarchical structure optimizes processes from resource allocation to energy management. A comparative analysis presented in Table \ref{table:Comparison inference of submissive} elucidates interactions' distinctions and synergies.

\begin{figure}[pt]
  \centering
  \includegraphics[width=3.49in]{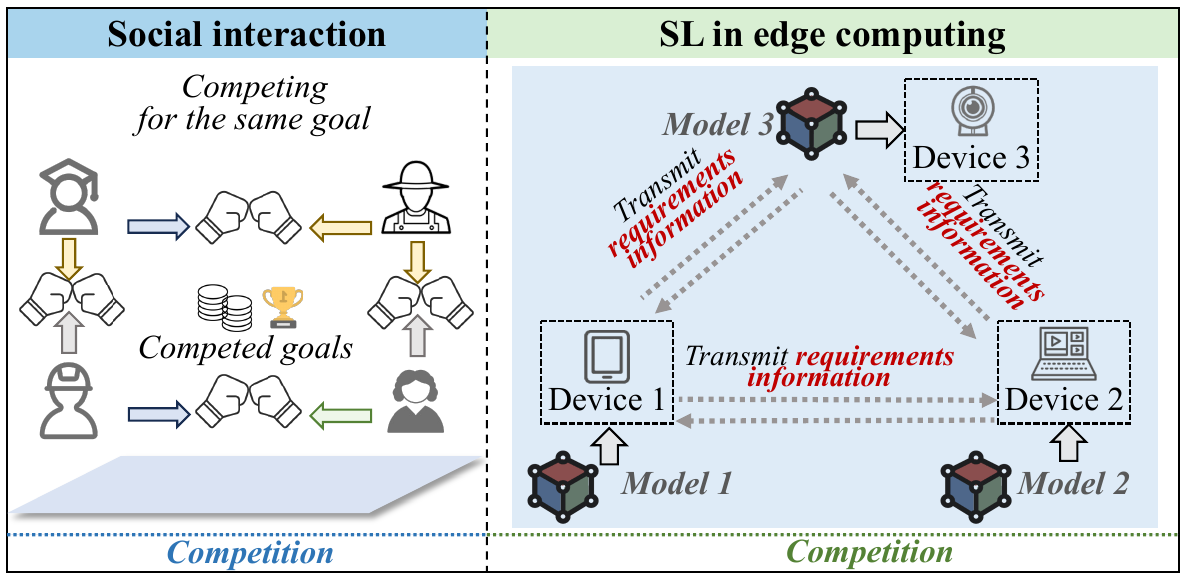}    
  \caption{Illustration of \textbf{competition} in socialized inference.}\label{Competition}
  \vspace{-0.5em} 
\end{figure}

\vspace{-0.5em}
\subsection{{Intra-layer Conflict Interaction in Socialized Inference}}
\subsubsection{\textbf{Competition}}
% In human societies, competition often fosters innovation but can cause discord during resource scarcity. 
In societies, competition can drive innovation but may also cause conflicts.
This duality is mirrored in EI's inference, where devices within the same layer vie for superiority as illustrated in Fig. \ref{Competition}. 
% The goal surpasses mere performance, aiming for real-time inference under tight constraints. Such competition sharpens inference techniques, but it also brings challenges: intensifying communication needs, elevating energy use, and increasing computational load. Especially in real-time scenarios, this could result in redundant calculations. 
Competitors in inference rely more on quick decisions and resource optimization than dataset dominance. The key lies in efficient responses without compromising EI's core objectives. Managing intra-layer competition effectively becomes essential to strike a balance between performance improvements and optimizing system resources and capabilities.
     
\begin{itemize}[leftmargin=*]
\item \textbf{Resource Allocation:} Intra-layer competition centers on resource allocation. Devices competing for limited computational resources may increase communication demands, adversely affecting communication latency. Additionally, increased competition often leads to higher energy demands, compromising the energy efficiency of the system. Fan \emph{et al.} \cite{DBLP:journals/tvt/FanHZSLTWL23} investigate networks where agents compete for resources, proposing a mechanism for efficient and fair allocation. Sun \emph{et al.} \cite{DBLP:journals/tmc/SunSLWC24} optimize for maximum network satisfaction through fair resource sharing.

\item \textbf{Task Offloading:} In edge-cloud collaborations, task offloading becomes another aspect of intra-layer competition. Devices jostle for offloading their tasks to available, capable nodes, potentially affecting latency and energy consumption. Offloading strategies can either promote energy efficiency or, if unchecked, exacerbate energy demands. Ji \emph{et al.} \cite{ji2022energy} contribute to this issue by focusing on energy-efficient offloading algorithms, considering latency tolerances and uncertainties inherent in such networks.

\item \textbf{Computing Efficiency:} The competition is also driving devices to bolster their computing efficiency. Devices vie to optimize algorithms, employ cutting-EC methods, or adopt strategies that promise efficiency. Such pursuits, while fostering innovation, could inadvertently increase computational overheads. Lim \emph{et al.} \cite{lim2019improving} provide insight into scheduling algorithms in allocation scenarios and offer effective strategies for task execution in competitive environments. Zhang \emph{et al.} \cite{zhang2022computational} expand on this by discussing computing schedules in EI that are energy-efficient.
\end{itemize}

{\color{black}The SL function Eq. \eqref{SL-function-define} for competition is $F([\text{devices}],$ $[\text{competition}], \varnothing, \text{inference}, \text{resources})$. The optimization objective is:
\vspace{-0.5em}
\begin{equation}
\begin{aligned}
\text{max} & \quad E = \lambda_1 R(C) + \lambda_2 T(C) + \lambda_3 Comp(C), \\
\text{s.t.} & \quad 0 \leq C \leq 1, \quad 0 \leq R(C), T(C), Comp(C) \leq 1, \\
& \quad \lambda_1 + \lambda_2 + \lambda_3 = 1,
\end{aligned}
\vspace{-0.5em}
\end{equation}
where $C$ is the level of competition, $R(C)$, $T(C)$, and $Comp(C)$ are the efficiency metrics for resource allocation, task offloading, and computation, respectively, and $\lambda_1$, $\lambda_2$, and $\lambda_3$ are the weights for the respective metrics.
}

The concept of competition is not only central to society but also critical within the domain of inference in EI. Devices within the same layer strive to excel in resource allocation, task offloading, and computing efficiency while maintaining a balance with the collective energy and computational needs of the system. The above works emphasize how important it is to manage this competition and offer methods for coordinating personal objectives with the overall objectives of EI in order to ensure reliable and effective conclusions.

{\color{black}
\textbf{Communication-aware Resource Allocation in Competitive Inference:}
In competitive inference scenarios, where multiple devices or agents compete for shared resources, communication efficiency plays a critical role in determining the overall system performance. The communication network topology, bandwidth allocation, and transmission strategies need to be carefully designed and optimized to ensure fair and efficient resource allocation among competing devices. Game-theoretic approaches and distributed optimization techniques \cite{lin2020distributed} can be applied to model and solve communication-aware resource allocation problems in competitive inference settings.}

\begin{figure}[pt]
  \centering
  \includegraphics[width=3.49in]{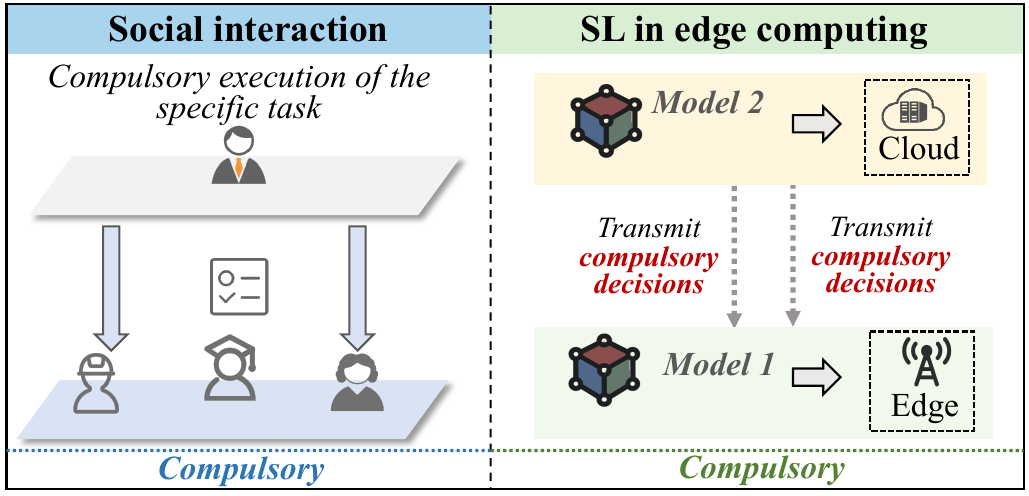}    
  \caption{Illustration of \textbf{compulsory} in socialized inference.}\label{Compulsory}
  % \vspace{-0.5em}
\end{figure}

\vspace{-1em}
\subsection{{Inter-layer Conflict Interaction in Socialized Inference}}
\subsubsection{\textbf{Unidirectional Competition - Compulsory}}
In social hierarchies, upper echelons dictate the actions of subordinate layers, establishing a unidirectional command chain, as depicted in Fig. \ref{Compulsory}. 
% This hierarchical model is paralleled in EI, where higher-level structures guide lower-level operations, promoting system uniformity. 
% In EI, higher-level structures guide lower-level operations, ensuring system uniformity.
Although this hierarchy centralizes decision-making and optimizes inference, it also generates issues such as communication delays, energy demands, and computing inefficiency. Mitigating these challenges necessitates collaborative strategies in resource allocation, task offloading, and scheduling, thereby enhancing resource efficiency, communication efficiency, and overall performance.
    
\begin{itemize}[leftmargin=*]
\item \textbf{Resource Allocation:}
The advantage in resource allocation is evident when higher strata enforce allocation decisions upon inferior layers. While it offers centralized control and enhanced resource optimization across the edge-cloud spectrum, it also brings about concerns related to communication latency, instigated by coordination overheads. 
Collaborative resource allocation algorithms, as studied by Jiang \emph{et al.} \cite{jiang2018resource}, can act as mitigators to combat this. Furthermore, energy conservation gives rise to the development of energy-centered allocation strategies that target efficient resource allocation and energy consumption.

\item \textbf{Task Offloading:}
In the edge-cloud collaboration, unidirectional compulsory is manifest when superior layers necessitate task offloading to their lower counterparts. This mechanism enhances centralized management and scalability but also raises latency issues. Almutairi \emph{et al.} \cite{almutairi2021novel} delve into task offloading, highlighting a fuzzy logic system to optimize the end-to-end service time. Proximity-based communication and energy-efficient offloading ensure tasks find their way to energy-optimized nodes.

\item \textbf{Computing Efficiency:} Unidirectional compulsory within layers during inference accentuates the edge-cloud collaborative system's computing efficiency. Although this compulsory can be demanding, it also catalyzes advancements in performance. Given the heterogeneity of computational resources, collaborative algorithms and ML approaches are needed. Maray \emph{et al.} \cite{DBLP:journals/iot/MarayMSB23} introduce a dynamic time-sensitive scheduling algorithm, while Fan \emph{et al.} \cite{fan2022collaborative} aim at optimizing the total task processing delay. Furthermore, Zhang \emph{et al.} \cite{zhang2023dependent} bring forth a mechanism optimizing the task completion time, emphasizing the limited storage space of the ES and its implications.
\end{itemize}

{\color{black}The SL function Eq. \eqref{SL-function-define} for compulsory interaction is $F([\text{server-devices}],\varnothing, [\text{conflict}], \text{inference}, \text{tasks})$. The optimization objective is:
% \vspace{-0.5em}
\begin{equation}
\begin{aligned}
{\text{max}} & \quad \delta_1 \cdot R(H) + \delta_2 \cdot T(H) + \delta_3 \cdot Comp(H), \\
\text{s.t.} & \quad 0 \leq H \leq 1,
\end{aligned}
\vspace{-0.5em}
\end{equation}
where $H$ is the level of directive from higher-level servers, $R(H)$, $T(H)$, and $Comp(H)$ are the efficiency metrics for resource allocation, task offloading, and computation under the higher-level directive, respectively, and $\delta_1$, $\delta_2$, and $\delta_3$ are the weights for the respective metrics.
}

\begin{table*}[t]
\color{black}
\centering
\caption{\color{black}Comparative Analysis of Inference Approaches Reflecting Conflict Interactions}
\begin{tabular}{|c|c|c|c|c|c|c|}
\hline
\textbf{Type}                                                              & \textbf{\begin{tabular}[c]{@{}c@{}}Social \\ Relationship\end{tabular}} & \textbf{Efficiency}                                                                                     & \textbf{Scalability}                                                                                  & \textbf{\begin{tabular}[c]{@{}c@{}}Communication\\ Overhead\end{tabular}}                                   & \textbf{\begin{tabular}[c]{@{}c@{}}Heterogeneity\\ Handling\end{tabular}}                         & \textbf{\begin{tabular}[c]{@{}c@{}}Cost\\ Efficiency\end{tabular}}                                             \\ \hline
\begin{tabular}[c]{@{}c@{}}Intra\\ -\\ layer\end{tabular}                  & \begin{tabular}[c]{@{}c@{}}Competition \\ \cite{DBLP:journals/tvt/FanHZSLTWL23,DBLP:journals/tmc/SunSLWC24,ji2022energy,lim2019improving,zhang2022computational} \end{tabular}                                                           & \begin{tabular}[c]{@{}c@{}}Optimized due to\\ intra-layer rivalry\\ fostering innovation\end{tabular}   & \begin{tabular}[c]{@{}c@{}}Moderate due to\\ potential resource\\ contention\end{tabular}             & \begin{tabular}[c]{@{}c@{}}Elevated due to\\ frequent resource-\\ negotiation\\ exchanges\end{tabular}      & \begin{tabular}[c]{@{}c@{}}Diverse due to\\ individual entities'\\ unique strategies\end{tabular} & \begin{tabular}[c]{@{}c@{}}Balanced due to\\ competitive\\ resource utilization\end{tabular}                   \\ \hline
\multirow{2}{*}{\begin{tabular}[c]{@{}c@{}}Inter\\ -\\ layer\end{tabular}} & \begin{tabular}[c]{@{}c@{}}Compulsory  \\ \cite{jiang2018resource,almutairi2021novel,DBLP:journals/iot/MarayMSB23,fan2022collaborative,zhang2023dependent} \end{tabular}                                                           & \begin{tabular}[c]{@{}c@{}}Centralized due to\\ top-down directives\\ optimizing processes\end{tabular} & \begin{tabular}[c]{@{}c@{}}High due to\\ structured and\\ directed resource\\ allocation\end{tabular} & \begin{tabular}[c]{@{}c@{}}Streamlined due to\\ clear directives\\ minimizing\\ back-and-forth\end{tabular} & \begin{tabular}[c]{@{}c@{}}Uniform due to\\ centralized standards\\ and guidelines\end{tabular}   & \begin{tabular}[c]{@{}c@{}}Economical due to\\ optimized and\\ directed resource\\ allocation\end{tabular}     \\ \cline{2-7} 
                                                                           & \begin{tabular}[c]{@{}c@{}}
Struggle \\ \cite{lee2020intelligent,kn2020energy,nguyen2023dependency,wang2022task,xavier2020collaborative,amer2022optimized} \end{tabular}                                                                & \begin{tabular}[c]{@{}c@{}}Adaptive due to\\ mutual adjustments\\ and negotiations\end{tabular}         & \begin{tabular}[c]{@{}c@{}}Dynamic due to\\ continuous\\ inter-layer\\ adjustments\end{tabular}       & \begin{tabular}[c]{@{}c@{}}Intensive due to\\ constant\\ bidirectional\\ negotiations\end{tabular}          & \begin{tabular}[c]{@{}c@{}}Varied due to\\ mutual influence\\ and diverse strategies\end{tabular} & \begin{tabular}[c]{@{}c@{}}Negotiable due to shared\\ responsibilities\\ and mutual\\ adjustments\end{tabular} \\ \hline
\end{tabular}
\label{table:Comparison inference of conflict}
\end{table*}

\begin{figure}[pt]
  \centering
  \includegraphics[width=3.49in]{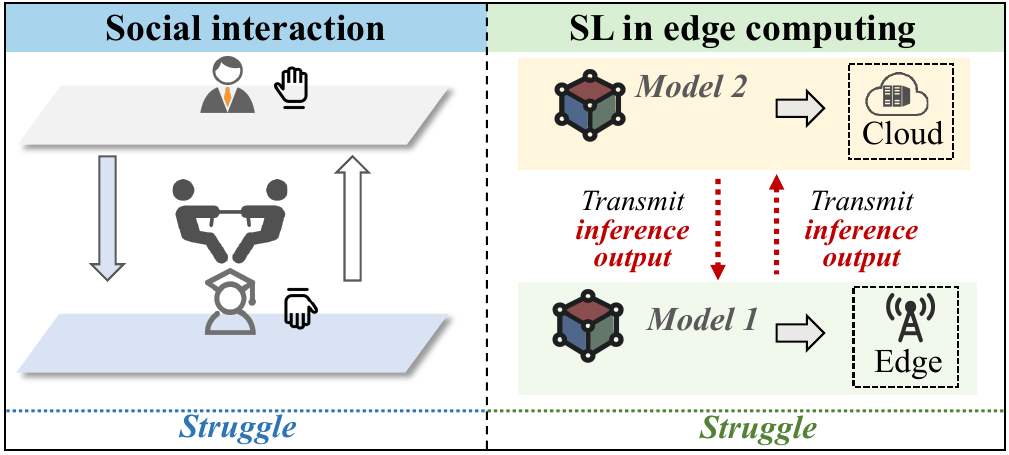}      
  \caption{Illustration of \textbf{struggle} in socialized inference.}\label{Struggle}
  \vspace{-0.5em}
\end{figure}
     
\subsubsection{\textbf{Bidirectional Competition - Struggle}}
In society, struggles involve complex interactions as people pursue objectives while reacting to others' actions (as shown in Fig. \ref{Struggle}). In inference, layers face bidirectional struggles over shared resources or conflicting goals.
% Just as societies undergo dynamics of competition and adaptation, EI inference mechanisms face contention as each layer seeks optimal performance while being influenced by the actions and reactions of others. 
Similar to how social struggles inspire innovation, these interactions might lead to optimal solutions.  However, unmanaged conflicts run the risk of inefficiencies.

\begin{itemize}[leftmargin=*]
\item \textbf{Resource Allocation:} Within EI, resources are finite, and the allocation dynamics mirror the social struggle for scarce resources. Each layer competes for these resources, driving a push towards greater operational efficiency. However, this struggle can also result in increased latency and energy consumption. Collaborative communication protocols, such as those proposed by Lee \emph{et al.} \cite{lee2020intelligent}, aim to enhance transparent communication and improve efficiency. Concurrently, energy consumption concerns are tackled by methods like those suggested by KN \emph{et al.} \cite{kn2020energy}, which seek to balance efficiency with assured QoS.

\item \textbf{Task Offloading:} The social dynamics of responsibility-sharing and allocation is reflected in the act of task offloading. Layers struggle with decisions regarding task offloading, often influenced by their intrinsic capabilities and external pressures. The challenge lies in balancing between local processing and offloading, each with its pros and cons. Nguyen \emph{et al.} \cite{nguyen2023dependency} offer an illustrative approach by promoting UAV collaboration, to achieve optimal offloading. Similarly, strategies presented by Wang \emph{et al.} \cite{wang2022task} promote a multi-tiered approach, echoing the social hierarchies and their interactions.

\item \textbf{Computing Efficiency:} Efficiency in computations, much like productivity in society, can be threatened by internal frictions or conflicts. Struggle interactions can give rise to redundant computations or misaligned task allocation. To address these issues, systems can emulate social cooperation frameworks. Algorithms, as proposed by Xavier \emph{et al.} \cite{xavier2020collaborative}, mimic social cooperation to ensure optimal resource utilization. Amer \emph{et al.} \cite{amer2022optimized} introduce a hierarchical approach to task scheduling, mirroring social hierarchies, with each level having its own set of responsibilities and interactions.
\end{itemize}

We can represent the bidirectional struggle with an optimization framework that captures dynamics between layers.
{\color{black}The SL function Eq. \eqref{SL-function-define} for struggle interaction is $F([\text{multi-layer}], \varnothing, [\text{struggle}], \text{inference}, \text{tasks})$. The optimization objective is:
% \vspace{-0.5em}
\begin{equation}
\begin{aligned}
{\text{max}} & \quad \eta_1 R(B) + \eta_2 T(B) + \eta_3 Comp(B), \\
\text{s.t.} & \quad 0 \leq B \leq 1,
\end{aligned}
\vspace{-0.5em}
\end{equation}
where $B$ is the level of bidirectional struggle, $R(B)$, $T(B)$, and $Comp(B)$ are the efficiency metrics for resource allocation, task offloading, and computation under bidirectional struggle, respectively. $\eta_1$, $\eta_2$, and $\eta_3$ are the weights for the respective metrics.
}

{\color{black}
\textbf{Communication Resource Competition in Struggle Interactions:}
In Struggle interactions, devices or agents from different layers compete for shared communication resources, which can lead to degraded inference performance. To mitigate this issue, effective communication resource allocation strategies need to be designed. One possible approach is to utilize game theory and optimization techniques to model and solve the communication resource allocation problem \cite{shi2020communication}. 

%分段
The objective is to fairly and efficiently allocate communication bandwidth and time slots among struggle devices, considering their inference task requirements and priorities. Additionally, hierarchical communication resource management frameworks can be employed to coordinate resource allocation across different layers, reducing conflicts and improving overall inference performance. By optimizing communication resource allocation in struggle interactions, socialized inference systems can achieve more stable and efficient inference.}

The bidirectional struggle in EI reflects the complex interdependencies familiar in society, emphasizing the continuous pursuit of resources, task offloading efficiency, and computing power. 
% This relentless pursuit, reminiscent of social evolution, catalyzes innovation despite inherent challenges. Recognizing these dynamics through a sociological lens is crucial for enhancing EI, yielding a system that is not only efficient but also resilient and adaptable.
Efficiency and complexity are stimulated by the struggle that exists within EI. EI's struggle, which is marked by rivalries at many levels, encourages creativity but makes energy management and communication more challenging. Table \ref{table:Comparison inference of conflict} describes conflict interactions in-depth and emphasizes how they affect the scalability and efficiency of the system.

\begin{figure*}[pt]
  \centering
  \includegraphics[width=\linewidth]{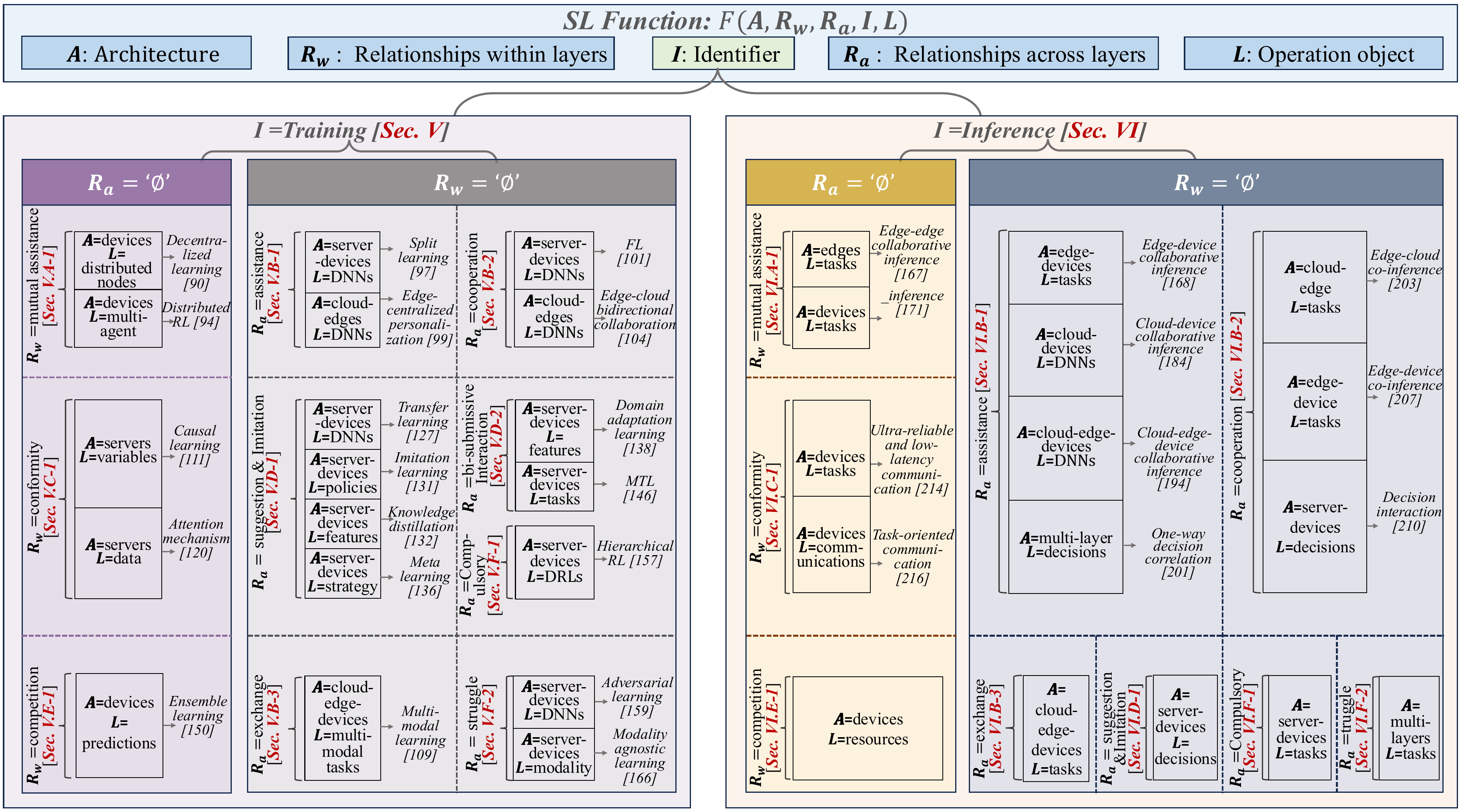}
  \caption{\color{black}
  The taxonomy graph of the unified SL function.
   It maps the diverse methods of socialized training and inference in EI to the variables $(\mathbf{A}, \mathbf{R}_w, \mathbf{R}_a, \mathbf{I}, \mathbf{L})$ defined in Eq. (1). This enables researchers to design customized and optimized SL schemes by selecting appropriate values for these variables, tailored to specific EI application requirements.}\label{SL-function}
  % \vspace{-0.5em}
\end{figure*}

% \vspace{-1em}
\subsection{Lessons Learned}
% In this section, we explored the intersection of social dynamics and EI inference. By comparing them with social interactions, we highlighted how social principles can improve decision-making and the overall adaptability of EI.
% \begin{itemize}[leftmargin=*]
% \item \emph{Emergence of interactivity}: EI is moving from isolated operations to increased collaboration, mirroring cooperative decision-making observed in social dynamics.
% \item \emph{Decentralization in decision-making}: EI promotes distributed inference, reflecting the trend towards decentralized decision-making in society.
% \item \emph{Latency-energy efficiency trade-off}: Balancing responsiveness and energy consumption is like the social trade-off between immediate benefits and resource conservation.
% \item \emph{Duality of collaboration and competition}: In EI, entities' collaborative and competitive behaviors can affect resource allocation and decision speed.
% \item \emph{Necessity for protocols and standardization}: Consistent protocols and standards are essential for smooth EI interactions, similar to social norms and codes.
% \item \emph{Security and privacy concerns}: As interactions and data exchanges grow, protecting sensitive information and system integrity becomes crucial.
% \item \emph{Advancements in algorithms and technologies}: Addressing these challenges requires developing algorithms and technologies that improve efficiency during inference.
% \end{itemize}

% Lessons learned section demonstrates how social rules concepts improve EI decision-making and flexibility. 
Lessons learned section highlights how social principles can improve decision-making and the overall adaptability of EI.
EI is moving from isolated operations to \textbf{increased collaboration}, mirroring cooperative decision-making observed in social dynamics. Similar to social governance trends, EI is moving toward \textbf{decentralized decision-making}. Similar to social resource management, the \textbf{latency-energy trade-off} balances energy savings with timely EI responses. Additionally, EI's \textbf{collaboration-competition dichotomy} influences how resources are allocated and how quickly decisions are made, reflecting social dynamics. 

To ensure seamless interactions within EI, \textbf{protocols and standardization} are necessary, as they align with social norms and regulations. Similar to social issues, protecting sensitive information requires a high priority for \textbf{security and privacy} as EI becomes more interconnected. The efficiency of EI depends on \textbf{advancing algorithms and technology} to address these issues. This technological evolution in EI parallels social advancements, suggesting a future where EI's complexity is adeptly aligned with the multifaceted nature of society.

{\color{black} 
In summary, the taxonomy graph of the SL function, presented in Fig. \ref{SL-function}, encapsulates the various aspects of socialized training and socialized inference discussed throughout Sections \ref{section6} and \ref{section7}, each mapped to the SL function defined in Eq. \eqref{SL-function-define}. The graph illustrates the diverse range of SL customization and optimization possibilities in EI scenarios. By carefully selecting the appropriate values for these variables, researchers and practitioners can design and implement SL schemes tailored to their specific EI application requirements. The taxonomy graph serves as a comprehensive reference for understanding the scope and potential of SL in EI systems, guiding future research and development efforts.}

\section{Tutorials for Implementing SL in EI}\label{section8}
% % We have examined the SL function and demonstrated its adaptability to diverse training and inference requirements in EI. 
% As shown in Fig. \ref{SL-function}, by modifying variables within SL function, we can design specific schemes that are tailored and optimized for distinct EI scenarios. %It leads to an integration that is not only more intelligent but also more efficient and communication-coherent.
% Furthermore, implementing SL in EI encompasses a range of technologies and standards. The integration of SL enhances user experiences across various platforms while also opening up new opportunities to empower EI services and applications. 
{\color{black}
To effectively realize SL functions and ensure their seamless integration into EI systems, the development and adoption of standardizations and platforms are crucial. Standardizations provide a common framework for implementing SL in EI, addressing critical aspects such as data formats, communication protocols, and security mechanisms. Platforms, on the other hand, offer the necessary software tools, libraries, and hardware components to support the deployment and execution of SL algorithms in EI environments. In this section, we discuss the key standardizations and platforms that enable the efficient implementation of SL in EI systems.
}

\subsection{{Standardizations for Implementing SL in EI}}
\subsubsection{Learning}
% The IEEE's FL standard IEEE 3652.1-2020 \cite{9382202} represents significant progress in standardizing SL within EI, developed by key stakeholders including 4Paradigm, Alipay, Huawei, and Tencent. This standard outlines FL's architecture, focusing on data, user, system aspects, scenarios, and performance metrics, for scaled and secure SL-driven EI applications. Moreover, standards like ISO/IEC TS 27570 \cite{ISO/IEC2021} and ISO/IEC DIS 27400 \cite{Guidelines2021} offer extensive guidelines for privacy protection and security in smart cities and IoT. ETSI EN 303 645 \cite{Cyber2020} provides specific cybersecurity baselines for consumer IoT, underlining the importance of adhering to standards for reliable and trustworthy SL-based EI solutions.
The IEEE's FL standard IEEE 3652.1-2020 \cite{9382202}, developed by industry leaders, sets a foundational architecture for SL in EI. 
{\color{black}The importance of this standard lies in its ability to provide a common framework for secure and scalable FL, which is a key enabler for SL in EI. By defining guidelines for data governance, model management and system architecture, IEEE 3652.1-2020 facilitates deployment of FL in various EI applications, such as smart healthcare and industrial IoT.}
It addresses critical aspects like data, user, and system requirements, enabling secure and scalable SL-driven applications. {\color{black}Further, general standards for privacy and security in IoT and smart cities, like ISO/IEC TS 27570 \cite{ISO/IEC2021} and ISO/IEC DIS 27400 \cite{Guidelines2021}, are important. These standards, along with consumer IoT cybersecurity baseline European Telecommunications Standards Institute (ETSI) EN 303 645 \cite{Cyber2020}, reinforce framework for trustworthy SL solutions in EI.}

\subsubsection{Computing}
% Modern wireless networks can employ either digital or analog modulation for computing functionalities. With digital modulation becoming predominant, MEC emerges as a pivotal solution for deploying EI systems \cite{DBLP:journals/comsur/WangHLNYC20}. Recognizing its potential, the ETSI ISG MEC has taken strides to standardize an system conducive to both edge-aware and edge-unaware applications, enriching the foundation for SL in EI applications \cite{Harmonizing-standards}. Meanwhile, 3GPP 5G specifications have set the groundwork for seamless EC, with designs that facilitate MEC's collabaration with 5G systems \cite{3rd-generation}. As 6G looms on the horizon, opportunities to unify communication and computing arise, merging MEC and 6G into a consolidated system optimized for EI. A future challenge remains in integrating AirComp functionalities for edge training in current networks, emphasizing the importance of marrying digital and analog modulation \cite{DBLP:journals/jsac/ShafiMSHZSTBW17}. The emphasis on SL, seamlessly incorporated into this evolving framework, will be instrumental in maturing EI systems, ensuring that they can adapt and evolve with social trends and user interactions.
In modern wireless networks, the shift towards digital modulation has propelled MEC as a vital component in deploying EI systems \cite{DBLP:journals/comsur/WangHLNYC20}. The ETSI ISG MEC aims to standardize a system conducive to EC, laying the foundation for SL in EI \cite{Harmonizing-standards}. The 3GPP 5G specifications lay the groundwork for EC integration \cite{3rd-generation}, and the advancement toward 6G opens prospects for unifying communication and computing, potentially integrating AirComp functionalities for enhanced edge training \cite{DBLP:journals/jsac/ShafiMSHZSTBW17}. SL's adaptive and responsive nature will be crucial in these evolving systems, ensuring their alignment with social trends and user interactions.

\subsection{{Platforms for Implementing SL in EI}}
\subsubsection{Software}
% With the advancement of EI, a burgeoning suite of software platforms has emerged to simulate and operationalize EI algorithms. While frameworks like TensorFlow Federated and PySyft lay foundational grounds for FL simulations, FedML \cite{DBLP:journals/corr/abs-2007-13518} takes it a step further, offering a versatile library to push the boundaries of FL in varied computing settings. FATE \cite{Fate}, developed by Webank’s AI Department, bridges the research-production gap, for the financial sector, harnessing secure computing and FL architectures. Additionally, EC frameworks like `KubeEdge' and AI-driven IoT solutions from Microsoft and NVIDIA enable real-time EI applications in sectors including retail and healthcare. With HarmonyOS \cite{HarmonyOS}, Huawei anchors EI in operating systems, allowing smoother interactions among smart devices. As we thread these advancements together, the integration of SL becomes paramount, ensuring that EI systems are continually refined through user interactions and social inputs.
The advancement of EI has pushed the development of a variety of software platforms to simulate and operationalize EI algorithms. Frameworks like TensorFlow Federated and PySyft provide the foundational groundwork for EI simulations. Platforms such as FedML \cite{DBLP:journals/corr/abs-2007-13518}, FATE \cite{Fate}, KubeEdge, and HarmonyOS \cite{HarmonyOS} extend these capabilities, offering versatile libraries and secure computing architectures for a broad range of EI applications, from retail to healthcare. 
{\color{black}For instance, the FedML platform offers a comprehensive framework for FL, supporting a wide range of algorithms, models, and datasets. It enables researchers and practitioners to easily develop and deploy FL solutions in EI environments, accelerating the adoption of SL in real-world applications. The platform's modular architecture and extensive documentation make it accessible to users with varying levels of expertise, fostering collaboration and innovation in the field of SL and EI.}
These platforms play a crucial role in bridging the research-production gap and enhancing real-time EI applications, setting the stage for the integration of SL, which further refines EI through continual user interaction and social inputs.

\begin{figure}[pt]
  \centering
  \includegraphics[width=\linewidth]{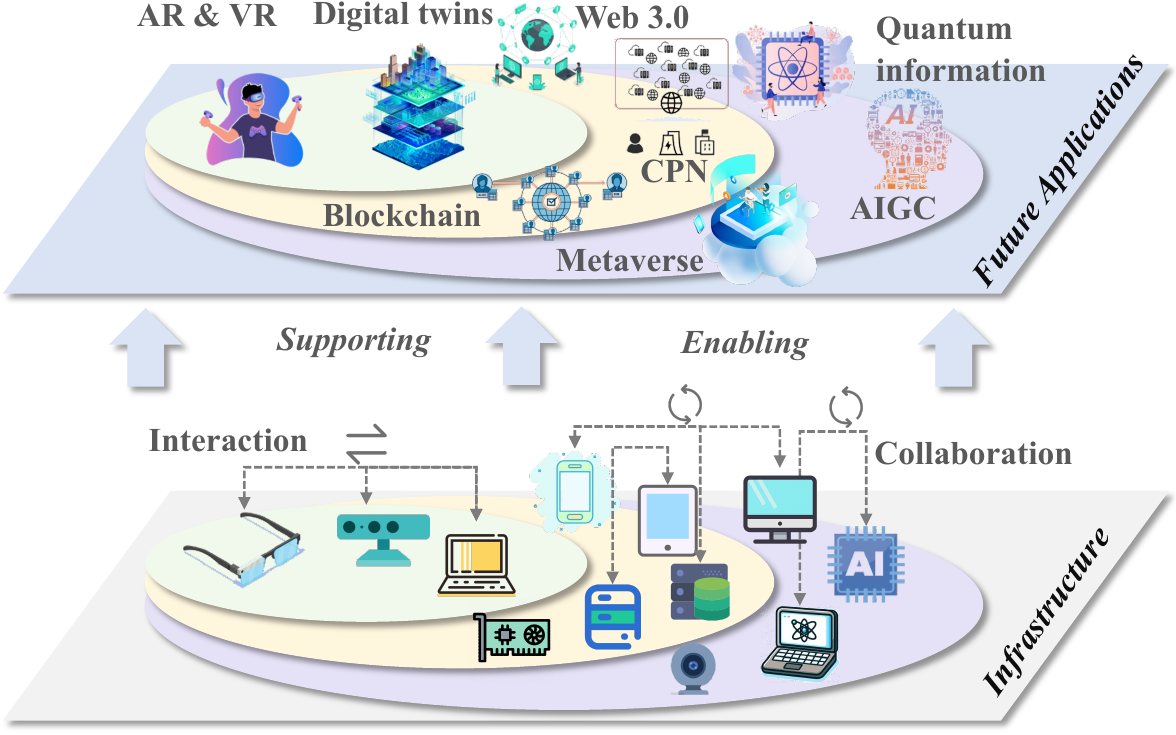}   
  \caption{Future applications of SL in EI.}\label{Applications of SL in EI}
  % \vspace{-0.5em}
\end{figure}

\subsubsection{Solver}
% As EI systems and wireless networks grow, resource allocation optimization is increasingly important. Tools like CVX \cite{CVX2014} facilitate large-scale convex optimization, converting problems into standard conic forms before utilizing robust solvers like MOSEK \cite{Mosek}. A notable advancement, the matrix stuffing technique \cite{DBLP:journals/tsp/Shi0OL15}, has expedited the modeling process by symbolically mapping problems to their conic forms, suggesting potential for symbolic conic transformation solvers. Moreover, Gurobi \cite{Gurobi} and MOSEK remain at the forefront for tackling mixed-integer conic programs. As we integrate these solvers into EI platforms, it becomes pivotal to embed SL paradigms, ensuring EI systems adapt based on user interactions. An illustration of this notion is ``Open-L2O” \cite{DBLP:journals/corr/abs-2103-12828}, blending the ``learning to optimize" framework to both assess and automatically craft algorithms, cementing the symbiotic relationship between solvers and SL in EI.
In the expanding landscape of EI and wireless networks, resource allocation optimization tools become increasingly critical. Tools like CVX \cite{CVX2014}, MOSEK \cite{Mosek}, and Gurobi \cite{Gurobi} are pivotal for large-scale convex optimization and handling mixed-integer conic programs. Techniques like the matrix stuffing technique \cite{DBLP:journals/tsp/Shi0OL15} accelerate the modeling process, enhancing the efficiency of symbolic conic transformation solvers. The integration of these solvers with SL paradigms on EI platforms is essential. An example is the ``Open-L2O” framework \cite{DBLP:journals/corr/abs-2103-12828}, which blends learning to optimize algorithms, showcasing how SL can be instrumental in adapting EI based on user interactions and feedback.

\subsubsection{Hardware}
% The efficiency of EI systems lies in the synergy between EI computing hardware and RF technologies. Computing hardware variants, from NVIDIA's GPUs to Google's TPUs, cater to specific EI needs, with comprehensive reviews available \cite{DBLP:journals/comsur/WangHLNYC20}. A notable stride in EI hardware design is the deep RL-assisted chip floorplanning, optimizing hardware for advanced EI tasks \cite{mirhoseini2021graph}. Alongside, RIS-enhanced FL systems stand out due to their adeptness at manipulating electromagnetic waves, fundamentally shaping the potential of EI platforms \cite{DBLP:journals/jsac/RenzoZDAYRT20}. The envisioned THz communication is poised to revolutionize EI by unifying sensing, communication, and learning \cite{amakawa2021white}. As we advance these technologies, integrating SL becomes crucial. SL ensures that as hardware evolves, systems remain adaptive, self-learning, and user-centric, resulting in continuously enhanced EI applications.
EI's performance hinges on the interplay between specialized computing hardware like NVIDIA's GPUs and Google's TPUs and RF technologies \cite{DBLP:journals/comsur/WangHLNYC20}. Advances such as deep RL-assisted chip floorplanning \cite{mirhoseini2021graph} and IRS-enhanced FL systems \cite{DBLP:journals/jsac/RenzoZDAYRT20} highlight the evolution in EI hardware design. Additionally, emerging THz communication \cite{amakawa2021white} is set to unify sensing, communication, and learning, further propelling EI capabilities. Incorporating SL is essential to ensuring that EI remains adaptable and user-centric, continually improving its efficiency and responsiveness.

\section{Future Applications and Future Research}\label{section9}
\subsection{{Future Applications of SL in EI}}\label{section9.A}
{\color{black}In Section \ref{I-C}, we provided a concise overview of the current application scenarios of SL in EI, highlighting the common characteristics and potential benefits across various domains. This section provides an in-depth analysis of SL in EI's \textbf{future applications}, exploring emerging domains and technologies. We comparatively study these application scenarios, discussing their unique requirements, challenges, and opportunities.
}
As illustrated in Fig. \ref{Applications of SL in EI}, the integration of SL into EI has an impact on multiple future application scenarios. This section examines various uses and offers guidance on how to put them into practice.

{\color{black}
\textbf{Digital Twins.}
% Digital twins are digital representations of physical entities that simulate real-world scenarios to aid in optimization. Cloud-edge collaborations, e.g., DITEN, enhance this by merging real and virtual systems \cite{lu2020communication}. Incorporating SL imbues digital twins with adaptability, enabling scenario-based evolution. Lu \emph{et al.}'s DTWN offloads data tasks with digital twins' help \cite{lu2020low}. Sun \emph{et al.}'s approach uses twins for task offloading; with SL, these twins could offer more adaptive insights \cite{sun2020reducing}. Similarly, a digital twin-enabled space-air network introduces potential for SL in network depictions \cite{sun2021dynamic}. In Bellavista \emph{et al.}'s work on resource allocation, SL guides digital twins in anticipating social needs, enhancing allocation efficiency \cite{bellavista2021application}.%数字孪生是物理实体的数字表示，可模拟现实世界的场景，帮助优化。云边合作（如 DITEN）通过合并真实和虚拟系统（cite{lu2020communication}）增强了这一点。将 SL 纳入数字孪生系统可赋予其适应性，从而实现基于场景的演进。Lu 的 DTWN 在数字双胞胎的帮助下卸载了数据任务。Sun ＆emph{et al.}的方法使用双胞胎进行任务卸载；通过 SL，这些双胞胎可以提供更具适应性的见解 \cite{sun2020reducing}。同样，数字双胞胎支持的空间-空气网络也为 SL 在网络描绘中的应用带来了潜力（cite{sun2021dynamic}）。在贝拉维斯塔（Bellavista \emph{et al.}）关于资源分配的研究中，SL 引导数字孪生预测社会需求，提高了分配效率（cite{bellavista2021application}）。
%20240630
Digital twins, as digital representations of physical entities, are evolving to incorporate SL for enhanced adaptability and optimization. Implementing complex digital twin models efficiently on resource-constrained edge devices is a key challenge. Future research should focus on developing lightweight, yet high-precision physical models and distributed training techniques. For instance, Lu \emph{et al.}'s DTWN \cite{lu2020low} could be extended to incorporate FL, enabling collaborative model updates across multiple edge nodes while preserving data privacy. Sun \emph{et al.}'s task offloading approach \cite{sun2020reducing} could be enhanced with RL-based decision-making to dynamically optimize offloading strategies based on network conditions and task characteristics. Future work in space-air networks \cite{sun2021dynamic} could explore hierarchical digital twins leveraging SL for multi-level optimization from individual device management to global network orchestration. Bellavista \emph{et al.}'s resource allocation research \cite{bellavista2021application} could be expanded to include predictive analytics, utilizing SL to anticipate and proactively adjust to resource demands.

\textbf{Web 3.0.}
% EI and Web 3.0 together drive decentralized applications (DApps) creation on the IoT edge \cite{lin2023unified}. A key feature of this fusion is real-time interaction through smart contracts and autonomous agents at the network's edge. Applied to Web 3.0-enhanced EI, SL can amplify the benefits. For instance, integrating SL into DApps development enables systems to adapt dynamically to learned behaviors and interactions, optimizing the user experience. Web 3.0's decentralization and SL's dynamic learning ensure data security and optimal use in EI environments prioritizing privacy \cite{10269669}. Industrial automation can leverage SL to optimize real-time machine-to-machine interactions, maintaining Web 3.0's robustness and reliability. Similarly, personalized healthcare can rapidly evolve. With SL-integrated EI in Web 3.0, health applications can learn from individual and global health data, providing tailored, efficient, and timely solutions.%EI和Web 3.0共同推动了物联网边缘去中心化应用（DApps）的创建。这种融合的一个关键特征是通过网络边缘的智能合约和自主代理进行实时交互。将 SL 应用于 Web 3.0 增强型 EI，可以放大其优势。例如，将 SL 集成到 DApps 开发中，可使系统动态适应所学行为和互动，优化用户体验。Web 3.0 的去中心化和 SL 的动态学习可确保数据安全，并在优先考虑隐私的 EI 环境中实现最佳使用（cite{10269669}）。工业自动化可以利用 SL 来优化机器与机器之间的实时交互，同时保持 Web 3.0 的稳健性和可靠性。同样，个性化医疗保健也可以快速发展。有了 Web 3.0 中的 SL 集成 EI，医疗应用可以从个人和全球健康数据中学习，提供量身定制、高效和及时的解决方案。
%20240630
The integration of EI and Web 3.0 promotes the creation of decentralized applications (DApps) on IoT edges \cite{lin2023unified}, which poses unique challenges for SL. Developing efficient consensus mechanisms in decentralized environments is a critical issue. Future research should investigate SL-enhanced consensus algorithms, like reputation-based systems where node credibility hinges on SL performance. Integrating SL in DApp development could enable adaptive smart contracts that evolve based on user interactions and network conditions. This necessitates novel contract languages that handle probabilistic and learning-based clauses. Privacy-preserving SL techniques require optimization on resource-constrained edge devices to ensure data security in EI environments \cite{10269669}. Research in industrial automation should focus on real-time SL algorithms for machine-to-machine interactions, employing online learning techniques to adapt to changing production conditions while upholding the robustness of Web 3.0.

\textbf{Computing Power Network (CPN).}
%The growth of AI emphasizes the need for an efficient CPN. EI's integration into CPN promises ultra-low latency and high bandwidth, especially when deployed near the base station \cite{DBLP:journals/pieee/ZhouCLZLZ19}. This positioning reduces end-user latency by bridging the gap between content sources and users. The union of SL and CPN within EI maximizes its potential. Through SL, the system can dynamically modify computational approaches, learning from node interactions. As industries within the Telecom Network seek customization \cite{DBLP:journals/tc/WangLWDZJ21}, EI with SL's algorithms can champion collaborative innovations. Moreover, the ETSI-defined EI platform's open APIs can harmoniously work with SL, making real-time adjustments from diverse data points. As CPN evolves towards the 6G micro-edge era, SL ensures smooth integrations between EI sites.
%20240630
Integrating EI into CPN presents optimization challenges for achieving ultra-low latency and high bandwidth, especially near base stations \cite{DBLP:journals/pieee/ZhouCLZLZ19}. Future studies should develop dynamic resource allocation algorithms that adjust in real time to SL demands. This approach requires predictive resource allocation techniques that utilize learned SL task patterns. Designing distributed learning algorithms that efficiently utilize heterogeneous computing resources across the CPN is a key challenge. As industries demand customization \cite{DBLP:journals/tc/WangLWDZJ21}, research must explore adaptive SL algorithms that dynamically adjust to varying computing resources and network conditions. The shift towards 6G micro-edge architectures demands novel SL frameworks that operate seamlessly across diverse EI sites, leveraging FL and others to balance data locality with global model optimization.

\textbf{Blockchain.}
% The integration of Blockchain and EI has synergized to optimize networking, computing, storage, and security in IoT \cite{8624417,huang2022blocksense}. Introducing SL to these systems offers tangible benefits. Liang \emph{et al.} leveraged blockchain with EI for obstacle detection in train autonomous circumambulate system \cite{DBLP:journals/tii/LiangZYM24}. Incorporating SL harnesses collective insights from blockchain interactions to personalize and enrich learning. Sharma \emph{et al.} emphasized IoT security enhancement through blockchain \cite{DBLP:journals/isci/SharmaNCFT23}. SL promotes collaborative security by predicting threats based on historical node interactions. Similarly, Desai \emph{et al.}'s objective of enhanced device responsiveness in secured systems \cite{desaimeasuring} can benefit from SL, which uses past blockchain transactions to anticipate and mitigate bottlenecks.%区块链与 EI 的整合协同优化了物联网中的网络、计算、存储和安全（cite{8624417,huang2022blocksense}。将 SL 引入这些系统可带来实实在在的好处。Liang \emph{et al.}利用区块链与EI，在列车自主环行系统中进行障碍物检测（\cite{DBLP:journals/tii/LiangZYM24}）。结合 SL 利用区块链交互的集体洞察力来个性化和丰富学习。Sharma \emph{et al.}强调了通过区块链增强物联网安全性（cite{DBLP:journals/isci/SharmaNCFT23}）。SL 根据历史节点互动预测威胁，从而促进协作安全。同样，Desai 等人提出的在安全系统中提高设备响应速度的目标也可以从 SL 中受益，SL 利用过去的区块链交易来预测和缓解瓶颈。
%20240630
Although the integration of Blockchain and EI optimizes IoT systems \cite{8624417,huang2022blocksense}, it presents challenges in efficiency and scalability. Future studies should develop SL-enhanced consensus mechanisms tailored for resource-constrained edge devices. Liang \emph{et al.}'s work on obstacle detection in train systems \cite{DBLP:journals/tii/LiangZYM24} could adopt FL to enable collaborative model updates while preserving data privacy across distributed nodes. Designing SL algorithms that leverage blockchain's immutability for secure, decentralized learning presents a key challenge. Building on Sharma \emph{et al.}'s IoT security framework \cite{DBLP:journals/isci/SharmaNCFT23}, future research should explore using distributed SL in blockchain-based anomaly detection to predict threats from historical node interactions. To enhance scalability in Desai \emph{et al.}'s system \cite{desaimeasuring}, research should focus on combining sharding techniques with SL to boost throughput in large-scale IoT networks.

\textbf{AIGC.}
Integrating SL into AIGC enhances its ability to adaptively learn from user interactions and collective insights, personalizing content, but also introduces unique challenges beyond traditional approaches \cite{tu2021ugc}. Future research should focus on SL-enhanced generative models that adapt to collective user behaviors at the edge. Building on Xu \emph{et al.}'s multi-layered framework \cite{xu2023unleashing}, developing SL algorithms that facilitate collaborative learning of GANs and VAEs across distributed edge nodes is necessary. Designing privacy-preserving SL techniques for personalized content generation, which potentially leverage FL with differential privacy, is a critical challenge. Research should focus on exploring SL-driven adaptive model partitioning that dynamically allocates tasks between different layers, based on social patterns and device capabilities. Furthermore, investigating SL-based content optimization techniques at the edge for various modalities, such as text \cite{li2018seq2seq}, images \cite{ren2021combiner}, and videos \cite{loeschcke2023text}, while balancing quality, latency and resource constraints, is crucial.

\textbf{Quantum Information.}
% With the help of qubits, quantum information uses the laws of quantum physics to increase processing capacity, which may be used to raise EI in a variety of ways. For instance, quantum computing can refine AI algorithms in EI by processing vast amounts of data, thus improving predictive accuracy. Quantum computing enhances AI algorithms in EI, improving predictive accuracy and resource optimization, especially in path planning \cite{li2020quantum}. This is particularly relevant in evolving quantum AI and in path optimizations. To further fortify EI, quantum communication safeguards edge interactions, establishing unparalleled data security \cite{DBLP:journals/tvt/NarottamaS23}.%在量子比特的帮助下，量子信息利用量子物理定律来提高处理能力，这可用于以多种方式提高 EI。例如，量子计算可以通过处理海量数据完善 EI 中的人工智能算法，从而提高预测准确性。量子计算可以增强 EI 中的人工智能算法，提高预测准确性和资源优化，尤其是在路径规划方面（cite{li2020quantum}）。这与量子人工智能的发展和路径优化尤为相关。为了进一步强化人工智能，量子通信保障了边缘交互，建立了无与伦比的数据安全性（cite{DBLP:journals/tvt/NarottamaS23}。
Integrating quantum computing principles with SL in EI presents unique challenges for enhancing distributed learning and security. Research should prioritize developing quantum-inspired SL algorithms to enhance collaborative learning efficiency in edge networks. For path planning \cite{li2020quantum}, quantum-inspired optimization techniques could enable edge devices to collaboratively solve complex problems more efficiently. Designing privacy-preserving SL protocols that leverage quantum-resistant cryptography to ensure long-term security against advancing quantum capabilities remains a key challenge. Investigating quantum key distribution (QKD) for securing SL data exchanges in edge networks is crucial \cite{DBLP:journals/tvt/NarottamaS23}. Future work should aim to integrate quantum-safe cryptography within SL frameworks, focusing on scalability and resilience in real-world edge environments.

\textbf{Metaverse.} 
% The Metaverse merges the virtual and real worlds, utilizing technologies like 5G, AI, and VR. It offers users experiences rooted in self-defined identities, socialization, entertainment, and learning \cite{10373900}. As the complexities of the metaverse grow, there is a trend of moving from cloud to EC to achieve better performance and lower latency. Younis \emph{et al.} pioneered this by integrating edge-cloud architecture into the metaverse, optimizing bandwidth and ensuring data security \cite{younis2020latency}.  
% Integrating SL can refine the EI-driven Metaverse. Using SL, the metaverse evolves based on user interactions, fostering an adaptable, virtual environment.
The Metaverse, which merges virtual and real worlds through 5G, AI, and VR technologies, presents unique challenges for EI and SL integration \cite{10373900}. As Metaverse complexity grows, the trend towards EC becomes crucial for improving performance and reducing latency. Extending Younis \emph{et al.}'s edge-cloud architecture \cite{younis2020latency}, future research should focus on SL-enhanced bandwidth optimization and data security in Metaverse environments. Key challenges include developing adaptive SL algorithms for dynamic edge-cloud resource allocation based on real-time user interactions and network conditions. Research into SL techniques that efficiently synchronize heterogeneous data types (visual, auditory, haptic) across distributed edge nodes in the Metaverse is needed. Furthermore, exploring SL-driven content adaptation mechanisms is crucial for developing virtual environments that evolve in response to collective user behaviors.

\textbf{AR and VR.}
% AR and VR revolutionize user interaction with the digital domain. AR blends the real and virtual, while VR immerses users in the virtual. Their need for low-latency interactions is addressed by edge cloud technology, which significantly bolsters their performance \cite{DBLP:journals/tmm/ZhangPLY23}. Mobile AR (MAR) emphasizes on-the-go access \cite{haynes2018mobile}. Given the intense data requests and latency sensitivity of MAR, hybrid architectures merging cloud, edge, and device components prove effective. In this setup, edges focus on real-time MAR engagement, with clouds handling complex, low-latency tasks \cite{yang2023cloud}. Introducing SL to EI in AR/VR fosters real-time adaptation, refining content based on interactions and feedback.%AR 和 VR 彻底改变了用户与数字领域的交互方式。AR融合了现实与虚拟，而VR则让用户沉浸在虚拟之中。边缘云技术满足了他们对低延迟交互的需求，大大提高了他们的性能（cite{DBLP:journals/tmm/ZhangPLY23}）。移动 AR（MAR）强调随身访问（cite{haynes2018mobile}）。鉴于移动 AR 的高强度数据请求和延迟敏感性，融合云、边缘和设备组件的混合架构被证明是有效的。在这种设置中，边缘侧重于实时 MAR 参与，云则处理复杂的低延迟任务。在AR/VR中将SL引入EI可促进实时适应，根据互动和反馈完善内容。
AR/VR require low-latency interactions, which are facilitated by edge-cloud technologies \cite{DBLP:journals/tmm/ZhangPLY23}. Future studies should develop SL-enhanced task allocation algorithms for hybrid cloud-edge-device architectures, focusing on Mobile AR \cite{haynes2018mobile}. Designing adaptive SL models that dynamically allocate computational tasks between edge and cloud, responding to real-time user interactions and network conditions, is a major challenge. Given intense data demands and latency sensitivity, investigating SL techniques for predictive content caching and rendering at the edge is crucial \cite{yang2023cloud}. Research into collaborative SL algorithms that facilitate real-time content adaptation across multiple AR/VR devices, leveraging collective user interactions to create personalized experiences, is needed. Furthermore, exploring SL-driven compression techniques for data streams could reduce bandwidth requirements while preserving QoE.
}

{\color{black}
\textbf{Comparative Analysis of SL in EI Application Scenarios.} 
Integrating SL in EI across various domains highlights both commonalities and unique challenges. While \emph{collaborative learning, adaptability, privacy preservation and others} are core principles of SL in EI, their implementation varies significantly depending on application-specific requirements. \emph{Data characteristics} and \emph{real-time processing demands} vary significantly across domains: digital twins and metaverse applications process complex, multi-modal data streams, whereas Web 3.0 and blockchain systems handle distributed, often encrypted data. This variability necessitates developing adaptive SL frameworks that can efficiently handle heterogeneous data types and processing requirements.

\emph{Device heterogeneity} presents another critical challenge, as the computational capabilities range from ESs in CPNs to resource-constrained IoT devices in AIGC scenarios. Future research should develop scalable SL algorithms capable of dynamically adjusting to diverse device capabilities, utilizing techniques like model compression, adaptive quantization, and intelligent task offloading. \emph{The objectives of SL} also differ across applications: AR/VR systems prioritize low-latency content adaptation, whereas quantum-enhanced EI focuses on secure, distributed computation.

Despite these differences, it is crucial to maintain SL's core principles across all applications. Future research should investigate unified SL frameworks that can adapt to diverse scenarios while ensuring collaborative learning, system adaptability, and privacy preservation. This approach could involve developing modular SL architectures with interchangeable components tailored to various application requirements, or exploring meta learning techniques that rapidly adapt to new domains with minimum fine-tuning. By addressing these challenges, SL in EI can evolve to meet the diverse needs of future applications while maintaining its fundamental advantages in collaborative, adaptive, and secure distributed learning.
}

{\color{black}
\subsection{{Open Problems and Future Research}}
% A wide spectrum of technologies has been proposed for addressing the various challenges to combining SL and EI. Although exciting and encouraging progress has been made, the research area of combined SL and EI still has some open problems that deserve more thorough investigations, which offers interesting topics for future research.
% The integration of SL and EI presents a broad spectrum of technologies addressing various challenges. 
Despite significant progress in integrating SL in EI, there remain open problems in this research area, offering directions for future exploration.

\subsubsection{\textbf{Strengthen Collaborative Decision-making}}
{\color{black}
One of the key challenges in integrating SL with EI is the enhancement of collaborative decision-making mechanisms. Future research should aim to develop SL algorithms capable of effectively simulating complex social decision-making processes. For instance, in smart city traffic management, MARL algorithms incorporating graph neural networks could effectively model collective vehicle and pedestrian behaviors. Privacy-preserving techniques are crucial. In healthcare, for example, FL with secure aggregation protocols and differential privacy could enable collaborative model training across hospitals without compromising patient data. Adaptive meta learning approaches, such as MAML, should be investigated for rapid policy adaptation in dynamic industrial settings. These advanced decision models, which balance individual privacy with collective intelligence, could significantly enhance the effectiveness of EI systems across various domains, from optimizing energy consumption in smart grids to enhancing robotic control in manufacturing.
}

\subsubsection{\textbf{Optimize Resource Allocation and Management}}
{\color{black}
Optimizing resource allocation and management in EI using SL principles remains a critical challenge. Future research should concentrate on developing SL-inspired strategies that leverage collective intelligence to dynamically optimize resources. For instance, in large-scale IoT networks, social influence-aware FL could optimize resource allocation across heterogeneous devices. By incorporating social rules analysis techniques, this approach could model device interactions and information propagation, enabling more efficient collaborative learning and resource sharing. Another promising direction is the development of socialized MARL algorithms for adaptive edge caching. These algorithms could learn from user behavior patterns and social connections to predict content popularity and optimize cache placement, significantly reducing latency and bandwidth usage. Moreover, addressing ethical concerns in resource allocation necessitates the development of fairness-aware SL algorithms that incorporate social choice theory concepts to ensure equitable resource allocation and respect for individual device preferences and capabilities.}

\subsubsection{\textbf{Enhance Collaborative Learning and Knowledge Sharing}}
{\color{black}
Advancing collaborative learning and information exchange within SL and EI is vital for creating an engaging and inclusive learning environment. Future research should develop advanced SL algorithms and platforms to facilitate seamless collaboration in EI. Socialized knowledge graphs incorporating social relationships and expertise levels can dynamically represent collective intelligence across edge devices, using adaptive attention mechanisms to prioritize information flow based on trust and relevance. Multi-agent transfer learning frameworks should be explored to promote active participation and diverse knowledge integration. This approach enables devices to share and adapt knowledge across domains based on social roles, using meta learning for rapid adaptation. Decentralized consensus algorithms inspired by social choice theory allow fair aggregation of diverse perspectives from edge devices while respecting individual preferences. Focusing on these SL-based algorithms and platforms can foster an engaging and inclusive learning environment, enhancing interaction and knowledge exchange within SL-enhanced EI systems.}

{\color{black}
\subsubsection{\textbf{Ensuring Scalability and Interoperability}}
%As EI systems continue to grow in complexity and heterogeneity, ensuring the scalability and interoperability of SL algorithms and frameworks becomes a critical challenge. Future research should discuss the challenges and research directions for developing SL algorithms and frameworks that can scale efficiently with the growing complexity and heterogeneity of EI systems. Researchers should explore innovative approaches for seamless interoperability between SL models and EI platforms, facilitating collaborative learning and decision-making across domains. Additionally, investigating techniques for optimizing the performance and resource utilization of SL algorithms in large-scale EI deployments is essential for their practical implementation. By addressing these challenges, future research can enable the effective integration of SL with EI systems, regardless of their scale or complexity.

Ensuring the scalability and interoperability of SL algorithms is critical as EI systems grow in complexity and heterogeneity. Future research should develop adaptive SL frameworks that dynamically adjust to the varying system scales and capabilities of devices. For instance, it is warranted to explore hierarchical SL models where edge devices form local social clusters for efficient knowledge sharing. Higher-level nodes could manage inter-cluster communication, potentially modeled by graph neural networks. To address interoperability, universal SL interfaces leveraging semantic web technologies could facilitate seamless integration across EI platforms, using ontology-based representations of SL processes. To optimize performance in large-scale deployments, developing incremental SL algorithms that allow continuous model updates without full retraining is essential. Exploring socialized load-balancing strategies, where tasks are distributed based on social roles and connectivity patterns and dynamically adapted using the RL algorithm, is also crucial.

\subsubsection{\textbf{Enhance Privacy Protection and Data Security}}

Addressing privacy protection and data security is crucial for integrating SL with EI. Future research should focus on devising innovative strategies to enhance privacy and security while preserving the collaborative nature of SL. These strategies should effectively safeguard personal data, employing advanced methods like context-aware encryption schemes that adapt based on the current learning phase and data importance. Exploring complex anonymization algorithms, including multi-dimensional privacy preservation techniques that protect both individual data and collective learning patterns, is necessary. Implementing robust access control mechanisms, such as adaptive multi-factor authentication systems based on the device's learning behavior and contribution patterns, is essential. These strategies should efficiently process data in dynamic EI environments to balance utility and privacy. Ensuring EI systems implementing SL are intelligent, collaborative, secure, and privacy-preserving is crucial for building trust and encouraging wider adoption in various applications.}
\section{Conclusion}\label{section10}
In this paper, we have explored the integration of SL and EI, two emerging paradigms that promise to revolutionize the fields of AI and EC. Specifically, we have discussed the background and fundamentals of SL, drawing inspiration from human social interactions and behaviors. We have also reviewed the existing concepts and paradigms related to SL, highlighting its distinctions and overlaps. Moreover, we have motivated the combination of SL and EI from two perspectives: how SL can address challenges faced by EI, and how EI can provide a suitable platform for SL implementation. To further illustrate this integration, we have presented three key aspects: socialized architecture, socialized training and socialized inference. Finally, we have investigated some future applications and tutorials of SL in EI. We have also discussed some open problems and identified future research directions on combining SL and EI. We believe that SL in EI has the potential to create more intelligent, adaptive, and robust systems that benefit various domains and applications.

% \vspace{-1em}
\bibliographystyle{IEEEtran}
\bibliography{IEEEabrv,mylib}

\begin{IEEEbiography}[{\includegraphics[width=1in,height=1.25in,clip,keepaspectratio]{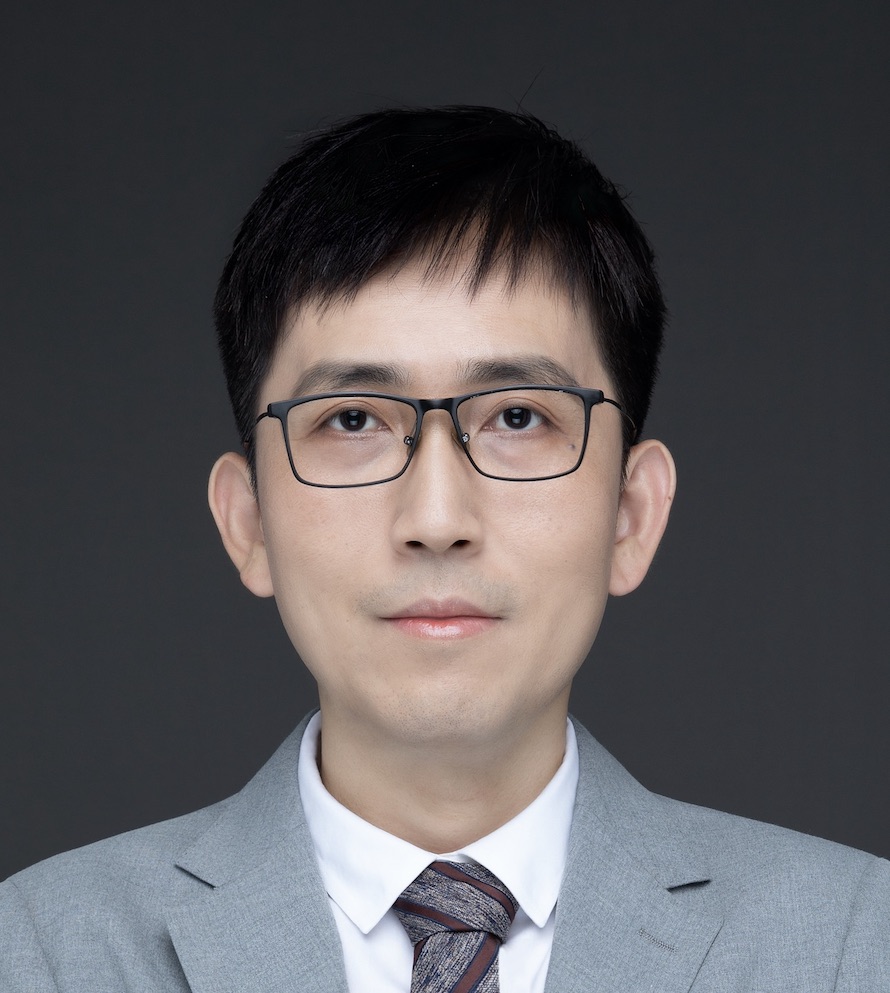}}]{Xiaofei Wang}
(Senior Member, IEEE) received the B.S. degree from Huazhong University of Science and Technology, China, and received M.S. and Ph.D. degrees from Seoul National University, Seoul, South Korea. He was a Postdoctoral Fellow with The University of British Columbia, Vancouver, Canada, from 2014 to 2016. He is currently a Professor with the College of Intelligence and Computing, Tianjin University, Tianjin, China. Focusing on the research of edge computing, edge intelligence, and edge systems, he has published more than 200 technical papers in IEEE JSAC, TCC, ToN, TWC, IoTJ, COMST, TMM, INFOCOM, ICDCS and so on. He has received the best paper awards of IEEE ICC, ICPADS, and in 2017, he was the recipient of the "IEEE ComSoc Fred W. Ellersick Prize", and in 2022, he received the "IEEE ComSoc Asia-Pacific Outstanding Paper Award".
\end{IEEEbiography}

\begin{IEEEbiography}
[{\includegraphics[width=1in,height=1.25in,clip,keepaspectratio]{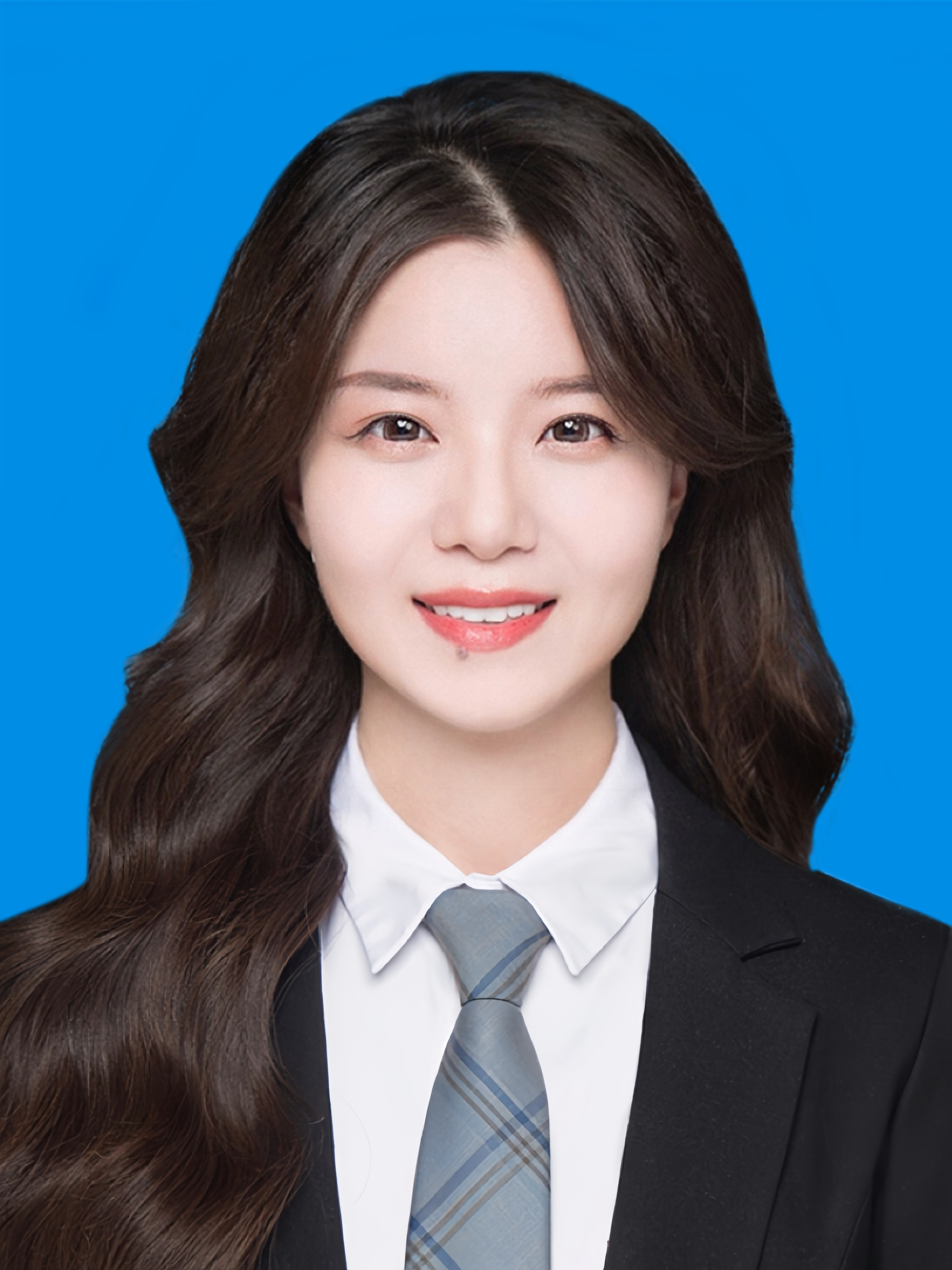}}]{Yunfeng Zhao} 
(Student Member, IEEE) received her B.S. degree from the College of Science, Nanjing Agricultural University, Jiangsu, China, in 2017, and the M.S. degree from the School of Mathematics, Tianjin University, Tianjin, China, in 2020.
She is currently pursuing the Ph.D. degree with the College of Intelligence and Computing, Tianjin University, Tianjin, China.
From October 2023 to September 2024, she visited Nanyang Technological University, Singapore, as a Visiting Scholar. 
Her current research interests include edge computing, edge intelligence, and distributed machine learning. %She received the B.S. National Scholarship of China in 2015, the Outstanding B.S. Graduates in 2017 and M.S. National Scholarship of China in 2019.
\end{IEEEbiography}

\begin{IEEEbiography}
[{\includegraphics[width=1in,height=1.25in,clip,keepaspectratio]{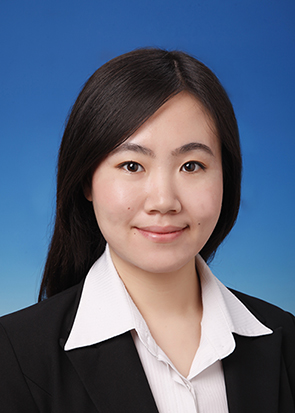}}]{Chao Qiu}
(Member, IEEE) received her B.S. degree in communication engineering from China Agricultural University, in 2013, and her Ph.D. in information and communication engineering from the Beijing University of Posts and Telecommunications, in 2019. She is currently an Associate Professor with the College of Intelligence and Computing, Tianjin University, China. From September 2017 to September 2018, she visited Carleton University, Ottawa, ON, Canada, as a Visiting Scholar. Her current research interests include edge computing, edge intelligence, and blockchain. 
\end{IEEEbiography}

\begin{IEEEbiography}
[{\includegraphics[width=1in,height=1.25in,clip,keepaspectratio]{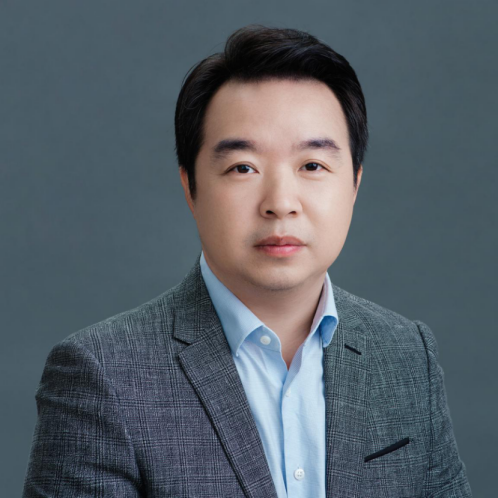}}]{Qinghua Hu}
(Senior Member, IEEE) received the B.S., M.S., and Ph.D. degrees from the Harbin Institute of Technology, Harbin, China, in 1999, 2002, and 2008, respectively. After that, he joined the Department of Computing at the Hong Kong Polytechnical University, Hong Kong, as a Postdoctoral Fellow. He was a Full Professor with Tianjin University, Tianjin, China, in 2012, where he is currently a Chair Professor and a Deputy Dean with the College of Intelligence and Computing. He is also with the Engineering
Research Center of City Intelligence and Digital Governance, Ministry of Education of the People’s Republic of China, Tianjin, and Haihe Lab of ITAI, Tianjin. His research interest is focused on uncertainty modeling, multimodality learning, incremental learning, and continual learning these years, funded by the National Natural Science Foundation of China and the National Key Research and Development Program of China. He has published more than 300 peer-reviewed papers in IEEE TRANSACTIONS ON KNOWLEDGE AND DATA ENGINEERING, IEEE TRANSACTIONS ON PATTERN ANALYSIS AND MACHINE INTELLIGENCE, and IEEE TRANSACTIONS ON NEURAL NETWORKS AND LEARNING SYSTEMS. Dr. Hu was a recipient of the Best Paper Award of ICMLC 2015 and ICME 2021. He is an Associate Editor of the IEEE TRANSACTIONS ON FUZZY SYSTEMS, Acta Automatica Sinica, and Acta Electronica Sinica.
\end{IEEEbiography}

\begin{IEEEbiography}[{\includegraphics[width=1in,height=1.25in,clip,keepaspectratio]{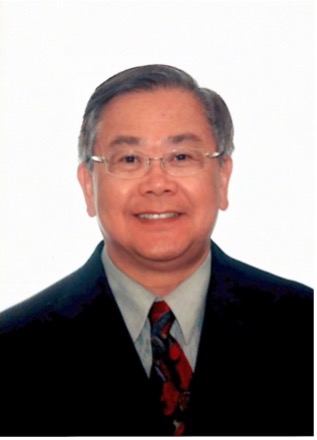}}]{Victor C. M. Leung} (Life Fellow, IEEE) is the Dean of the Artificial Intelligence Research Institute and a Professor of Engineering at Shenzhen MSU-BIT University, China, a Distinguished Professor of Computer Science and Software Engineering at Shenzhen University, China, and an Emeritus Professor of Electrical and Computer Engineering and Director of the Laboratory for Wireless Networks and Mobile Systems at the University of British Columbia (UBC), Canada. His research is in the broad areas of wireless networks and mobile systems, and he has published widely in these areas. His published works have together attracted more than 65,000 citations. He is named in the current Clarivate Analytics list of “Highly Cited Researchers”. Dr. Leung is serving on the editorial boards of the IEEE Transactions on Green Communications and Networking, IEEE Transactions on Computational Social Systems, and several other journals. He received the 1977 APEBC Gold Medal, 1977-1981 NSERC Postgraduate Scholarships, IEEE Vancouver Section Centennial Award, 2011 UBC Killam Research Prize, 2017 Canadian Award for Telecommunications Research, 2018 IEEE TCGCC Distinguished Technical Achievement Recognition Award, and 2018 ACM MSWiM Reginald Fessenden Award. He co-authored papers that were selected for the 2017 IEEE ComSoc Fred W. Ellersick Prize, 2017 IEEE Systems Journal Best Paper Award, 2018 IEEE CSIM Best Journal Paper Award, and 2019 IEEE TCGCC Best Journal Paper Award.  He is a Life Fellow of IEEE, and a Fellow of the Royal Society of Canada (Academy of Science), Canadian Academy of Engineering, and Engineering Institute of Canada. 
\end{IEEEbiography}

\newpage
\vspace{11pt}
\vfill

\end{document}